%% file: main.tex
\documentclass[twocolumn]{aastex63}

\usepackage{natbib}
\usepackage{enumitem}
\usepackage{graphicx}
\usepackage{amsmath}
\usepackage{bm}
\usepackage{graphics}
\usepackage{url}
\usepackage{epsfig}
\usepackage{ulem}
\usepackage{verbatim}
\usepackage{float}
\usepackage{lastpage}
\usepackage{amsmath}

\def\lsim{\hbox{\rlap{\raise 0.425ex\hbox{$<$}}\lower 0.65ex\hbox{$\sim$}}}
\def\gsim{\hbox{\rlap{\raise 0.425ex\hbox{$>$}}\lower 0.65ex\hbox{$\sim$}}}

\def\arcsec{\hbox{$^{\prime\prime}$}}
\def\arcdeg{\mbox{$^\circ$}}

\newcommand{\hst}{\protect\hbox{$H\!ST$}}

\def\I{\,\textsc{i}}

\received{}
\revised{}
\accepted{}

\submitjournal{ApJ}

\shorttitle{GW170817 with {\it HST}}
\shortauthors{C. D. Kilpatrick}

\begin{document}

\title{Deep {\it Hubble Space Telescope} Observations of GW170817: Complete Light Curves and the Properties of the Galaxy Merger of NGC~4993}

\correspondingauthor{C. D. Kilpatrick}
\email{ckilpatrick@northwestern.edu}

\input{affiliation}
\input{authors}

\begin{abstract}
We present the complete set of {\it Hubble Space Telescope} imaging of the binary neutron star merger GW170817 and its optical counterpart AT~2017gfo.  Including deep template imaging in F814W, F110W, F140W, and F160W at 3.4~years post-merger, we re-analyze the full light curve of AT~2017gfo across 12 bands from 5--1273 rest-frame days after merger.  We obtain four new detections of the short $\gamma$-ray burst (GRB) 170817A afterglow from 109--170 rest-frame days post-merger.  These detections are consistent with the previously observed $\beta=-0.6$ spectral index in the afterglow light curve with no evidence for spectral evolution.  We also analyze our limits in the case of novel late-time optical and IR emission signatures, such as a kilonova afterglow or infrared dust echo, but find our limits are not constraining in these contexts.  We use the new data to construct deep optical and infrared stacks, reaching limits of $M=-6.3$ to $-4.6$~mag, to analyze the local environment around AT~2017gfo and low surface brightness features in its host galaxy NGC~4993.  We rule out the presence of any globular cluster at the position of AT~2017gfo to $2.3 \times 10^{4}~L_{\odot}$, including those with the reddest $V-H$ colors.  Finally, we analyze the substructure of NGC~4993 in deep residual imaging, and find shell features which extend up to 71.8\arcsec\ (14.2~kpc) from the center of the galaxy. We find that the shells have a cumulative stellar mass of $6.3\times10^{8}~M_{\odot}$, roughly 2\% the total stellar mass of NGC~4993, and mass-weighted ages of $>$3~Gyr.  We conclude that it was unlikely the GW170817 progenitor system formed in the galaxy merger, which based on dynamical signatures and the stellar population in the shells mostly likely occurred 220--685 Myr ago.
\end{abstract}


\section{Introduction}\label{sec:intro}

The discovery and localization of the binary neutron star (NS) merger GW170817 \citep{Abbott17:gw,Abbott17:mma} from a gravitational wave (GW) signal and its optical counterpart enabled the first detailed study of this rare phenomenon.  The electromagnetic counterpart to this event was initially identified from a short $\gamma$-ray burst \citep[GRB;][]{GW170817:fermi} called GRB\,170817A, confirming the hypothesis that NS mergers are sources of these high-energy astrophysical phenomena and launch relativistic jets \citep{Lattimer76,Li98,Metzger10}.  GRB\,170817A was followed 11~hours later by the discovery of a kilonova, or an optical transient powered by the radioactive decay of $r$-process elements synthesized in the merger's neutron-rich ejecta \citep[called AT~2017gfo;][]{Coulter17,Chornock17,Cowperthwaite17,Drout17,Kilpatrick17,Tanvir17,Troja17}.  When the counterpart became visible 111~days after merger and after Sun constraint, its optical and infrared (IR) emission appeared dominated by a broadband, synchrotron-powered afterglow from a relativistic and structured jet \citep{Alexander18,Lazzati18,Lyman18,Margutti18,Mooley18,Troja18,Fong19}, also observable at early times in both the radio and X-ray bands \citep{Alexander17,Haggard17,Hallinan17,Margutti17,Troja17}.

{\it Hubble Space Telescope} (\hst) imaging and spectroscopy of GW170817 contributed significantly to analysis of the kilonova \citep{Cowperthwaite17,Kasliwal17,Tanvir17,Troja17}, GRB afterglow \citep{Lyman18,Fong19,Lamb19}, and its S0 host galaxy NGC~4993 \citep{Blanchard17,Coulter17,Kilpatrick17,Levan17,Palmese17,Pan17}.  Despite being the closest known NS merger to date at $\approx$40~Mpc, optical and near-IR emission from GW170817 faded below the detection thresholds of the largest ground-based telescopes due to a combination of poor observability, intrinsic faintness ($>$26~mag) and contaminating light from its bright host galaxy. Thus, all optical detections and the most constraining upper limits on AT~2017gfo at $>$100 days have been enabled by \hst\ (\autoref{fig:color-image}).  This large, homogeneous set of high-resolution imaging, spanning serendipitous archival imaging from months before the merger, to exhaustive follow-up campaigns years after detection of the GW170817, uniquely probes its optical light curve, local environment, and faint features in the host galaxy.

{\it HST} optical light curves enabled constraints on the bulk energetics, ejecta velocity, and opacity of kilonova ejecta, which was observed in distinct ``blue'' and ``red'' components at early and late-times, respectively \citep{Cowperthwaite17,Tanvir17,Troja17}.  The latter component indicates that kilonovae produce a significant mass of lanthanides in their ejecta and may account for the bulk of heavy element production in the ``third peak'' of $r$-process production \citep{Kasen17,Metzger17}.

Combined with contemporaneous data obtained in the radio and X-ray bands, the \hst\ data of AT~2017gfo beyond $\approx 100$~days primarily probe the jetted, relativistic outflow from the NS merger.  These light curves exhibited a constant spectral index for the first $\approx$900~days with a peak at 160~days followed by a relatively rapid decline \citep{Margutti17,Lyman18,Lamb19,Fong19}.  Modeling of these data are best described by a structured jet with a relatively narrow, collimated core (3--5$^{\circ}$) and a wider-angle outflow moving at slower velocity \citep{Alexander17,Margutti17,Wu18,Hajela19,Wu19,Margutti20}.  However, variations in the spectral index at 1234~days post-merger as seen in recent X-ray detections suggests that the structured jet is evolving or some new emission component, such as a relativistic shock from the slower-moving kilonova ejecta, is beginning to dominate the afterglow light curve \citep{Hajela20,Balasubramanian21,Hajela21,Troja21}. With the exception of \hst/F606W data \citep{Fong19}, optical and near-IR measurements of the faint afterglow luminosities were performed without deep template images, introducing uncertain contributions from the host galaxy.

The joint set of \hst\ observations also enabled the highest resolution analysis of the local environment around AT~2017gfo, globular clusters proximate to its merger site, and the global properties of NGC~4993.  Some of the most intriguing features of this galaxy are the shells of gas and stars extending $\approx$13~kpc from the center of the galaxy \citep{Blanchard17,Palmese17,Ebrova20}, whereas the projected separation of AT~2017gfo is only 2.2~kpc.  These shells likely indicate a merger between NGC~4993 and a less massive galaxy within the past few hundred million years \citep{Palmese17,Ebrova20}.  This inference is based on the relatively limited amount of \hst\ imaging that was available within 2~years from the event.  Detailed analysis of all \hst\ imaging obtained since then could reveal additional, fainter shells interior or exterior to the primary structure. There is currently over 140~ks of wide-band \hst\ imaging from the ultraviolet (UV) to near-IR (\autoref{tab:observations}), and a more detailed analysis of the shell structures' overall morphology, luminosity, and colors can constrain the time since they were accreted, their mass, and age of the associated stellar population.

Here we present all \hst\ observations of AT~2017gfo and its host galaxy obtained to date, including new data: 14~orbits of Wide Field Camera 3 (WFC3) F814W, F110W, F140W, and F160W imaging obtained from Jan. to Feb.\ 2021. Throughout this paper, we assume a line-of-sight Milky Way reddening to AT~2017gfo and NGC~4993 of $E(B-V)=0.109$~mag \citep{Schlafly11}.  All magnitudes given throughout this paper are on the AB magnitude system \citep{Oke83}.  All dates and times are given in Coordinated Universal Time (UTC).

\section{Observations}\label{sec:obs}

We retrieved all observations of GW170817/AT2017gfo from the MAST archive across all filters, spanning from around 5 to 1285 days after merger (observer frame; note that in \autoref{tab:observations} and for figures showing the light curve we provide the epoch in rest-frame days), as well as the pre-merger \hst/ACS image from 2017 Apr.\ 28. These observations used 12 different filters from the UV to the IR bands, and they include data from the Advanced Camera for Surveys (ACS) Wide Field Camera (WFC) and the Wide Field Camera 3 (WFC3) UVIS and IR detectors. In particular, we present new WFC3 observations from 2021 Jan. 4 to Feb.\ 22 under Program 15886 (PI: Fong) in the F814W, F110W, F140W, and F160W filters (see \autoref{tab:observations}). The primary purpose of these late-time observations is to serve as ``template images'' for the image subtraction procedure (discussed in \autoref{sec:photometry}) in these bands. The new templates comprise 3--4 orbits in each band, for a cumulative on-source exposure time of 7.8--10.4~ks in each visit.

\input{observations} 

Starting with the {\tt flc} (for WFC3/UVIS and ACS/WFC) and {\tt flt} (for WFC3/IR) files, we reduced all observations using the {\tt hst123} analysis and reduction code\footnote{\url{https://github.com/charliekilpatrick/hst123}} as described in \citet[][]{Kilpatrick21}.  We first combine images from every unique visit and band as listed in \autoref{tab:observations}. We used {\tt astrodrizzle} to optimally stack and regrid each WFC3/UVIS observation to a pixel scale of 0.05\arcsec~pix$^{-1}$ and each WFC3/IR observation to 0.064\arcsec~pix$^{-1}$ with {\tt driz\_sep\_pixfrac=0.8} and {\tt final\_pixfrac=0.8}.  We also used {\tt drizzlepac.photeq} to ensure a uniform photometric zeropoint across both WFC3/UVIS chips before image combination.  In this way, photometric precision is preserved in the combined images, which are known to exhibit 0.4\% root-mean-square photometric variation compared with the original flux-calibrated {\it HST} images \citep[see discussion in][]{Fruchter97,Fruchter02,McMaster08}. 

\begin{figure*}[!t]
    \centering
    \includegraphics[width=0.98\textwidth]{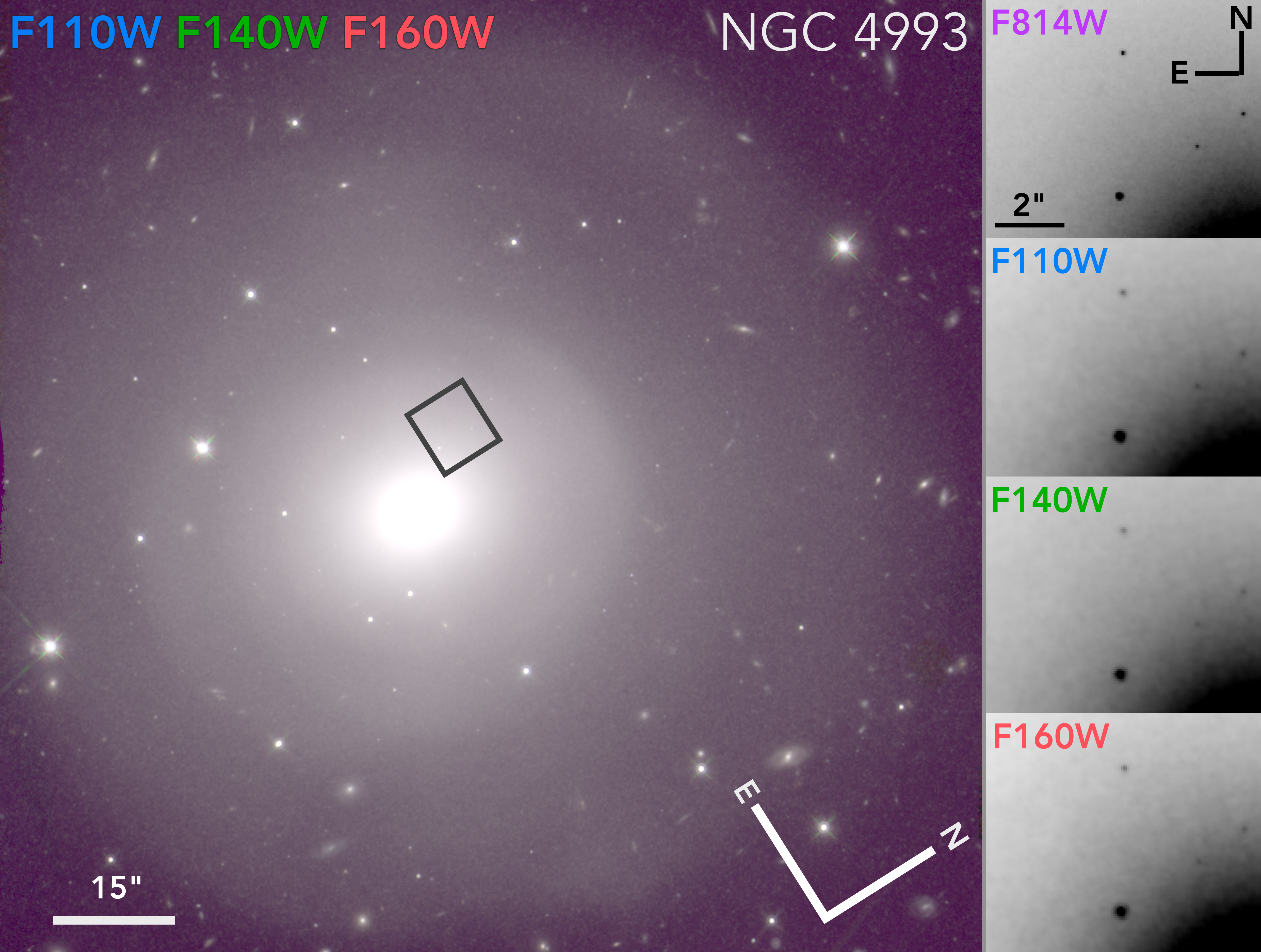}
    \caption{Three-color {\it HST} WFC3/IR imaging of a 120.9$\times$116.8~arcsec$^{2}$ region covering site of GW170817 in F160W (red), F140W (green), and F110W (blue), comprised of all data taken to date in these bands.  The black box represents the region highlighted by the sub-panels centered on AT2017gfo, a 9.1$\times$7.9~arcsec$^{2}$ region in F814W, F110W, F140W, and F160W from the latest epochs in each band (obtained on 22 Feb, 7 Feb, 4 Jan, and 6 Jan 2021, respectively, as described in \autoref{tab:observations}).}
    \label{fig:color-image}
\end{figure*}

We show an IR three-color image (F110W, F140W, F160W) comprised of the template images, and single-filter zoomed versions centered on the location of GW170817/AT2017gfo in the F814W, F110W, F140W, F160W filters in \autoref{fig:color-image}. The lack of any apparent source at the transient location means that they can be adequately used for templates against which we can subtract any earlier imaging (\autoref{sec:photometry}).

In addition to creating stacks for each unique visit, we create ``deep stacks'' by combining every available image observed in the F606W, F814W, F110W, F140W, and F160W filters (bottom five rows of \autoref{tab:observations}). We use these ``deep stacks'' of NGC~4993 for our analysis of the galaxy structure and environment (\autoref{sec:shells}).

\subsection{Photometry and Image Subtractions for AT~2017gfo}\label{sec:photometry}

We performed photometry in every {\tt flc}/{\tt flt} frame (for ACS/WFC, WFC3/UVIS, and WFC3/IR where appropriate) using {\tt dolphot} \citep{dolphot}.  Our reductions followed standard recommendations for each imager as described in {\tt dolphot} \citep{dolphot} and {\tt hst123} \citep{Kilpatrick21}.  We use this photometry for imaging without late-time templates or images where AT~2017gfo is detected at $>$20$\sigma$, which comprise every detection between 22--29 Aug.\ 2017 and is reported in \autoref{tab:observations}.

For the remaining data in which AT~2017gfo is detected at $<$20$\sigma$, comprising observations obtained from 6 Dec.\ 2017 through 14 Aug. 2018, we report photometry derived from subtracted imaging, which comes primarily from \hst/WFC3.  Our F606W photometry, including all imaging of AT~2017gfo observed by ACS during this time period, is taken from \citet{Fong19}, which follows a similar procedure to image subtraction described below.

Subtracting early- and late-time imaging stacks requires an understanding of the point-spread function (PSF) shape in both epochs.  WFC3 has a relatively stable PSF with little change in shape observed over 11~yr of operation.  The primary PSF variations are ``breathing'' modes due to thermal expansion over the orbital period of the spacecraft.  The effect of these changes in the full-width at half-maximum (FWHM) of the PSF is largest at bluer wavelengths, with at most 0.3\% variation redward of 8000~\AA\ \citep{WFC3}.  Otherwise, the WFC3 PSF in this wavelength range is well approximated by a Gaussian profile (before pixelation).  Therefore, the difference between two WFC3 frames observed in the same filter and instrumental configuration is dominated by the current position angle (PA) of the spacecraft on the sky, which in turn affects the PA of any non-axially symmetric components of the PSF, such as the location of diffraction spikes (as in \autoref{fig:color-image}, where the PA was $\approx$110$^{\circ}$).  Thus in order to match PSF shape between imaging observed at different epochs, we use a relatively simple, Gaussian convolution kernel.

\begin{figure*}
    \centering
    \includegraphics[width=\textwidth]{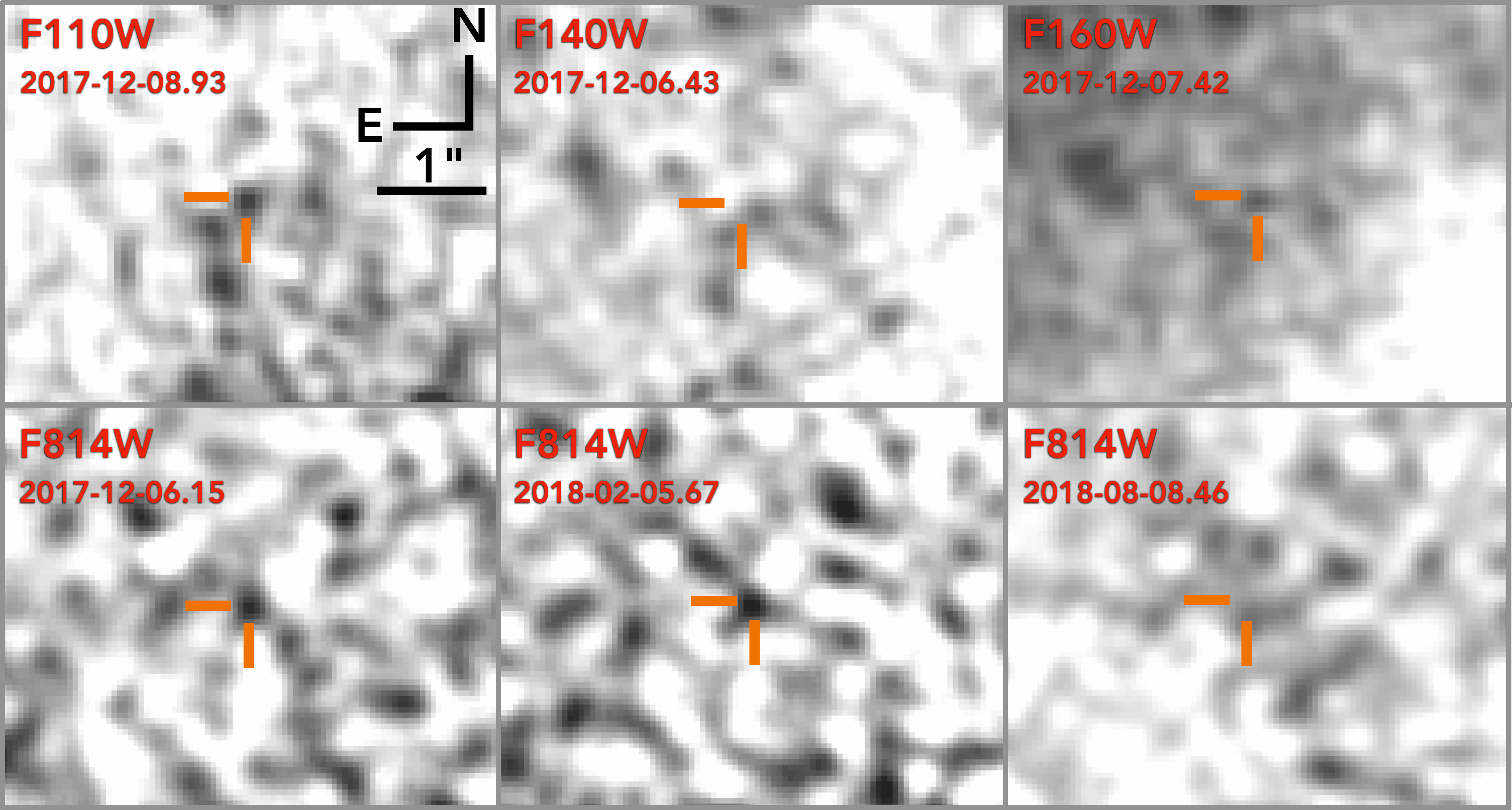}
    \caption{Difference imaging centered on the location of AT~2017gfo in F814W, F110W, F140W, and F160W from after 2017 Dec 6.  The template imaging used in subtractions consists of stacked imaging obtained after 2021 Jan 4 (Program GO-15886, PI Fong; see \autoref{tab:observations}).  We performed forced PSF photometry at the location of AT~2017gfo (noted with orange lines in all frames) as described in \autoref{sec:photometry} and obtain detections of AT~2017gfo in F110W (upper left), F160W (upper right), F814W on 2017 Dec.\ 6 (lower left), and F814W on 2018 Feb.\ 5 (lower middle).  In the remaining two subtractions, we calculate upper limits at the site of AT~2017gfo as described in \autoref{sec:limits}.}
    \label{fig:diff-image}
\end{figure*}

For all F814W, F110W, F140W, and F160W imaging, which comprise our late-time imaging where AT~2017gfo is detected at $<$20$\sigma$, we follow a similar procedure to \citet{Fong19} using {\tt hotpants} \citep{hotpants} to subtract our template images from the combined imaging in each visit.  The specific parameters used in each subtraction were varied in order to reduce the root-mean-square residuals from stars observed close to AT~2017gfo, but in general our default parameters are {\tt bgo=0.1}, {\tt ko=0.05}, and {\tt nsx=nsy=5}.  In all cases, we convolve and normalize the input image to the  template image (i.e., {\tt c=t}, {\tt n=t}).  These parameters are similar across both WFC3/UVIS and WFC3/IR imaging.  We show example subtractions from all four bands in \autoref{fig:diff-image}.

After image subtraction, we derive photometry for AT~2017gfo in all four bands by empirically reconstructing the instrumental PSF from isolated stars in the original science image using the {\tt python}-based tool {\tt photutils}.  We then perform forced PSF photometry at the position of AT~2017gfo, which is derived by aligning each frame to early-time imaging of the kilonova.  Photometric uncertainties were calculated from the $\chi^{2}$ of the profile fit following methods described in \citet{Stetson87}.  All detections of the afterglow from subtracted imaging are reported in \autoref{tab:observations}.

\subsection{Upper Limits on Emission near AT~2017gfo}\label{sec:limits}

For several epochs, we do not detect any significant ($>$3$\sigma$) emission when performing forced photometry at the location of AT~2017gfo.  In these cases, we place upper limits on the presence of any optical or near-IR counterpart with the {\tt FakeStars} procedure in {\tt dolphot}.  Following procedures described in \citet{Kilpatrick17}, we injected 1,000 sources into magnitude bins of 0.1~mag from 20--30~mag (i.e., 100,000 sources in total) around the location of AT~2017gfo in that frame.  We varied the position of each source randomly by drawing a Gaussian random variable centered at (x,y)=(0,0) and with a Gaussian FWHM=$\mathrm{PSF}/\sqrt{\mathrm{theoretical~S/N}}$ for that source.  We then determined the magnitude threshold at which $>$99.7\% of sources were recovered at $>$3$\sigma$, which we consider to be the limiting magnitude for that visit as reported in \autoref{tab:observations}.

We repeated this procedure independently both for the template images and deep stacks.  In the latter case, instead of injecting sources at the location of AT~2017gfo, we instead injected them randomly within a 2\arcsec\ radius of that position.  In this way, we avoid biasing our magnitude limit with residual flux from AT~2017gfo in each of the frames we stacked.  Thus we consider the limiting magnitude for the deep stacks to be the threshold at which we can detect point-like sources near AT~2017gfo, rather than AT~2017gfo itself.  All of these limits are reported in \autoref{tab:observations}.

\section{The Complete {\it HST} Light Curve of the GW170817 Counterpart}

We confirm previous {\it HST} detections of AT~2017gfo observed after 2017 Dec.\ 6 \citep[e.g., in][]{Lyman18,Margutti18,Troja18,Lamb19,Ryan20} in our difference imaging.  In addition, the use of deep template imaging in our subtractions uncovers four new detections (\autoref{fig:diff-image} and \autoref{tab:observations}) in F814W (2017 Dec.\ 6 and 2018 Feb.\ 6), F110W (2017 Dec.\ 9), and F160W (2017 Dec.\ 7).  Below we compare the {\it HST} light curve of AT~2017gfo to models of kilonova and GRB afterglow emission.  We also consider emission components not previously observed in the optical or near-IR but that can be constrained by our new limits, including late-time changes in the afterglow spectral index \citep[as in][]{Balasubramanian21,Hajela21,Troja21} and an infrared dust echo \citep{Lu21}.  All model magnitudes, wavelengths and dates discussed are given in the rest-frame accounting for the redshift and luminosity distance to NGC~4993 \citep[$z=0.00980$ and 40.7~Mpc as in][]{Cantiello18} and assuming a merger time of 2017 Aug.\ 17 12:41:04 \citep{Abbott17:gw}.

\subsection{The Kilonova Light Curve Before 2017 August 30}\label{sec:kilonova}

\begin{figure*}
    \centering
    \includegraphics[width=\textwidth]{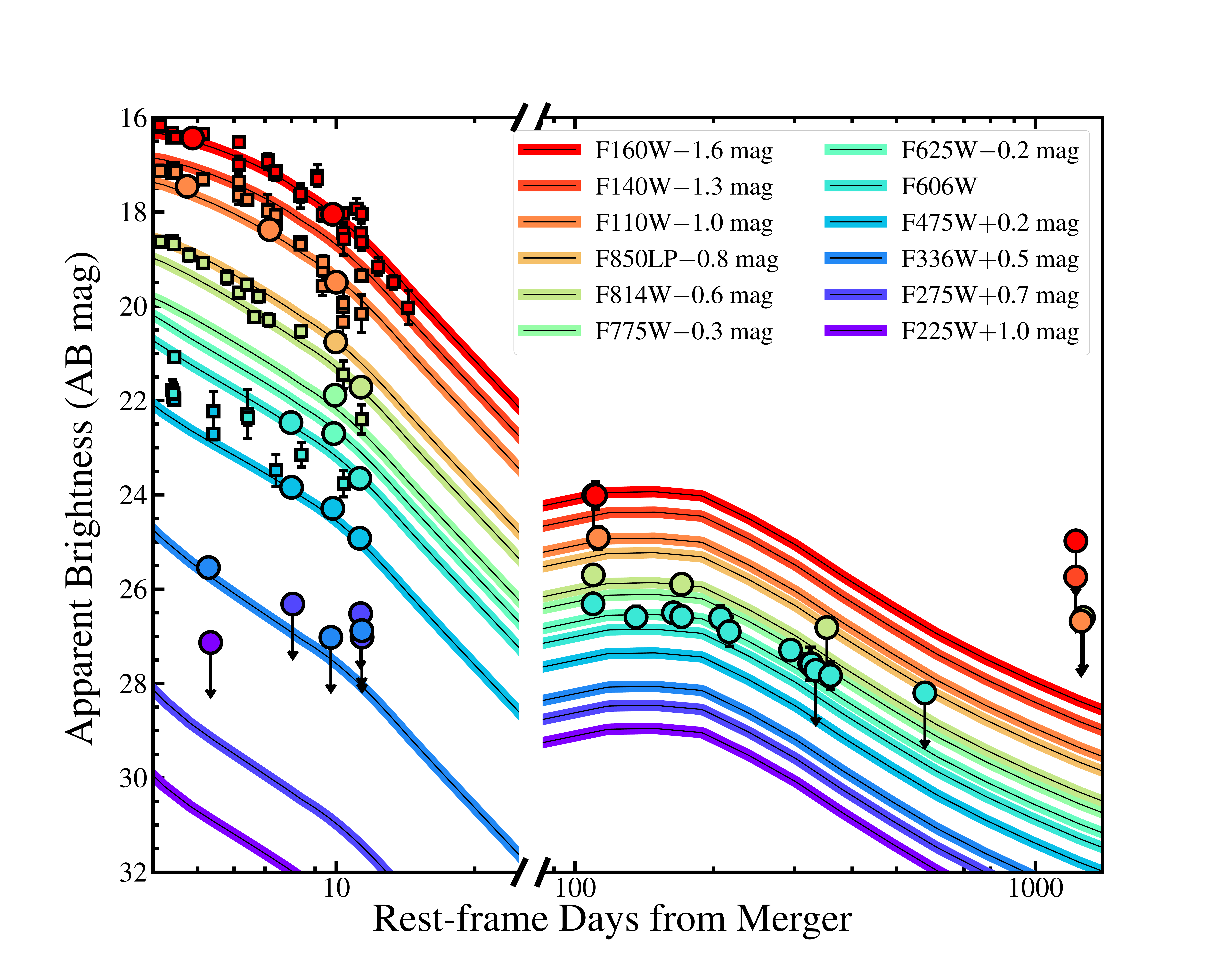}
    \caption{Our {\it HST} photometry (circles) of AT~2017gfo from \autoref{tab:observations}.  All limits and detections are calculated in this publication apart from F606W detections from after 6 Dec.\ 2017, which come from \citet{Fong19}.  For comparison to the early-time \hst\ observations, we plot contemporaneous detections of AT~2017gfo (squares) from \citet{Villar17} obtained in  filters with comparable filter transmission to those on ACS/WFC, WFC3/UVIS, and WFC3/IR, including $H$ (F160W), $J$ (F110W), $I$ (F814W), $V$ (F606W), and $g$ \citep[F475W; originally presented in][]{Arcavi17,Andreoni17,Coulter17,Cowperthwaite17,Diaz17,Drout17,Evans17,Kasliwal17,Pian17,Smartt17,Tanvir17,Troja17,Utsumi17}.}
    \label{fig:lightcurve}
\end{figure*}

The kilonova emission of AT~2017gfo dominated the observed UV, optical and near-IR light at rest-frame epochs of $\lesssim 12$~days (2017 Aug.\ 30), after which the field went into Solar conjunction. The kilonova of AT~2017gfo was inferred to have multiple components to its UV to near-IR light curve following a thermal, radioactively-heated prescription as in \citet{Arnett82}.  These two components are commonly referred to as ``blue'' and ``red''  \citep[see, e.g.,][]{Metzger10,Barnes13,Kasen13,Barnes16}, based on a rapidly-declining component with opacity $\kappa\approx1$~cm$^{2}$~g$^{-1}$ observed that dominated the spectral energy distribution (SED) within 2~rest-frame days of merger and a longer-lived component with $\kappa\approx5$--$10$~cm$^{2}$~g$^{-1}$ observed at later times \citep{Arcavi17,Andreoni17,Cowperthwaite17,Drout17,Lipunov17,Kasliwal17,Kilpatrick17,McCully17,Metzger17,Nicholl17,Smartt17,Soares-Santos17,Tanvir17,Utsumi17,Valenti17,Villar17}. The characterization of the kilonova at these very early epochs was led by ground-based campaigns and the Neil Gehrels {\it Swift} Observatory ({\it Swift}), but at UV wavelengths AT~2017gfo had faded below the threshold detectable by {\it Swift} once \hst\ started to observe (\autoref{fig:lightcurve}).

The earliest set of \hst\ observations occurred at $\approx4.7-5.3$ days after merger, spanning the UV to near-IR bands. First, these observations enabled the only detection at UV wavelengths (F336W) at $>$3~days from merger \citep[whereas most of the UV detections before this epoch came from {\it Swift}/UVOT as in][]{Cowperthwaite17,Drout17,Evans17}, providing an important anchor on the blue component. Subsequent \hst\ observations also provided the deepest and latest constraints on this blue component emission. Second, \hst\ probed the more slowly-evolving, red kilonova emission with multi-band observations continuing to $\approx 12$~days.

To characterize the red and blue kilonova components of AT~2017gfo in the early-time {\it HST} data before 2017 August 30, we adopt the kilonova SEDs of \citet{Kasen17}.  As in \citet{Kilpatrick17}, we use two models to characterize both kilonova components, consisting of a blue component with an ejecta mass $M_{\mathrm{ej}}=0.025~M_{\odot}$ of ejecta moving with an ejecta velocity $v_{k}=0.25~c$.  The ejecta composition is broadly characterized by a lanthanide fraction $X_{\mathrm{lan}}$ defined as the ratio of the mass of heavy lanthanide species ($Z$$=$57--71) to the total ejecta mass \citep[see, e.g.,][]{Kasen13,Barnes13}.  For the blue kilonova model, this corresponds to a radial gradient in $X_{\mathrm{lan}}$ with $\log(X_{\mathrm{lan}})=-6$ in the outer layers and increasing to a higher but still low $\log(X_{\mathrm{lan}})=-4$ in the center \citep[see][for details]{Kasen17,Kilpatrick17}.  The red kilonova model is simply characterized by $M_{\mathrm{ej}}=0.035~M_{\odot}$, $v_{k}=0.15~c$, and $\log(X_{\mathrm{lan}})=-2$. The parameters used here for the two kilonova components are broadly consistent with other studies and the corresponding light curves are well matched to the \hst\ data.

\begin{figure*}
    \centering
    \includegraphics[width=0.7\textwidth]{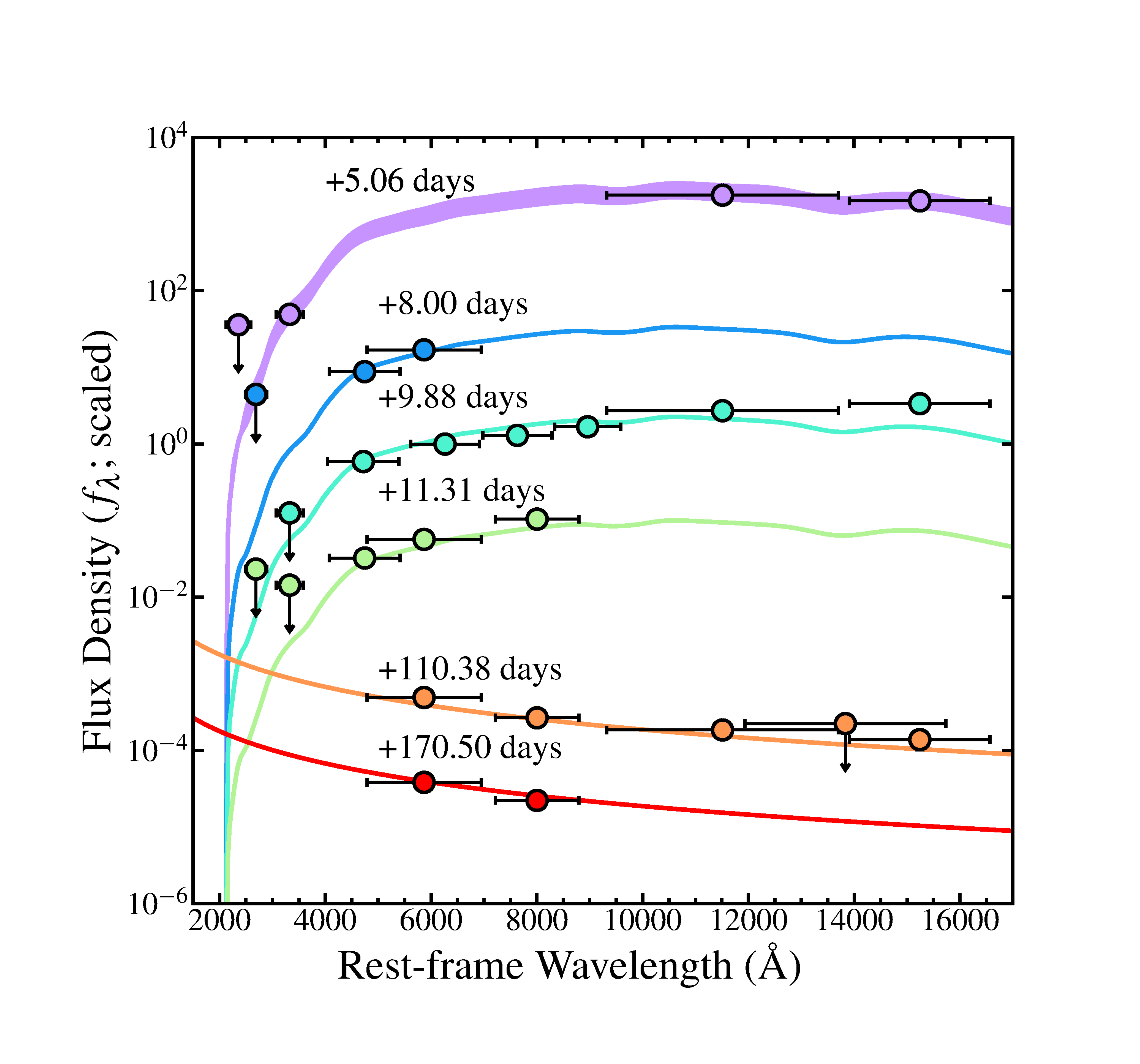}
    \vspace{-0.3in}
    \caption{The spectral energy distribution of the kilonova (rest-frame days 5.06--11.31) and GRB afterglow (110.38--170.50 days) components of AT~2017gfo as constrained by {\it HST} detections and upper limits (circles) and described in \autoref{sec:kilonova} and \autoref{sec:afterglow}.  The horizontal error bars correspond to the equivalent rectangular width of the corresponding filter as described in \citet{SVO12}.  We overplot the average kilonova and GRB afterglow models for data obtained within $\pm$0.5~day of the average day given next to each model.  For the first model at 5.06~days (violet), there are two kilonova models from \citet{Kasen17} within this time range, which are plotted as a shaded region between the brighter (upper) and fainter (lower) model.}
    \label{fig:sed}
\end{figure*}

We also combined the time-varying SEDs from these models and smoothed them over a window of 100~\AA\ while masking out spectral bins where the flux drops to zero.  We then passed the predicted spectrum at each time through the filter transmission functions for each ACS/WFC, WFC3/UVIS and WFC3/IR band.  We overplot these models for the bands in which AT~2017gfo was observed on the left side of \autoref{fig:lightcurve}.   In addition, in \autoref{fig:sed} we show example SEDs at times when there were more than two broadband {\it HST} constraints on AT~2017gfo over a span of 1~day.

As shown in previous work on AT~2017gfo \citep{Arcavi17,Cowperthwaite17,Drout17,Kilpatrick17,Smartt17,Villar17}, the best-fitting model suggests that the total luminosity transitioned from blue- to red-kilonova dominated at around 3 rest-frame days from merger.  Thus all {\it HST} observations, which start at 4.74~rest-frame days, occurred at a time when the vast majority of kilonova emission were well-characterized by our red kilonova model.  

The only exceptions are the optical and UV bands where a rapidly-declining blue tail of the kilonova spectral energy distribution was observed from around 5 to 11 rest-frame days from merger (\autoref{fig:lightcurve} and \autoref{fig:sed}).  While the {\it HST} observations probed timescales when the red kilonova component was dominant, they also provided the only constraints on the rapidly-declining blue tail.  These data can uniquely probe this component of the kilonova at later times and its physical origin, which is still largely unsolved but may be due to energy injection from a long-lived NS remnant that lowers the electron fraction in the bluer ejecta \citep{Lippuner17}, or possibly from accretion outflows from a disk that forms around the merger \citep{Miller19}.  

\subsection{The GRB Afterglow Light Curve After 2017 December 6}\label{sec:afterglow}

After the field once again became observable with \hst\ at $>100$ rest-frame days from merger, the optical and near-IR emission from AT~2017gfo was dominated by GRB afterglow  \citep{Lyman18,Mooley18,Troja18,Fong19,Lamb19}.  Novel to this work are the late-time templates described in \autoref{sec:obs}, which enabled four new detections in F814W, F110W, and F160W. To compare our updated photometry and upper limits from AT~2017gfo at these epochs, we compare its {\it HST} light curve to the afterglow model based on an off-axis relativistic structured jet and presented in \citet{Hajela19}.  We adopt the updated parameters of \citet{Hajela21} for a relativistic structured jet viewed at an angle of $\theta_{\mathrm{obs}}=23^{\circ}$ and interstellar medium density $n_{0}=0.01$~cm$^{-3}$.  We choose these models for comparison over other afterglow models \citep[e.g., {\tt JetFit} models in][with $\theta_{\rm obs}\approx30^{\circ}$]{Wu18,Wu19} because the predicted observation angle is consistent with independent constraints from superluminal motion in the relativistic jet \citep[$\approx$20$^{\circ}$ in][]{Mooley18}.

The resulting optical and near-IR light curves are shown on the right side of \autoref{fig:lightcurve} with the corresponding spectral energy distributions in \autoref{fig:sed}.  These models are relatively good fits to the observed {\it HST} data, with minimal inverse-variance weighted average residuals of 0.1~mag compared with measurement uncertainties in each detection of 0.15--0.29~mag.

Consistent with the findings of \citet{Fong19}, \citet{Lamb19}, and \citet{Hajela19}, we find no evidence for a change in spectral shape across the optical and near-IR spectral energy distribution (\autoref{fig:sed}).  Our best constraints come from the afterglow light curve at 109.6 and 170.5 rest-frame days from merger, with two and three detections over a span of $\approx$2~days, respectively.  In both cases, the observations are consistent with a constant spectral index of $f_{\nu}\propto\nu^{-0.6}$, reinforcing the broader constraints from radio to X-ray observations \citep{Hajela19} and a lack of any synchrotron curvature at intermediate frequencies.

\subsection{Constraints on a Kilonova Afterglow}\label{sec:kn-afterglow}

Recent detections of AT~2017gfo by {\it Chandra} at 0.3--10~keV suggest that the spectral index of the afterglow is changing even as it fades \citep{Hajela20,Balasubramanian21,Hajela21,Troja21}.  The cause of this change in spectral index has been attributed to changes in the density of the circumburst medium, shock velocity, or microphysical parameters  where the afterglow originates \citep{Granot18} or a new emission source arising from interaction between the slower kilonova ejecta and circumburst medium \citep[called a ``kilonova afterglow,'' e.g.,][]{Hajela19}.  

Although we do not detect AT~2017gfo in the latest {\it HST} observations, in principle these limits can be used to constrain the presence and nature of a kilonova afterglow \citep{Kilpatrick21a}.  Following a similar analysis to \citet{Hajela21}, we use each of our {\it HST} optical limits to constrain the optical to X-ray spectral index $\beta_{OX}$ (where $F_{\nu}\propto\nu^{-\beta_{OX}}$) implied by the 0.3--10~keV X-ray detection of AT~2017gfo at $\approx$1234~days, which had $F_{X}=2.47\substack{+0.62\\-0.91}\times10^{-15}$~erg~cm$^{-2}$~s$^{-1}$.  Based on our optical limits, the strongest constraints on $\beta_{OX}$ comes from the F110W limit, which implies that $\beta_{OX}\lesssim0.85$.  

This is moderately more constraining than the spectral index implied by F140W presented in \citet{Hajela21} but significantly steeper than the spectral index implied by an afterglow spectrum with no evolution ($\propto \nu^{-0.6}$) let alone the shallower spectrum implied by radio measurements \citep{Balasubramanian21}.  Without $>$2~mag rebrightening in the overall light curve of AT~2017gfo, it is unlikely that new optical observations will be constraining for the future evolution of this specific emission model.

\subsection{Constraints on an Infrared Dust Echo}\label{sec:dust-echo}

Our new late-time constraints on AT~2017gfo can provide unique constraints on IR dust echoes from the GRB afterglow \citep[similar to those observed for supernovae and tidal disruption events, e.g.,][]{Graham83,Jiang21}.  IR dust echoes may be observed from the interaction between UV emission from the GRB afterglow and a sufficiently dense shell of dust proximate to the NS merger.  As demonstrated in \citet{Lu21}, a shell of dust surrounding the merger would be sublimated up to some radius ($r_{\rm sub}$) at which point radiation from the afterglow can no longer heat dust grains above their sublimation temperature.  For an off-axis afterglow with a viewing angle $\theta_{\rm obs}$, the timescale on which an IR dust echo from the heated dust grains becomes visible is therefore $t = r_{\rm sub}/c~(1 - \cos~\theta_{\rm obs})$, which for $\theta_{\rm obs}=20^{\circ}$ and $r_{\rm sub}=6$~pc could be as long as 1.2~yr.

\begin{figure}
    \centering
    \includegraphics[width=0.49\textwidth]{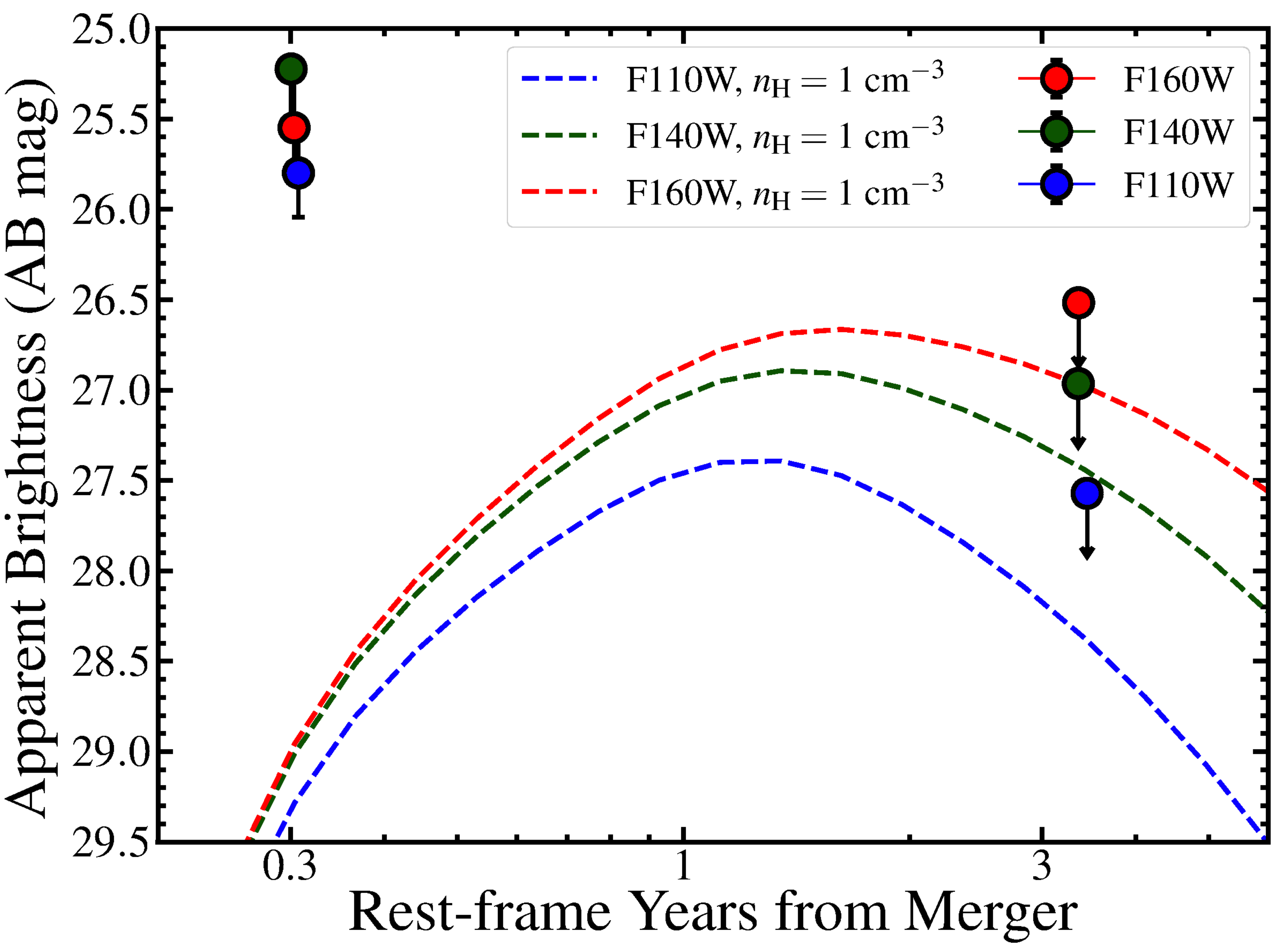}
    \caption{Model light curves for infrared dust echoes assuming a circum-merger density of $n_{\rm H}=1$~cm$^{-3}$, $\theta_{\rm obs}=20^{\circ}$ and in F110W, F140W, and F160W.  We compare these light curves to our detections and limits on AT~2017gfo at $\approx$0.3 and 3 rest-frame years post-merger.}
    \label{fig:dust-echo}
\end{figure}

We show our limits compared with expected light curves for IR dust echoes in F110W, F140W, and F160W in \autoref{fig:dust-echo}.  The comparison light curves are constructed from the model presented in \citet{Lu21}\footnote{From code available at \url{https://github.com/wenbinlu/dustecho}.}.  They are parameterized primarily by the density of gas in the circum-merger medium, which was measured to be $n_{\rm H}\approx$10$^{-5}$--$10^{-2}$~cm$^{-3}$ depending on the exact afterglow model and microphysical parameters assumed \citep{Margutti18,Wu18,Hajela19,Wu19}.  We note that we have assumed that the afterglow has initial UV luminosity of $L_{UV}=3\times10^{47}$~erg~s$^{-1}$ that peaks on a timescale of $t_{\rm max}=300$~s as in \citet{Lu21}.  For this model, our limits do not extend below the predicted flux for IR dust echoes propagating through an environment with $n_{\rm H}=1$~cm$^{-3}$, and since $L_{\nu} \propto n_{\rm H}$ we do not constrain this emission mechanism for realistic conditions around GW170817.  However, we note that the predicted light curve peaks on later timescales in redder bands, and so observations with the {\it James Webb Space Telescope} may be able to detect a IR dust echo for GW170817 in the future.

\section{Limits on Stars and Unresolved Stellar Clusters}\label{sec:limit-analysis}

\begin{figure}
    \centering
    \includegraphics[width=0.49\textwidth]{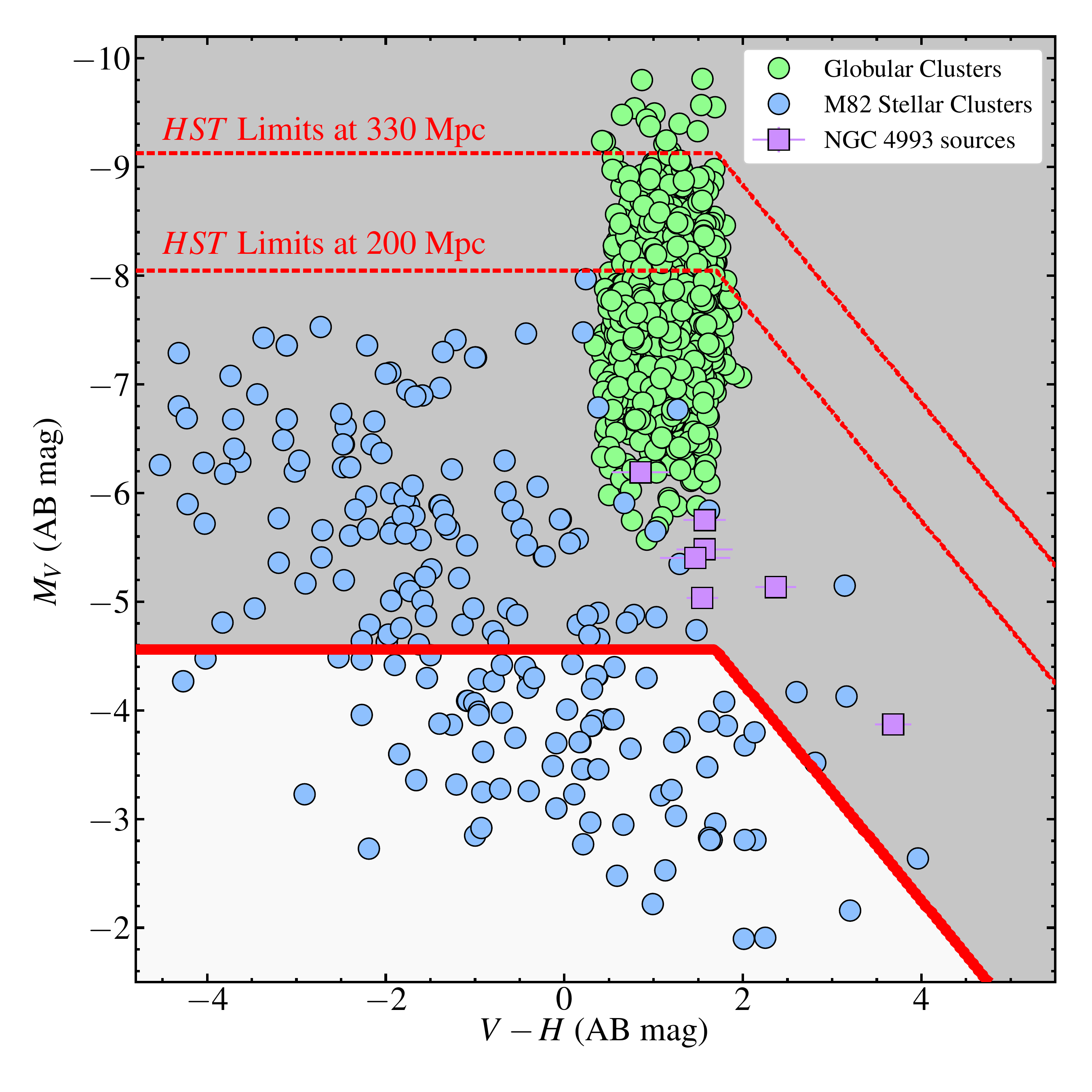}
    \caption{Our F606W ($V$-band) absolute magnitude and F606W-F160W ($V-H$) color limits at the distance of NGC~4993 (grey region is ruled out).  For comparison, we show the corresponding limit assuming a distance of 200~Mpc and 330~Mpc (dashed red lines).  We place these limits in the context of photometry of globular clusters modeled from those in NGC~1399 \citep[green circles; see description in \autoref{sec:limit-analysis} and][]{Blakeslee12}, stellar clusters in the starburst galaxy M82 \citep[blue circles; here we use F555W photometry as a proxy for $V$-band from][]{Li15}, and sources within 2\arcsec\ ($\approx$380~pc) of the site of AT~2017gfo (purple squares).}
    \label{fig:limits}
\end{figure}

We take advantage of the deep stacks which reach limits of 26.8--28.8~mag to place stringent limits on stellar sources and unresolved clusters. Previous F606W imaging ruled out bluer globular clusters (GCs) with $L \gtrsim 6.7 \times 10^{3}\,L_{\odot}$ (\citealt{Fong19}; see also \citealt{Lamb19}), well below the peak of the F606W GC luminosity function (GCLF) determined for NGC~4993 \citep{Lee18}. However, observations of nearby quiescent galaxies find that GC populations exhibit color bimodalities \citep{Larsen01}, with red clusters corresponding to a metal-rich population \citep{BrodieStrader06}. Indeed, $V-H \approx 0.3-2$~mag was derived for the giant elliptical galaxy NGC~1399 \citep{Blakeslee12}. 

At the distance of NGC~4993, our deep limit (\autoref{sec:limits}) of $m_{\rm F160W}\gtrsim$26.8~mag corresponds to $M_{\rm F160W} \gtrsim$$-$6.2~mag corrected for Milky Way extinction. To directly compare this to an $H$-band luminosity function and place constraints on an unresolved red GC, we create a representative GCLF by sampling the $V_{\rm F606}-H_{\rm F160}$ GC color distribution of NGC~1399 \citep{Blakeslee12}, and apply this color correction to the NGC~4993 GCLF in F606W \citep{Lee18}. Our limits rule out redder clusters to $L\gtrsim 2.3 \times 10^{4}~L_{\odot}$, or $M\gtrsim 4.6 \times 10^{4}~M_{\odot}$ assuming a standard mass-to-light ratio of $2M_{\odot}/L_{\odot}$; only $\approx 0.05\%$ of mass in red GCs are below this limit. Although this is nominally less constraining in luminosity than the previous F606W limit, we have placed an independent and deep constraint on reddened clusters.

To place these limits in the context of sources identified near GW170817, we obtain photometry of 7 point-like sources within a 2\arcsec\ radius (380~pc) of the site of AT~2017gfo as shown in \autoref{fig:limits}.  These objects were selected from F160W to be point-like \citep[{\tt dolphot} object type 1 for ``bright stars''][]{dolphot}.  Given these selection criteria, they tend to be redder than most cluster candidates in nearby galaxies and fainter than evolved massive stars \citep[e.g., from][]{Drout09,Li15}.  The brightest two of these sources are consistent with the low end of the GCLF, but otherwise they all appear too faint to be GCs.

We also show our limits placed at a distance of 200~Mpc and 330~Mpc, which are approximately the predicted detection ranges for binary NS mergers during Observing Run 4 and Observing Run 5 \citep{LIGO15,LIGO20}.  High-resolution optical mapping of future binary NS mergers using {\it HST}-like sensitivity can probe part of the GCLF and address their potential origin in GCs \citep{Pooley03,Rodriguez16}.

\section{Shell Structure around NGC~4993}\label{sec:shells}

\begin{figure*}
\includegraphics[width=\textwidth]{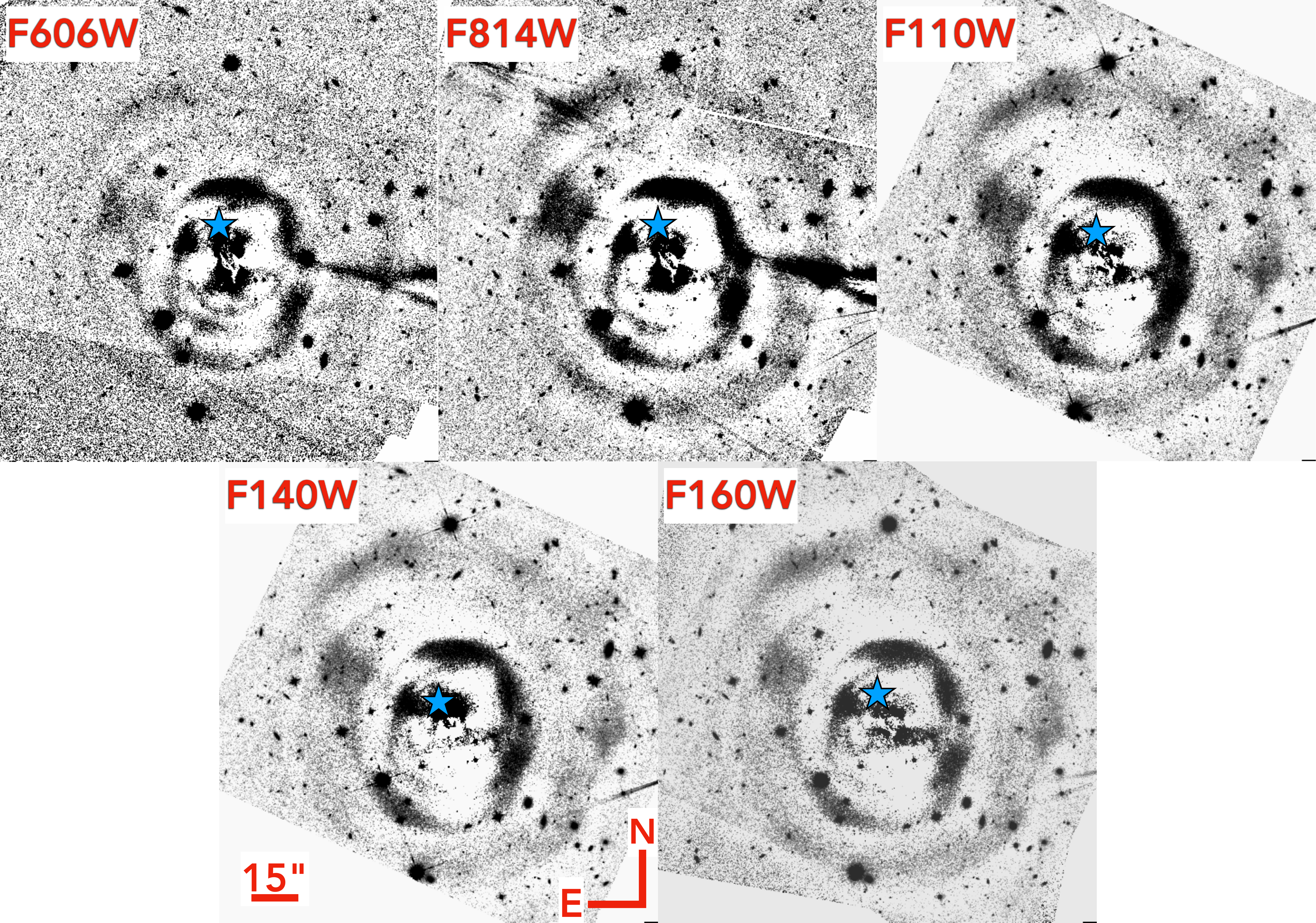}
\caption{Our {\tt GALFIT} subtracted deep stacks of NGC~4993 showing the galaxy's shell structure in F606W, F814W, F110W, F140W, and F160W.  All images are centered on AT~2017gfo and represent the same 140\arcsec$\times$147\arcsec\ region.  We indicate the location of AT~2017gfo in each frame with a blue star.  Note that in all frames, light from the diffraction pattern of a bright star (USNOA2 0600-15448796; $m_{V}=$12.6~mag) is visible on the right-hand side and is unassociated with any shell emission.  We mask out these features in our analysis of the shell features as described in \autoref{sec:shell-id}.}
\label{fig:galfit}
\end{figure*}

\subsection{{\tt GALFIT} Surface Brightness Modeling}\label{sec:galfit}

Our deep stacks allow us to characterize the morphological substructure of NGC~4993 in great detail. To analyze this structure, we used the complete set of F606W, F814W, F110W, F140W, and F160W imaging, consisting of 63.9, 18.5, 20.0, 12.6, and 14.5~ks, respectively, at the site of GW170817.  Based on the kilonova and afterglow light curves at the time each individual exposure was obtained, we expect AT~2017gfo to have an exposure-weighted magnitude of 23.6, 25.9, 22.3, 26.7, and 21.5~mag in our stacks.  This is nominally above the deep stack detection thresholds from \autoref{sec:limits} in every band.  However, the combined kilonova and afterglow flux is much fainter than the galaxy as a whole and we mask out emission from point sources, and so we do not expect this flux to significantly affect the quality of the model or analysis described below.

In order to isolate the shell structure, we use the {\tt GALFIT} software package \citep{GALFIT} to fit two-dimensional surface brightness profiles to the smooth galaxy light from NGC~4993 with a S\'{e}rsic model. A S\'{e}rsic profile is parameterized by S\'{e}rsic index $n$, effective radius $r_e$, position angle (PA) and ellipticity. When performing these fits, we mask out bad pixels and light from stars and galaxies in the field using the segmentation map of sources derived by combining the {\tt astrodrizzle} image mask and running {\tt SExtractor} \citep{sextractor} to identify point-like sources.  Due to the large spatial extent of NGC~4993 in these images, we use {\tt DETECT\_THRESH} = 15 and {\tt BACK\_SIZE} = 16 (size of the mesh in pixels over which the background is estimated) to prevent masking flux from the galaxy itself.  We also perform a PSF deconvolution using the PSF model described in \autoref{sec:obs}.

First, we model the surface brightness profile of NGC~4993 in each of the F606W, F814W, F110W, F140W and F160W deep stacks. We allow the centroid of the galaxy and aforementioned parameters to vary for a single S\'{e}rsic profile. Next, we undertake a two-component fit to better identify and characterize the large-scale substructure. For the primary component, we fix the center position from the previous profile and use the other fitted parameters as input values for a new fit. For the secondary component, we use a new S\'{e}rsic profile with the same fixed center position and all other parameters free to vary. The addition of a secondary component results in an improved residual map and $\chi^2$ (by a factor 2--8 in each band) compared to the single S\'{e}rsic case. F606W is the only band that is not well fit by a double S\'ersic profile. We find that the inclusion of a S\'ersic profile modified by a Fourier mode 2 (that distorts the shape of the 2D ellipsoid) for the secondary profile improves the fitting compared to standard S\'ersic profiles in F606W. In the near-IR bands (F110W, F140W and F160W), the fit is further improved by the addition of a third, PSF-like component at the center to model the presence of a weak AGN.

Across all bands, we find that the primary component is characterized by $n\approx2.9$--4.4, (representative of a de Vaucouleurs profile for elliptical galaxies), and $r_e \approx 14$--18\arcsec, resulting in $r_e\approx2.9-3.5$~kpc at the distance of NGC~4993. The PA evolves from optical to near-IR, suggesting the presence of different superimposed stellar populations, as also found in \citet{Palmese17} using an independent data set. The other primary component results are also consistent with fits previously performed in the literature of a single S\'ersic profile \citep{Blanchard17,Levan17,Palmese17}. The secondary component aims at modeling the core of the galaxy, which is not properly accounted for by the primary component.  Inclusion of this secondary component enables a better fit to the galaxy profile and the detection of shell features closer ($<$20\arcsec) to the center of NGC~4993. Modeling of this core component is more challenging in the optical bands, where dust lane obscuration complicates the geometry of the galaxy more than in the redder bands. In the near-IR bands F110W, F140W, and F160W, the fitting prefers a cored, less cuspy component (i.e. a low S\'ersic index $n_2$, see \autoref{table:galfit}) close to a Gaussian or exponential disk.

The residual images, as a result of subtracting the best-fitting light profile, are shown in \autoref{fig:galfit} and the {\tt GALFIT} parameters for each frame are given in \autoref{table:galfit}.

\begin{deluxetable*}{lccccccccc}
\savetablenum{2}
\tabletypesize{\footnotesize}
\centering
\tablecolumns{10}
\tabcolsep0.06in
\def\arraystretch{0.94}
\tablecaption{{\tt GALFIT} Parameters for NGC~4993}
\tablehead{
\colhead{Band} &
\colhead{$n_1$} &
\colhead{$n_2$} &
\colhead{$r_{e1}$} &
\colhead{$r_{e2}$} &
\colhead{$(b/a)_1$} &
\colhead{$(b/a)_2$} &
\colhead{PA$_1$} &
\colhead{PA$_2$} &
\colhead{$m_{\mathrm{PSF}}$}\\
&
&
&
(kpc) &
(kpc) &
&
&
(deg) &
(deg) &
(AB mag)
}
\startdata
    F606W & 3.99  & 0.08   & 3.53 & 0.15 & 0.870 & 0.757 & -12.2  &  75.2 & -- \\ 
    F814W & 4.80  & 0.13   & 4.40 & 0.12 & 0.848 & 0.695 &  -9.4  & 54.2 & -- \\
    F110W & 4.15  & 0.18   & 3.06 & 0.12 & 0.820 & 0.854 &  -7.6  &  40.0 & 10.59 \\
    F140W & 3.47  & 0.95   & 2.97 & 0.17 & 0.829 & 0.834 &  -8.1  &   8.3 & 10.06 \\
    F160W & 4.38  & 0.71   & 2.78 & 0.13 & 0.820 & 0.762 &  -7.0  &  30.4 & 10.76 \\
\enddata
\tablecomments{Morphological parameters estimates for a double S\'ersic profile in the different bands using {\tt GALFIT}. The subscripts 1 and 2 refer to the first and second S\'ersic component, respectively.  In the infrared bands, we include an additional PSF component to account for the presence of a weak AGN as described in \autoref{sec:galfit}.  This component is characterized solely by the integrated magnitude ($m_{\mathrm{PSF}}$) provided here in AB mag.  The effective radius ($r_e$) for both components is provided in kpc assuming $D=40.7$~Mpc to NGC~4993.  The model ellipticity is parameterized by the ratio between the semi-minor and semi-major axes of the model ellipsoid ($b/a$). The effective radii $r_e$ are provided in pixels, the position angles (PA) in degrees east of north.}\label{table:galfit}
\end{deluxetable*}

\subsection{Identification of Shell Features around NGC~4993 and Analysis of their Stellar Populations}\label{sec:shell-id}

NGC~4993 was classified as an S0 galaxy and part of the group LGG~332 \citep{Garcia93} with at least two visible shells \citep[it is also called MC1307-231;][]{MalinCarter83}.  As previously shown in both ground-based \citep{Palmese17} and {\it HST} observations \citep{Blanchard17,Levan17,Ebrova20}, NGC~4993 has clear shell structure in all of our deep {\it HST} stacked imaging (\autoref{fig:galfit}).  The apparently random distribution of shells around NGC~4993 is a characteristic of so-called ``Type II'' shell galaxies likely indicating a deviation from a perfectly radial orbit for the merger \citep[as opposed to Type I shell galaxies with axisymmetric shells distributed in a double cone pattern, e.g.,][]{Hernquist88, Sanderson13, Bilek15}.  \citet{Ebrova20} identify ten distinct shell features along the major photometric axis of the optical emission from NGC~4993 using stacked F814W imaging.  

Here we use our {\tt GALFIT}-subtracted residual imaging to systematically identify shell features at all position angles around NGC~4993.  However, the quality of our {\tt GALFIT} model results in large residuals and radial derivatives in those residuals close ($<$10\arcsec) to the center of NGC~4993.  We ignore these features in the analysis below, and our detection of shells at the smallest radial separations from NGC~4993 only extends to features observed at the approximate radius of AT~2017gfo in \autoref{fig:galfit} (10.6\arcsec).  The shell features are the most prominent in the near-IR, and so we use the WFC3/IR F160W stacked frame with the largest IR spatial footprint to systematically identify shells around NGC~4993.

From the residual F160W stacked image, we also mask out emission due to stars and background galaxies using our {\tt sextractor} segmentation map of point-like features.  To validate that the final shell image was relatively free of emission other than the shells, we visually inspected the masked image to ensure that there were no clearly detectable point-like features.  We then filled in all masked pixels flagged by the {\tt sextractor} segmentation image with the median value for all non-masked pixels within a 2\arcsec\ radius of each masked pixel.  Finally, we rebinned each frame into a grid of 1\arcsec$\times$1\arcsec\ pixels representing the median pixel value in each cell of the grid rescaled by the average number of pixels per 1~arcsec$^{2}$ cell.  We repeat this process in each band, binning by 244~pixels for the WFC3/IR frames with 0.064\arcsec~pix$^{-1}$ and by 400~pixels for the ACS and WFC3/UVIS frames with 0.05\arcsec~pix$^{-1}$.

Next we performed aperture photometry using {\tt photutils} to estimate the surface brightness of the shells in each cell of our grid.  For F160W, these surface brightnesses range from 21.0--30.0~mag~arcsec$^{-2}$ with brighter shells tending to be at smaller projected separations from NGC~4993.  Finally, we segmented the gridded map of shell surface brightnesses into 24 individual shell features using a method analogous to {\tt CLUMPFIND} \citep{clumpfind} by identifying groups of local maxima in 4\arcsec$\times$4\arcsec\ rectangular apertures.  From the 24 local maxima we identified, we segmented the remaining pixels into features by identifying adjacent pixels whose surface brightnesses are within 0.3~mag of the local maximum.  We iterated on the previous step until no remaining pixels could be found within 0.3~mag of the previous step. 

\begin{figure}
    \centering
    \includegraphics[width=0.49\textwidth]{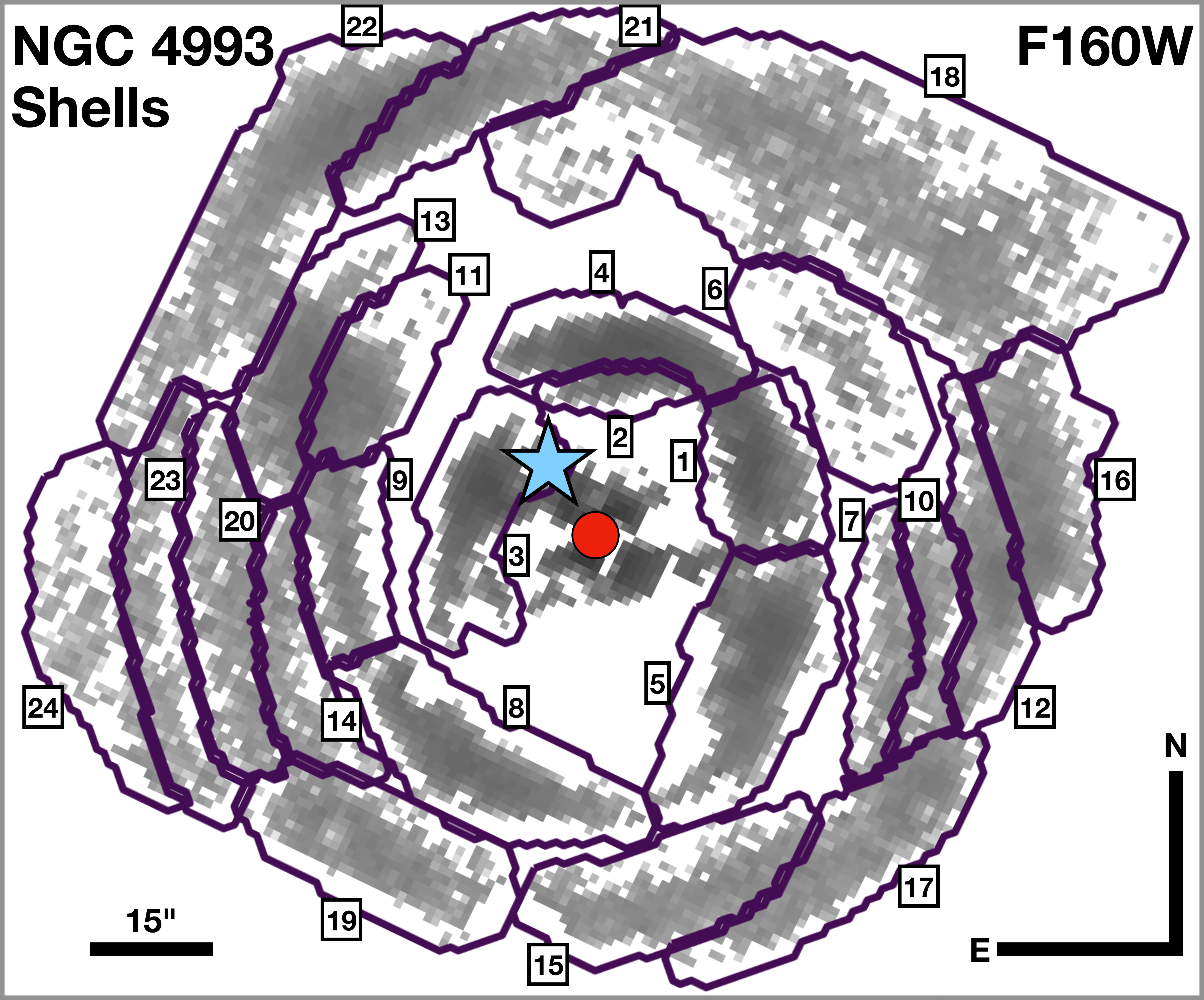}
    \caption{{\tt GALFIT}-subtracted F160W image of NGC~4993 with masking and binning into 1\arcsec$\times$1\arcsec\ cells as described in \autoref{sec:shell-id}.  We segment this image into twenty four shell features ordered from smallest to largest projected separation from the center of NGC~4993 (red circle), corresponding to the features given in \autoref{tab:shells}.  The location of AT~2017gfo is indicated with a blue star.}
    \label{fig:shell-id}
\end{figure}

We emphasize that these shell features are segmented arbitrarily and do not necessarily correspond to distinct ``shells'' as defined in shell-type galaxies or previous analysis of NGC~4993 \citep[e.g.,][]{MalinCarter83,Garcia93,Ebrova20}.  Thus we refer to these structures (which appear to form a double spiral structure in \autoref{fig:shell-id}) as ``shell features'' for convenience.  We order each shell feature from largest to smallest projected separation in \autoref{tab:shells} with that order overlaid on the F160W image in \autoref{fig:shell-id}.  These features do not have a one-to-one correspondence with the ten shells identified in \citet{Ebrova20} in part because we use F160W as opposed to F814W, we consider shell features at all position angles around NGC~4993 rather than those only along the major photometric axis, and our imaging is significantly deeper.  However, we are confident that this census of shell emission is complete to the depth of our masked residual F160W image.

\begin{deluxetable*}{lccccccccc}
\savetablenum{3}
\tabletypesize{\footnotesize}
\centering
\tablecolumns{4}
\tabcolsep0.06in
\def\arraystretch{0.94}
\tablecaption{Shell Locations, Photometry, and Stellar Population Properties}
\tablehead {
\colhead{\#} &
\colhead{$r_{\mathrm{sep}}$} &
\colhead{PA} &
\colhead{$\bar{S}_{\mathrm{F160W}}$} &
\colhead{F606W$-$F160W} &
\colhead{F814W$-$F160W} &
\colhead{F110W$-$F160W} &
\colhead{F140W$-$F160W} &
\colhead{$\log(M_{*}/M_{\odot})$} &
\colhead{$\log(T_{*}/\mathrm{yr})$} \\
&
(arcsec) &
(deg) &
(mag~arcsec$^{-2}$) &
(mag) &
(mag) &
(mag) &
(mag) &
&
}
\startdata
1 & 15.8 & 104.1 & 23.98 & 2.05 & 0.83 & 0.05 & 0.06 & 7.87$\substack{+0.12\\-0.12}$ & 9.79$\substack{+0.13\\-0.96}$ \\
2 & 17.3 & 32.7 & 24.22 & 1.71 & 0.96 & 0.23 & 0.01 & 7.45$\substack{+0.09\\-0.09}$ & 9.73$\substack{+0.10\\-0.70}$ \\
3 & 19.9 & 316.2 & 24.10 & 2.35 & 1.04 & 0.15 & 0.13 & 7.89$\substack{+0.15\\-0.15}$ & 9.82$\substack{+0.17\\-1.21}$ \\
4 & 21.3 & 37.2 & 23.71 & 1.51 & 0.91 & 0.08 & -0.13 & 7.52$\substack{+0.21\\-0.21}$ & 9.77$\substack{+0.24\\-1.43}$ \\
5 & 22.1 & 176.2 & 24.56 & 2.00 & 1.05 & -0.03 & -0.08 & 7.66$\substack{+0.19\\-0.19}$ & 9.80$\substack{+0.21\\-1.20}$ \\
6 & 31.2 & 92.8 & 27.42 & 1.28 & 0.78 & 0.27 & -0.03 & 6.02$\substack{+0.17\\-0.17}$ & 9.76$\substack{+0.25\\-1.51}$ \\
7 & 35.0 & 156.9 & 26.15 & 1.30 & 0.78 & 0.09 & 0.32 & 6.41$\substack{+0.17\\-0.17}$ & 9.74$\substack{+0.23\\-1.08}$ \\
8 & 36.5 & 255.3 & 25.00 & 1.97 & 0.94 & 0.27 & -0.01 & 7.68$\substack{+0.14\\-0.14}$ & 9.79$\substack{+0.16\\-0.90}$ \\
9 & 37.4 & 302.8 & 25.59 & 2.03 & 1.01 & 0.01 & -0.00 & 6.92$\substack{+0.18\\-0.18}$ & 9.79$\substack{+0.22\\-1.24}$ \\
10 & 38.0 & 148.7 & 26.09 & 1.50 & 0.79 & 0.30 & 0.26 & 6.63$\substack{+0.12\\-0.12}$ & 9.78$\substack{+0.16\\-0.76}$ \\
11 & 38.6 & 330.7 & 25.07 & 2.50 & 0.94 & 0.19 & -0.11 & 7.50$\substack{+0.15\\-0.15}$ & 9.81$\substack{+0.17\\-1.13}$ \\
12 & 43.2 & 138.0 & 25.47 & 2.18 & 0.88 & 0.15 & -0.04 & 7.37$\substack{+0.10\\-0.10}$ & 9.82$\substack{+0.12\\-0.88}$ \\
13 & 45.4 & 330.6 & 25.13 & 1.74 & 0.84 & 0.22 & 0.02 & 7.35$\substack{+0.12\\-0.12}$ & 9.82$\substack{+0.14\\-1.10}$ \\
14 & 46.2 & 290.2 & 26.30 & 3.02 & 0.94 & 0.44 & 0.42 & 6.58$\substack{+0.11\\-0.11}$ & 9.84$\substack{+0.14\\-0.88}$ \\
15 & 46.4 & 213.2 & 26.10 & 2.03 & 0.92 & 0.62 & 0.45 & 7.19$\substack{+0.12\\-0.12}$ & 9.82$\substack{+0.14\\-0.86}$ \\
16 & 47.9 & 125.2 & 25.58 & 1.93 & 0.93 & 0.50 & 0.30 & 7.29$\substack{+0.11\\-0.11}$ & 9.84$\substack{+0.13\\-0.81}$ \\
17 & 47.9 & 182.7 & 25.74 & 2.40 & 0.92 & 0.69 & 0.41 & 7.20$\substack{+0.14\\-0.14}$ & 9.83$\substack{+0.17\\-0.94}$ \\
18 & 50.0 & 71.8 & 26.44 & 2.14 & 0.88 & 0.36 & 0.18 & 7.27$\substack{+0.13\\-0.13}$ & 9.81$\substack{+0.15\\-1.02}$ \\
19 & 51.9 & 258.9 & 25.99 & 1.40 & 0.85 & 0.61 & 1.13 & 6.75$\substack{+0.12\\-0.12}$ & 9.82$\substack{+0.16\\-0.98}$ \\
20 & 52.2 & 295.5 & 26.39 & 1.80 & 0.92 & 0.26 & 0.16 & 6.53$\substack{+0.15\\-0.15}$ & 9.61$\substack{+0.20\\-0.64}$ \\
21 & 56.0 & 26.4 & 25.47 & 2.24 & 0.88 & 0.19 & 0.02 & 7.28$\substack{+0.12\\-0.12}$ & 9.83$\substack{+0.15\\-0.83}$ \\
22 & 59.8 & 211.8 & 25.68 & 2.42 & 0.99 & 0.58 & 0.41 & 7.64$\substack{+0.10\\-0.10}$ & 9.85$\substack{+0.12\\-0.75}$ \\
23 & 61.0 & 299.2 & 26.78 & 1.34 & 0.85 & 0.60 & 0.26 & 6.25$\substack{+0.13\\-0.13}$ & 9.83$\substack{+0.18\\-1.14}$ \\
24 & 68.0 & 294.3 & 27.48 & 1.40 & 0.91 & 0.40 & 0.16 & 5.39$\substack{+0.11\\-0.11}$ & 9.66$\substack{+0.17\\-1.07}$ \\
\enddata
\tablecomments{Properties of the 24 shell features we present in \autoref{sec:shells}.  Here shell \# corresponds to the features shown in \autoref{fig:shell-id}.  The projected separation ($r_{\mathrm{sep}}$) and position angle (PA; east of north) are averaged over each 1\arcsec$\times$1\arcsec\ cell in the F160W map and weighted by the F160W flux.  Both quantities are computed with respect to the center of NGC~4993 in our deep F160W image (red circle in \autoref{fig:shell-id}), which we measure to be RA=13:09:47.70, Dec=-23:23:02.305 (J2000) in the {\it Gaia} DR2 astrometric frame \citep{Lindegren18}.  $\bar{S}_{\mathrm{F160W}}$ is the average surface brightness across each shell in F160W.  We also indicate the average colors of each shell, weighted by the flux in F160W, in F606W, F814W, F110W, and F140W with respect to F160W.  $M_{\star}$ and $T_{\star}$ represent the total stellar mass and median mass-weighted stellar age, respectively, as determined from our {\tt Prospector} fits \autoref{sec:prospector} and averaged across each shell feature.}\label{tab:shells}
\end{deluxetable*}

The location of each shell, its photometric properties, and stellar population properties of each shell derived from {\tt Prospector} (see discussion in \autoref{sec:prospector}) are given in \autoref{tab:shells}.  The total stellar mass in the shell features is $6.3\substack{+2.6\\-1.5}\times10^{8}~M_{\odot}$ or approximately 1.4--2.5\% of the total mass in stars depending on the stellar mass estimate in NGC~4993 as a whole \citep[$3.0$--$4.5\times10^{10}~M_{\odot}$ in][]{Blanchard17,Palmese17,Ebrova20}.

\begin{figure}
    \centering
    \includegraphics[width=0.49\textwidth]{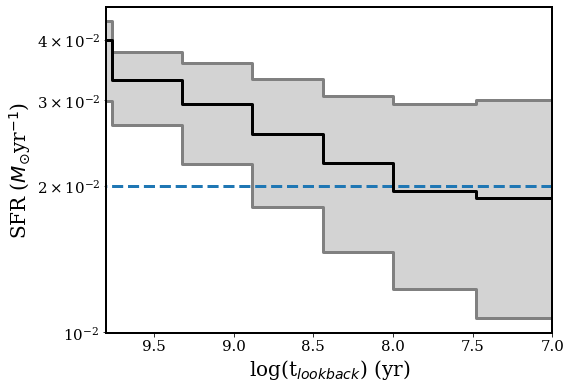}
    \caption{The star formation history averaged across all shell features in NGC~4993.  We include the median star formation rate (SFR) in $M_{\odot}~\mathrm{yr}^{-1}$ (black line) as well as the limits indicated by the 16th and 84th percentile most likely SFR (grey region), roughly corresponding to our 1$\sigma$ uncertainty.  The SFR declines monotonically with time from its peak rate of $\approx0.04~M_{\odot}~\mathrm{yr}^{-1}$ and begins to level off in the two youngest bins with ages of $<$262~Myr.  For reference, we draw a dashed line at the point where the predicted SFR is 1/2 the peak rate where this leveling off begins.}
    \label{fig:sfh}
\end{figure}

We consistently find a median mass-weighted age across the shell features of $>3$~Gyr, similar to the smooth galaxy light profile of $\approx 5-11$~Gyr. We note that these statistics are limited by our constraints on the spectral energy distribution blueward of F606W.  However, at most 1.3\% of all the stars in the shell features are $<262$~Myr based on the non-parametric star formation history described in \autoref{sec:prospector}.  The maximum star formation rate averaged across all of the shell features is approximately 0.04~$M_{\odot}~\mathrm{yr}^{-1}$ in the oldest ($>$4.7~Gyr) age bin (\autoref{fig:sfh}).  This is on the low side for the star formation main sequence at $z>1$ \citep[e.g.,][]{Noeske07} but still within $\approx$2$\sigma$ of expectations.  The star formation rate declines monotonically until it levels off in the youngest two age bins at the current day rate of $\approx0.02~M_{\odot}~\mathrm{yr}^{-1}$, roughly half of its peak rate.

\subsection{Properties of the Galaxy Merger of NGC~4993}

We analyzed the time since first impact of the merger for the shells identified around NGC~4993 in the context of the shell radius-age relations derived in \citet{Ebrova20} based on the radial profile model from \citet{Palmese17}.  This relation assumes that accreted galaxies plunge into their host galaxies on radial trajectories and stars stripped from the infalling galaxies then move along close-to radial orbits.  Thus the shells correspond to overdensities of stars near the apocenters of their orbits \citep[see also][]{Quinn84,Dupraz86}.  Due to the energy gradient in the satellite galaxy, the kinetic energy of the stars forming the shells, and thus the apocenter radii move outward with time, allows us to place a lower limit on the time from accretion by calculating the time it would take stars to reach the outermost shell in the gravitational potential of NGC~4993.

From the largest projected radius at which we detect any shell emission, we can calculate the lookback time since the merger, where a lookback time of zero corresponds to the redshift of NGC~4993. The outermost shell radius of 71.8\arcsec (or 14.2~kpc) corresponds to a minimum lookback time of 220~Myr, comparable to the 200~Myr estimate in \citet{Ebrova20}.  We do not detect any significant emission beyond this radius in our deep stacked F160W image, and so we are confident that there are no shell features to deep luminosity and mass limits ($<10^{6}~M_{\odot}$) based on our {\tt Prospector} analysis.

The star formation history derived in \autoref{sec:prospector} for the shell features levels off after our 262--685~Myr bin, and 2.5$\substack{+2.5\\-1.3}$\% of the total stellar mass is formed at this time.  We infer from this finding that the merger occurred at most 685~Myr ago, largely depleting the secondary galaxy of star formation material and resulting in a small fraction of younger stars in the shell structure. This finding and the small population of stars  with the youngest ages are still consistent with a relatively ``dry'' merger as noted by \citet{Levan17}. The presence of some additional gas from the merger is  also indicated by the weak AGN observed in {\it Chandra} and radio imaging \citep{Blanchard17} and in the WFC3/IR bands.  Our limit on the time since the merger is corroborated by upper limit derived from dynamical models performed in \citet{Ebrova20}, which suggest that the number and separation between the inner shells is consistent with a merger timescale of $<$600~Myr.  Although our limit on the age is less constraining, our independent method directly probes the stellar population that is likely formed in the merger, and so we consider the upper age limit to be $<$685~Myr below.

Based on the morphological substructure of NGC~4993 and the lower and upper limits on the time from merger we derived, we consider the true origin of the BNS progenitor of GW170817. Using the total estimated stellar mass of the shell features as a proxy for the secondary galaxy in the merger (i.e., the smaller galaxy that was accreted by NGC~4993), we find a minimum mass ratio of $\approx $1:50, classifying the past event as a minor merger. We note that it is possible that we are not completely characterizing the stellar mass of the secondary component due to geometric orientation of the shells and our view in projection, a small fraction of the secondary's stars not being located in the observed shells (i.e., not located close to the apocenter of their orbit), and the limits of our {\it HST} observations preclude detection of the lowest surface brightness features. Still, none of these reasons would account for a major increase in the mass ratio.

From simulations, the stellar mass ratios of shell-forming features are typically at least 1:10, but somewhat lower ratios are occasionally observed \citep[see e.g.][Figure 8]{Pop17}. The incidence of shells is also more common in isolated galaxies and known to decrease in groups or rich clusters \cite{Colbert2001}, and NGC~4993 is part of a group \citep{Garcia93}.

We conclude that based on the stellar mass in shells, it is highly improbable that the BNS progenitor of GW170817 originated from the low-mass secondary galaxy, and instead originated from the primary. However, based on the typical delay times of BNS mergers which extend to several Gyr, and the minimum time since merger of 220~Myr, it is fully plausible that the BNS progenitor was formed long before merger. In this scenario, the galaxy-scale merger may have affected the ultimate trajectory, orbit, and merger timescale of the progenitor of GW170817.

\section{Conclusions}

We present the full set of \hst\ observations of GW170817/AT~2017gfo obtained to date including new template observations obtained from 2021 Jan. 4 to Feb. 22.  The full \hst\ data set, representing over 140~ks of wide-band observations obtained in a range of filters and instruments, is one of the deepest sets of observations of $\lesssim$40~Mpc galaxy ever obtained with \hst.  These observations have enabled four new detections of the non-thermal afterglow from 2017 Dec 7 to 2018 Feb 6 as well as new limits on the presence of optical and near-IR sources around AT~2017gfo and extended shell emission in NGC~4993 as a whole.

The new detections of the non-thermal GRB afterglow, which span from 109--170 rest-frame days post-merger, remain consistent with an unchanging spectral index of $\beta\approx-0.6$.  However, similar constraints on the evolution of NS merger counterparts out to later times ($>$100~days) can yield insight into their origin and the processes driving their emission mechanisms, including kilonova afterglows \citep{Hajela21}, magnetar-boosted events \citep{Fong21a}, IR dust echoes \citep{Lu21}, and variations in the circum-merger density \citep[e.g.,][]{RamirezRuiz19}.

The limits on sources near AT~2017gfo derived from the full \hst\ data set are also significantly constraining for nearby GCs, which have been suggested as a potential origin for the binary NS progenitor system of the merger \citep{Belczynski18,Baillot19,Ye20}.  We can detect all GCs, including those with the reddest colors, down to 4.6$\times$10$^{4}~M_{\odot}$, representing the large majority of the GCLF.  Future binary NS mergers, especially those near the distance limit of the LIGO during Observing Runs 4 and 5, will likely not have such strong limits on the presence of coincident GCs.  However, even at 200~Mpc, limits similar to those for AT~2017gfo can rule out 27\% of the GCLF, precluding an origin from the most massive and luminous GCs.

The total \hst\ data set also enables the most complete sample of low surface brightness features around NGC~4993, and in particular the shell structure first noted in \citet{MalinCarter83}.  These data provide an observational blueprint for a Type II shell galaxy, that can be studied irrespective of its connection to BNS mergers. Using F606W through F160W imaging, we are able to identify shell structure out to $\approx$71.8\arcsec, the most complete census of such emission around NGC~4993 to date.  Fitting this photometry with {\tt Prospector} stellar population models, we constrain the total stellar mass in the shells to be 6.3$\times$10$^{8}~M_{\odot}$, approximately 2\% of that in NGC~4993 as a whole, with a mass-weighted stellar age across all of the shells $>3$~Gyr.  The geometry of the shells supports an age from 220--600~Myr \citep{Ebrova20}, and the star formation history of the shells support a maximum time since merger of ${\sim}685$~Myr. Given the lack of evidence for a very young stellar component in the shells, we consider it unlikely that the progenitor of AT~2017gfo originated in this stellar population.

\acknowledgments

We thank W. Lu for providing the IR dust echo models presented in this paper, and Jay Strader for helpful discussions.
Support for program \#15886 was provided by NASA through a grant from the Space Telescope Science Institute, which is operated by the Association of Universities for Research in Astronomy, Inc., under NASA contract NAS 5-26555.
The Fong Group at Northwestern acknowledges support by the National Science Foundation under grant Nos. AST-1814782, AST-1909358 and CAREER grant No. AST-2047919.
A.~Hajela is partially supported by a Future Investigators in NASA Earth and Space Science and Technology (FINESST) award \#\,80NSSC19K1422.
Some of the data used in this paper were obtained from the Mikulski Archive for Space Telescopes (MAST), which is operated by NASA.  These data come from programs GO-14270 (PI Levan), GO-14607 (PI Troja), GO-14771 (PI Tanvir), GO-14804 (PI Levan), SNAP-14840 (PI Bellini), GO-14850 (PI Troja), GO-15329 (PI Berger), GO-15346 (PI Kasliwal), GO-15482 (PI Lyman), and GO-15606 (Co-PIs Margutti \& Fong).
K.~D.~A.~acknowledges support provided by NASA through the NASA Hubble Fellowship grant HST-HF2-51403.001 awarded by the
Space Telescope Science Institute, which is operated by the Association of Universities for Research in Astronomy, Inc., for NASA, under contract NAS5-26555.\\

\bigskip

\vspace{2mm}
\facilities{{\it HST} (ACS, WFC3)}
\vspace{2mm}
\software{{\tt astropy} \citep{astropy},
          {\tt dolphot} \citep{dolphot},
          {\tt drizzlepac} \citep{drizzlepac},
          {\tt hotpants} \citep{hotpants},
          {\tt photutils} \citep{photutils}}
          
\clearpage

\appendix

\section{Shell Stellar Population Modeling}\label{sec:prospector}

We constrained the properties of the stellar population in the NGC~4993 shells using the stellar population inference code {\tt Propsector} \citep{Leja17,Johnson21}.  This analysis is based on the broad-band photometry from each of our deep stacked and {\tt GALFIT}-subtracted images in \autoref{sec:shells}.  Following the masking and gridding procedure described in \autoref{sec:shell-id}, we gridded each subtracted image into 4\arcsec$\times$4\arcsec\ cells.  We then performed photometry and identified 268 such cells in which there was flux detected at $>$3$\sigma$ in the F606W, F814W, F110W, F140W, and F160W frames.  This set of photometry formed the basis for 268 independent {\tt Prospector} fits.

We fit this photometry using a non-parametric star formation history (SFH) with seven age bins, with the first two spaced from 0--30~Myr and 30--100~Myr, and the remaining five log-spaced in time with an upper limit at the age of the Universe (13.63~Gyr) at the redshift of NGC~4993.  We assumed an initial mass function from \citet{Chabrier03}.  In addition, we adopt a continuity prior as described in \citet{Fong21} and originally presented in \citet{Leja19a} such that the SFH does not sharply deviate from a flat distribution, but otherwise we do not place any constraints on SFH.

We used this SFH model to fit for a total stellar mass ($\log M_{\star}/M_{\odot}$) and metallicity ($\log Z_{\star}/Z_{\odot}$).  We also include two dust components to model the attenuation of light due to dust in stellar birth clouds and affecting only young stars ({\tt dust1}) and attenuation of light in the diffuse interstellar medium ({\tt dust2}, parameterized as {\tt dust2/dust1}).  Both parameters can be interpreted as the additional optical depth due to each dust component \citep[see, e.g.,][]{Conroy13,Kriek13,Price14}.  This model is a power-law perturbation from the dust attenuation curve in \citet{Calzetti00}, and so we fit a power-law index for the attenuation curve for {\tt dust1} and {\tt dust2}, referred to as $\delta$, for which we adopt a flat prior from $-1.0$ to 0.4.

For all 268 {\tt Prospector} fits, we assume the fixed distance and redshift given above and our input photometry was corrected for Milky Way extinction.  We performed the fit in each of the 268 cells by jointly fitting all five bands in that cell with the nested sampling routine {\tt dynesty} \citep{Speagle20}.  The in-band magnitudes are inferred for each WFC3/UVIS or WFC3/IR band using {\tt Python-fsps} \citep{Conroy09,Conroy10}.  We adopt an error floor of 5\% in the photometric uncertainties to avoid overfitting to the shell flux in each cell, some of which have photometric uncertainties $<$0.01~mag.

To combine the output quantities for each cell into individual shell features as indicated in \autoref{tab:shells} and \autoref{fig:shell-id}, we concatenated the sampled parameters for the {\tt Prospector} fit in each cell.  Each stellar mass in the concatenated samples was rescaled by the total stellar mass inferred across all of the cells we included in that shell feature.  

Finally, we note that the 4\arcsec$\times$4\arcsec\ cells do not span the same solid angle as the 1\arcsec$\times$1\arcsec\ cells we used to segment the shell emission in \autoref{sec:shell-id}.  This is due to the fact that we require a $>$3$\sigma$ detection in all five photometric bands, and the shells are detected at higher significance in the F160W band.  Therefore, we rescaled the total stellar mass in each shell feature by the ratio between the F160W flux in the 1\arcsec$\times$1\arcsec\ map to the same flux in the 4\arcsec$\times$4\arcsec\ map.  This ratio is 1.1--2.5 for each shell, with features at larger projected separations tending to have larger ratios.  The final stellar mass inferred for each shell feature is given in \autoref{tab:shells} along with the median mass-weighted age.

\bibliography{main}
\bibliographystyle{aasjournal}

\end{document}

%% file: affiliation.tex
\newcommand{\NU}{\affiliation{Center for Interdisciplinary Exploration and Research in Astrophysics (CIERA) and Department of Physics and Astronomy, Northwestern University, Evanston, IL 60208, USA}}

\newcommand{\CfA}{\affiliation{Center for Astrophysics\:$|$\:Harvard \& Smithsonian, 60 Garden St. Cambridge, MA 02138, USA}}

\newcommand{\FermiLab}{\affiliation{Cosmic Physics Center, Fermi National Accelerator Laboratory, PO Box 500, Batavia, IL 60510-5011, USA}}

\newcommand{\KICP}{\affiliation{Kavli Institute for Cosmological Physics, The University of Chicago, Chicago, IL 60637, USA}}

\newcommand{\PSUAA}{\affiliation{Department of Astronomy \& Astrophysics, The Pennsylvania State University, University Park, PA 16802, USA}}

\newcommand{\ICDS}{\affiliation{Institute for Computational \& Data Sciences, The Pennsylvania State University, University Park, PA, USA}}

\newcommand{\IGC}{\affiliation{Institute for Gravitation and the Cosmos, The Pennsylvania State University, University Park, PA 16802, USA}}

\newcommand{\UCBerkeley}{\affiliation{Department of Astronomy, University of California, Berkeley, CA 94720-3411, USA}}

\newcommand{\Einstein}{\altaffiliation{NASA Einstein Fellow}}   

%% file: authors.tex
\author[0000-0002-5740-7747]{Charles D. Kilpatrick}
\NU

\author[0000-0002-7374-935X]{Wen-fai Fong}
\NU

\author[0000-0003-0526-2248]{Peter K. Blanchard}
\NU

\author[0000-0001-6755-1315]{Joel~Leja}
\PSUAA\ICDS\IGC

\author[0000-0002-2028-9329]{Anya E. Nugent}
\NU

\author[0000-0002-6011-0530]{Antonella Palmese}
\FermiLab\KICP

\author[0000-0001-8340-3486]{Kerry Paterson}
\NU

\author[0000-0003-2539-8206]{Tjitske Starkenburg}
\NU

\author[0000-0002-8297-2473]{Kate D. Alexander}
\Einstein\NU

\author[0000-0002-9392-9681]{Edo Berger} 
\CfA

\author[0000-0002-7706-5668]{Ryan Chornock}
\UCBerkeley

\author[0000-0003-2349-101X]{Aprajita Hajela}
\NU

\author[0000-0002-8297-2473]{Raffaella Margutti}
\UCBerkeley

%% file: observations.tex
\begin{deluxetable*}{cccllccc}
\savetablenum{1}
\tabletypesize{\footnotesize}
\centering
\tablecolumns{9}
\tabcolsep0.06in
\def\arraystretch{0.93}
\tablecaption{{\it Hubble Space Telescope} Photometry of the Ultraviolet, Optical, and Near-Infrared Counterpart to GW170817\label{tab:observations}}
\tablehead {
\colhead{Start Date} &
\colhead{MJD} &
\colhead{Rest-frame Epoch} &
\colhead{Instrument} &
\colhead{Filter} &
\colhead{Exposure Time} &
\colhead{Magnitude} &
\colhead{Magnitude Error} \\
(UTC) &
&
(days) &
&
&
(s) &
(AB mag) &
(mag)
}
\startdata
2017 Apr 28.15\tablenotemark{\scriptsize a} & 57871.15321 & $-$110.29 &    ACS/WFC &  F606W &   348.00 & $>$27.20 &    -- \\
2017 Aug 22.32 & 57987.31734 &  4.74 &    WFC3/IR &  F110W &   297.69 &    18.456 & 0.002 \\
2017 Aug 22.45 & 57987.44978 &  4.87 &    WFC3/IR &  F160W &   297.69 &    18.033 & 0.002 \\
2017 Aug 22.86 & 57987.85647 &  5.28 &  WFC3/UVIS &  F336W &  1089.00 &    25.04 & 0.104 \\
2017 Aug 22.92 & 57987.91564 &  5.33 &  WFC3/UVIS &  F225W &  1089.00 & $>$26.13 &    -- \\
2017 Aug 24.77 & 57989.76747 &  7.17 &    WFC3/IR &  F110W &   297.69 &    19.376 & 0.003 \\
2017 Aug 25.58 & 57990.57922 &  7.97 &  WFC3/UVIS &  F606W &   452.00 &    22.467 & 0.019 \\
2017 Aug 25.60 & 57990.60205 &  8.00 &  WFC3/UVIS &  F475W &   520.00 &    23.639 & 0.026 \\
2017 Aug 25.65 & 57990.65157 &  8.04 &  WFC3/UVIS &  F275W &   620.00 & $>$25.62 &    -- \\
2017 Aug 27.36 & 57992.35973 &  9.74 &  WFC3/UVIS &  F336W &  2200.00 & $>$26.52 &    -- \\
2017 Aug 27.45 & 57992.44996 &  9.83 &    WFC3/IR &  F160W &  1607.12 &    19.650 & 0.003 \\
2017 Aug 27.45 & 57992.45483 &  9.83 &    ACS/WFC &  F475W &  1395.00 &    24.085 & 0.021 \\
2017 Aug 27.50 & 57992.50148 &  9.88 &    ACS/WFC &  F625W &   890.00 &    22.899 & 0.108 \\
2017 Aug 27.56 & 57992.56367 &  9.94 &    ACS/WFC &  F775W &   740.00 &    22.186 & 0.069 \\
2017 Aug 27.60 & 57992.60180 &  9.98 &    ACS/WFC & F850LP &   680.00 &    21.557 & 0.027 \\
2017 Aug 27.63 & 57992.62637 & 10.00 &    WFC3/IR &  F110W &  1309.43 &    20.490 & 0.004 \\
2017 Aug 28.88 & 57993.88414 & 11.25 &  WFC3/UVIS &  F475W &  1120.00 &    24.720 & 0.057 \\
2017 Aug 28.89 & 57993.88553 & 11.25 &  WFC3/UVIS &  F606W &   768.00 &    23.651 & 0.031 \\
2017 Aug 28.94 & 57993.93794 & 11.30 &  WFC3/UVIS &  F275W &  1709.00 & $>$25.82 &    -- \\
2017 Aug 28.95 & 57993.95272 & 11.31 &  WFC3/UVIS &  F814W &   560.00 &    22.316 & 0.030 \\
2017 Aug 29.01 & 57994.00675 & 11.37 &  WFC3/UVIS &  F336W &   946.00 & $>$26.38 &    -- \\
2017 Aug 29.02 & 57994.01757 & 11.38 &  WFC3/UVIS &  F275W &  1089.00 & $>$26.32 &    -- \\
2017 Dec 06.02 & 58093.01904 & 109.42 &  WFC3/UVIS &  F606W &  2264.00 &    26.310 & 0.190 \\
2017 Dec 06.15 & 58093.15206 & 109.55 &  WFC3/UVIS &  F814W &  2400.00 &    26.298 & 0.150 \\
2017 Dec 06.43 & 58093.42907 & 109.82 &    WFC3/IR &  F140W &  4793.86 & $>$25.30 &    -- \\
2017 Dec 07.42\tablenotemark{\scriptsize b} & 58094.41545 & 110.80 &    WFC3/IR &  F160W &  4808.67 & 25.612 &    0.288 \\
2017 Dec 08.93 & 58095.93259 & 112.30 &    WFC3/IR &  F110W &  7635.25 & 25.908 &    0.244 \\
2018 Jan 01.57 & 58119.57043 & 135.71 &    ACS/WFC &  F606W &  2120.00 &    26.590 & 0.230 \\
2018 Jan 29.72 & 58147.71752 & 163.59 &  WFC3/UVIS &  F606W &  2372.00 &    26.500 & 0.190 \\
2018 Feb 05.67 & 58154.66997 & 170.47 &  WFC3/UVIS &  F814W &  2400.00 &    26.495 & 0.194 \\
2018 Feb 05.74 & 58154.73617 & 170.54 &  WFC3/UVIS &  F606W &  2400.00 &    26.580 & 0.220 \\
2018 Mar 14.62 & 58191.62273 & 207.06 &  WFC3/UVIS &  F606W &  2432.00 &    26.610 & 0.260 \\
2018 Mar 23.89 & 58200.89223 & 216.24 &    ACS/WFC &  F606W &  2120.00 & $>$26.90 &    -- \\
2018 Jun 10.32 & 58279.32202 & 293.91 &  WFC3/UVIS &  F606W &  5220.00 &    27.290 & 0.200 \\
2018 Jul 11.75\tablenotemark{\scriptsize b} & 58310.74890 & 325.04 &  WFC3/UVIS &  F606W &  14070.00 & $>$27.58 &    -- \\
2018 Jul 20.35 & 58319.35419 & 333.56 &    ACS/WFC &  F606W &  2120.00 & $>$27.72 &    -- \\
2018 Aug 08.46 & 58338.45574 & 352.47 &  WFC3/UVIS &  F814W &  5220.00 & $>$27.41 &    -- \\
2018 Aug 14.94 & 58344.93633 & 358.89 &  WFC3/UVIS &  F606W & 14070.00 &    27.830 & 0.290 \\
2019 Mar 24.67\tablenotemark{\scriptsize b} & 58566.67373 & 578.48 & ACS/WFC & F606W & 6728.00 & $>$28.40 & -- \\
2021 Jan 04.98 & 59218.97751 & 1224.45 &    WFC3/IR &  F140W &  7823.49 & $>$27.04 &    -- \\
2021 Jan 06.17 & 59220.17390 & 1225.63 &    WFC3/IR &  F160W &  7823.49 & $>$26.58 &    -- \\
2021 Feb 07.75 & 59252.74630 & 1257.89 &    WFC3/IR &  F110W & 10423.49 & $>$27.68 &    -- \\
2021 Feb 22.62 & 59267.61813 & 1272.62 &  WFC3/UVIS &  F814W &  7940.00 & $>$27.20 &    -- \\ \hline
-- & -- & -- & WFC3+ACS & F606W & 64212.00 & $>$28.80\tablenotemark{\scriptsize c} & -- \\
-- & -- & -- & WFC3/UVIS& F814W & 18520.00 & $>$28.60\tablenotemark{\scriptsize c} & -- \\
-- & -- & -- & WFC3/IR  & F110W & 19963.56 & $>$28.07\tablenotemark{\scriptsize c} & -- \\
-- & -- & -- & WFC3/IR  & F140W & 12617.34 & $>$27.58\tablenotemark{\scriptsize c} & -- \\
-- & -- & -- & WFC3/IR  & F160W & 14536.96 & $>$26.81\tablenotemark{\scriptsize c} & -- \\
\enddata
Upper limits correspond to $3\sigma$ confidence.
\tablenotetext{a}{This is the only pre-merger epoch of \hst\ imaging \citep[see discussion for upper limits in][]{Kilpatrick17}.}
\vspace{-0.06in}
\tablenotetext{b}{We combine imaging in the same band and obtained 1--7~days apart: 2017 Dec 6 and 2017 Dec 8 (F160W), 2018 Jul 10 and 2018 Jul 13 (F606W), 2019 Mar 21 and 2019 Mar 27 \citep[F606W; see also][]{Fong19}.}
\vspace{-0.06in}
\tablenotetext{c}{Limits on a source within 2\arcsec\ of GW170817 across all \hst\ imaging as discussed in \autoref{sec:limits}.}
\end{deluxetable*}

%% file: main.bbl
\begin{thebibliography}{}
\expandafter\ifx\csname natexlab\endcsname\relax\def\natexlab#1{#1}\fi
\providecommand{\url}[1]{\href{#1}{#1}}
\providecommand{\dodoi}[1]{doi:~\href{http://doi.org/#1}{\nolinkurl{#1}}}
\providecommand{\doeprint}[1]{\href{http://ascl.net/#1}{\nolinkurl{http://ascl.net/#1}}}
\providecommand{\doarXiv}[1]{\href{https://arxiv.org/abs/#1}{\nolinkurl{https://arxiv.org/abs/#1}}}

\bibitem[{{Abbott} {et~al.}(2017{\natexlab{a}}){Abbott}, {Abbott}, {Abbott},
  {Acernese}, {Ackley}, {Adams}, {Adams}, {Addesso}, {Adhikari}, {Adya},
  {Affeldt}, {Afrough}, {Agarwal}, {Agathos}, {Agatsuma}, {Aggarwal}, {Aguiar},
  {Aiello}, {Ain}, {Ajith}, {Allen}, {Allen}, {Allocca}, {Altin}, {Amato},
  {Ananyeva}, {Anderson}, {Anderson}, {Angelova}, {Antier}, {Appert}, {Arai},
  {Araya}, {Areeda}, {Arnaud}, {Arun}, {Ascenzi}, {Ashton}, {Ast}, {Aston},
  {Astone}, {Atallah}, {Aufmuth}, {Aulbert}, {AultONeal}, {Austin},
  {Avila-Alvarez}, {Babak}, {Bacon}, {Bader}, {Bae}, {Bailes}, {Baker},
  {Baldaccini}, {Ballardin}, {Ballmer}, {Banagiri}, {Barayoga}, {Barclay},
  {Barish}, {Barker}, {Barkett}, {Barone}, {Barr}, {Barsotti}, {Barsuglia},
  {Barta}, {Barthelmy}, {Bartlett}, {Bartos}, {Bassiri}, {Basti}, {Batch},
  {Bawaj}, {Bayley}, {Bazzan}, {B{\'e}csy}, {Beer}, {Bejger}, {Belahcene},
  {Bell}, {Berger}, {Bergmann}, {Bernuzzi}, {Bero}, {Berry}, {Bersanetti},
  {Bertolini}, {Betzwieser}, {Bhagwat}, {Bhandare}, {Bilenko}, {Billingsley},
  {Billman}, {Birch}, {Birney}, {Birnholtz}, {Biscans}, {Biscoveanu}, {Bisht},
  {Bitossi}, {Biwer}, {Bizouard}, {Blackburn}, {Blackman}, {Blair}, {Blair},
  {Blair}, {Bloemen}, {Bock}, {Bode}, {Boer}, {Bogaert}, {Bohe}, {Bondu},
  {Bonilla}, {Bonnand}, {Boom}, {Bork}, {Boschi}, {Bose}, {Bossie},
  {Bouffanais}, {Bozzi}, {Bradaschia}, {Brady}, {Branchesi}, {Brau}, {Briant},
  {Brillet}, {Brinkmann}, {Brisson}, {Brockill}, {Broida}, {Brooks}, {Brown},
  {Brown}, {Brunett}, {Buchanan}, {Buikema}, {Bulik}, {Bulten}, {Buonanno},
  {Buskulic}, {Buy}, {Byer}, {Cabero}, {Cadonati}, {Cagnoli}, {Cahillane},
  {Calder{\'o}n Bustillo}, {Callister}, {Calloni}, {Camp}, {Canepa},
  {Canizares}, {Cannon}, {Cao}, {Cao}, {Capano}, {Capocasa}, {Carbognani},
  {Caride}, {Carney}, {Carullo}, {Casanueva Diaz}, {Casentini}, {Caudill},
  {Cavagli{\`a}}, {Cavalier}, {Cavalieri}, {Cella}, {Cepeda},
  {Cerd{\'a}-Dur{\'a}n}, {Cerretani}, {Cesarini}, {Chamberlin}, {Chan}, {Chao},
  {Charlton}, {Chase}, {Chassande-Mottin}, {Chatterjee}, {Chatziioannou},
  {Cheeseboro}, {Chen}, {Chen}, {Chen}, {Cheng}, {Chia}, {Chincarini},
  {Chiummo}, {Chmiel}, {Cho}, {Cho}, {Chow}, {Christensen}, {Chu}, {Chua},
  {Chua}, {Chung}, {Chung}, {Ciani}, {Ciolfi}, {Cirelli}, {Cirone}, {Clara},
  {Clark}, {Clearwater}, {Cleva}, {Cocchieri}, {Coccia}, {Cohadon}, {Cohen},
  {Colla}, {Collette}, {Cominsky}, {Constancio}, {Conti}, {Cooper}, {Corban},
  {Corbitt}, {Cordero-Carri{\'o}n}, {Corley}, {Cornish}, {Corsi}, {Cortese},
  {Costa}, {Coughlin}, {Coughlin}, {Coulon}, {Countryman}, {Couvares}, {Covas},
  {Cowan}, {Coward}, {Cowart}, {Coyne}, {Coyne}, {Creighton}, {Creighton},
  {Cripe}, {Crowder}, {Cullen}, {Cumming}, {Cunningham}, {Cuoco}, {Dal Canton},
  {D{\'a}lya}, {Danilishin}, {D'Antonio}, {Danzmann}, {Dasgupta}, {Da Silva
  Costa}, {Dattilo}, {Dave}, {Davier}, {Davis}, {Daw}, {Day}, {De}, {DeBra},
  {Degallaix}, {De Laurentis}, {Del{\'e}glise}, {Del Pozzo}, {Demos}, {Denker},
  {Dent}, {De Pietri}, {Dergachev}, {De Rosa}, {DeRosa}, {De Rossi}, {DeSalvo},
  {de Varona}, {Devenson}, {Dhurandhar}, {D{\'\i}az}, {Dietrich}, {Di Fiore},
  {Di Giovanni}, {Di Girolamo}, {Di Lieto}, {Di Pace}, {Di Palma}, {Di Renzo},
  {Doctor}, {Dolique}, {Donovan}, {Dooley}, {Doravari}, {Dorrington},
  {Douglas}, {Dovale {\'A}lvarez}, {Downes}, {Drago}, {Dreissigacker},
  {Driggers}, {Du}, {Ducrot}, {Dudi}, {Dupej}, {Dwyer}, {Edo}, {Edwards},
  {Effler}, {Eggenstein}, {Ehrens}, {Eichholz}, {Eikenberry}, {Eisenstein},
  {Essick}, {Estevez}, {Etienne}, {Etzel}, {Evans}, {Evans}, {Factourovich},
  {Fafone}, {Fair}, {Fairhurst}, {Fan}, {Farinon}, {Farr}, {Farr},
  {Fauchon-Jones}, {Favata}, {Fays}, {Fee}, {Fehrmann}, {Feicht}, {Fejer},
  {Fernandez-Galiana}, {Ferrante}, {Ferreira}, {Ferrini}, {Fidecaro},
  {Finstad}, {Fiori}, {Fiorucci}, {Fishbach}, {Fisher}, {Fitz-Axen},
  {Flaminio}, {Fletcher}, {Fong}, {Font}, {Forsyth}, {Forsyth}, {Fournier},
  {Frasca}, {Frasconi}, {Frei}, {Freise}, {Frey}, {Frey}, {Fries}, {Fritschel},
  {Frolov}, {Fulda}, {Fyffe}, {Gabbard}, {Gadre}, {Gaebel}, {Gair},
  {Gammaitoni}, {Ganija}, {Gaonkar}, {Garcia-Quiros}, {Garufi}, {Gateley},
  {Gaudio}, {Gaur}, {Gayathri}, {Gehrels}, {Gemme}, {Genin}, {Gennai},
  {George}, {George}, {Gergely}, {Germain}, {Ghonge}, {Ghosh}, {Ghosh},
  {Ghosh}, {Giaime}, {Giardina}, {Giazotto}, {Gill}, {Glover}, {Goetz},
  {Goetz}, {Gomes}, {Goncharov}, {Gonz{\'a}lez}, {Gonzalez Castro},
  {Gopakumar}, {Gorodetsky}, {Gossan}, {Gosselin}, {Gouaty}, {Grado}, {Graef},
  {Granata}, {Grant}, {Gras}, {Gray}, {Greco}, {Green}, {Gretarsson}, {Groot},
  {Grote}, {Grunewald}, {Gruning}, {Guidi}, {Guo}, {Gupta}, {Gupta}, {Gushwa},
  {Gustafson}, {Gustafson}, {Halim}, {Hall}, {Hall}, {Hamilton}, {Hammond},
  {Haney}, {Hanke}, {Hanks}, {Hanna}, {Hannam}, {Hannuksela}, {Hanson},
  {Hardwick}, {Harms}, {Harry}, {Harry}, {Hart}, {Haster}, {Haughian}, {Healy},
  {Heidmann}, {Heintze}, {Heitmann}, {Hello}, {Hemming}, {Hendry}, {Heng},
  {Hennig}, {Heptonstall}, {Heurs}, {Hild}, {Hinderer}, {Ho}, {Hoak}, {Hofman},
  {Holt}, {Holz}, {Hopkins}, {Horst}, {Hough}, {Houston}, {Howell}, {Hreibi},
  {Hu}, {Huerta}, {Huet}, {Hughey}, {Husa}, {Huttner}, {Huynh-Dinh}, {Indik},
  {Inta}, {Intini}, {Isa}, {Isac}, {Isi}, {Iyer}, {Izumi}, {Jacqmin}, {Jani},
  {Jaranowski}, {Jawahar}, {Jim{\'e}nez-Forteza}, {Johnson},
  {Johnson-McDaniel}, {Jones}, {Jones}, {Jonker}, {Ju}, {Junker}, {Kalaghatgi},
  {Kalogera}, {Kamai}, {Kandhasamy}, {Kang}, {Kanner}, {Kapadia}, {Karki},
  {Karvinen}, {Kasprzack}, {Kastaun}, {Katolik}, {Katsavounidis}, {Katzman},
  {Kaufer}, {Kawabe}, {K{\'e}f{\'e}lian}, {Keitel}, {Kemball}, {Kennedy},
  {Kent}, {Key}, {Khalili}, {Khan}, {Khan}, {Khan}, {Khazanov}, {Kijbunchoo},
  {Kim}, {Kim}, {Kim}, {Kim}, {Kim}, {Kim}, {Kimbrell}, {King}, {King},
  {Kinley-Hanlon}, {Kirchhoff}, {Kissel}, {Kleybolte}, {Klimenko}, {Knowles},
  {Koch}, {Koehlenbeck}, {Koley}, {Kondrashov}, {Kontos}, {Korobko}, {Korth},
  {Kowalska}, {Kozak}, {Kr{\"a}mer}, {Kringel}, {Krishnan}, {Kr{\'o}lak},
  {Kuehn}, {Kumar}, {Kumar}, {Kumar}, {Kuo}, {Kutynia}, {Kwang}, {Lackey},
  {Lai}, {Landry}, {Lang}, {Lange}, {Lantz}, {Lanza}, {Larson},
  {Lartaux-Vollard}, {Lasky}, {Laxen}, {Lazzarini}, {Lazzaro}, {Leaci},
  {Leavey}, {Lee}, {Lee}, {Lee}, {Lee}, {Lee}, {Lehmann}, {Lenon}, {Leon},
  {Leonardi}, {Leroy}, {Letendre}, {Levin}, {Li}, {Linker}, {Littenberg},
  {Liu}, {Liu}, {Lo}, {Lockerbie}, {London}, {Lord}, {Lorenzini}, {Loriette},
  {Lormand}, {Losurdo}, {Lough}, {Lousto}, {Lovelace}, {L{\"u}ck}, {Lumaca},
  {Lundgren}, {Lynch}, {Ma}, {Macas}, {Macfoy}, {Machenschalk}, {MacInnis},
  {Macleod}, {Maga{\~n}a Hernandez}, {Maga{\~n}a-Sandoval}, {Maga{\~n}a
  Zertuche}, {Magee}, {Majorana}, {Maksimovic}, {Man}, {Mandic}, {Mangano},
  {Mansell}, {Manske}, {Mantovani}, {Marchesoni}, {Marion}, {M{\'a}rka},
  {M{\'a}rka}, {Markakis}, {Markosyan}, {Markowitz}, {Maros}, {Marquina},
  {Marsh}, {Martelli}, {Martellini}, {Martin}, {Martin}, {Martynov}, {Marx},
  {Mason}, {Massera}, {Masserot}, {Massinger}, {Masso-Reid}, {Mastrogiovanni},
  {Matas}, {Matichard}, {Matone}, {Mavalvala}, {Mazumder}, {McCarthy},
  {McClelland}, {McCormick}, {McCuller}, {McGuire}, {McIntyre}, {McIver},
  {McManus}, {McNeill}, {McRae}, {McWilliams}, {Meacher}, {Meadors}, {Mehmet},
  {Meidam}, {Mejuto-Villa}, {Melatos}, {Mendell}, {Mercer}, {Merilh},
  {Merzougui}, {Meshkov}, {Messenger}, {Messick}, {Metzdorff}, {Meyers},
  {Miao}, {Michel}, {Middleton}, {Mikhailov}, {Milano}, {Miller}, {Miller},
  {Miller}, {Millhouse}, {Milovich-Goff}, {Minazzoli}, {Minenkov}, {Ming},
  {Mishra}, {Mitra}, {Mitrofanov}, {Mitselmakher}, {Mittleman}, {Moffa},
  {Moggi}, {Mogushi}, {Mohan}, {Mohapatra}, {Molina}, {Montani}, {Moore},
  {Moraru}, {Moreno}, {Morisaki}, {Morriss}, {Mours}, {Mow-Lowry}, {Mueller},
  {Muir}, {Mukherjee}, {Mukherjee}, {Mukherjee}, {Mukund}, {Mullavey}, {Munch},
  {Mu{\~n}iz}, {Muratore}, {Murray}, {Nagar}, {Napier}, {Nardecchia},
  {Naticchioni}, {Nayak}, {Neilson}, {Nelemans}, {Nelson}, {Nery}, {Neunzert},
  {Nevin}, {Newport}, {Newton}, {Ng}, {Nguyen}, {Nguyen}, {Nichols}, {Nielsen},
  {Nissanke}, {Nitz}, {Noack}, {Nocera}, {Nolting}, {North}, {Nuttall},
  {Oberling}, {O'Dea}, {Ogin}, {Oh}, {Oh}, {Ohme}, {Okada}, {Oliver},
  {Oppermann}, {Oram}, {O'Reilly}, {Ormiston}, {Ortega}, {O'Shaughnessy},
  {Ossokine}, {Ottaway}, {Overmier}, {Owen}, {Pace}, {Page}, {Page}, {Pai},
  {Pai}, {Palamos}, {Palashov}, {Palomba}, {Pal-Singh}, {Pan}, {Pan}, {Pang},
  {Pang}, {Pankow}, {Pannarale}, {Pant}, {Paoletti}, {Paoli}, {Papa}, {Parida},
  {Parker}, {Pascucci}, {Pasqualetti}, {Passaquieti}, {Passuello}, {Patil},
  {Patricelli}, {Pearlstone}, {Pedraza}, {Pedurand}, {Pekowsky}, {Pele},
  {Penn}, {Perez}, {Perreca}, {Perri}, {Pfeiffer}, {Phelps}, {Piccinni},
  {Pichot}, {Piergiovanni}, {Pierro}, {Pillant}, {Pinard}, {Pinto}, {Pirello},
  {Pitkin}, {Poe}, {Poggiani}, {Popolizio}, {Porter}, {Post}, {Powell},
  {Prasad}, {Pratt}, {Pratten}, {Predoi}, {Prestegard}, {Prijatelj},
  {Principe}, {Privitera}, {Prix}, {Prodi}, {Prokhorov}, {Puncken}, {Punturo},
  {Puppo}, {P{\"u}rrer}, {Qi}, {Quetschke}, {Quintero}, {Quitzow-James},
  {Raab}, {Rabeling}, {Radkins}, {Raffai}, {Raja}, {Rajan}, {Rajbhandari},
  {Rakhmanov}, {Ramirez}, {Ramos-Buades}, {Rapagnani}, {Raymond}, {Razzano},
  {Read}, {Regimbau}, {Rei}, {Reid}, {Reitze}, {Ren}, {Reyes}, {Ricci},
  {Ricker}, {Rieger}, {Riles}, {Rizzo}, {Robertson}, {Robie}, {Robinet},
  {Rocchi}, {Rolland}, {Rollins}, {Roma}, {Romano}, {Romano}, {Romel}, {Romie},
  {Rosi{\'n}ska}, {Ross}, {Rowan}, {R{\"u}diger}, {Ruggi}, {Rutins}, {Ryan},
  {Sachdev}, {Sadecki}, {Sadeghian}, {Sakellariadou}, {Salconi}, {Saleem},
  {Salemi}, {Samajdar}, {Sammut}, {Sampson}, {Sanchez}, {Sanchez},
  {Sanchis-Gual}, {Sandberg}, {Sanders}, {Sassolas}, {Sathyaprakash},
  {Saulson}, {Sauter}, {Savage}, {Sawadsky}, {Schale}, {Scheel}, {Scheuer},
  {Schmidt}, {Schmidt}, {Schnabel}, {Schofield}, {Sch{\"o}nbeck}, {Schreiber},
  {Schuette}, {Schulte}, {Schutz}, {Schwalbe}, {Scott}, {Scott}, {Seidel},
  {Sellers}, {Sengupta}, {Sentenac}, {Sequino}, {Sergeev}, {Shaddock},
  {Shaffer}, {Shah}, {Shahriar}, {Shaner}, {Shao}, {Shapiro}, {Shawhan},
  {Sheperd}, {Shoemaker}, {Shoemaker}, {Siellez}, {Siemens}, {Sieniawska},
  {Sigg}, {Silva}, {Singer}, {Singh}, {Singhal}, {Sintes}, {Slagmolen},
  {Smith}, {Smith}, {Smith}, {Somala}, {Son}, {Sonnenberg}, {Sorazu},
  {Sorrentino}, {Souradeep}, {Spencer}, {Srivastava}, {Staats}, {Staley},
  {Steinke}, {Steinlechner}, {Steinlechner}, {Steinmeyer}, {Stevenson},
  {Stone}, {Stops}, {Strain}, {Stratta}, {Strigin}, {Strunk}, {Sturani},
  {Stuver}, {Summerscales}, {Sun}, {Sunil}, {Suresh}, {Sutton}, {Swinkels},
  {Szczepa{\'n}czyk}, {Tacca}, {Tait}, {Talbot}, {Talukder}, {Tanner},
  {T{\'a}pai}, {Taracchini}, {Tasson}, {Taylor}, {Taylor}, {Tewari}, {Theeg},
  {Thies}, {Thomas}, {Thomas}, {Thomas}, {Thorne}, {Thorne}, {Thrane},
  {Tiwari}, {Tiwari}, {Tokmakov}, {Toland}, {Tonelli}, {Tornasi},
  {Torres-Forn{\'e}}, {Torrie}, {T{\"o}yr{\"a}}, {Travasso}, {Traylor},
  {Trinastic}, {Tringali}, {Trozzo}, {Tsang}, {Tse}, {Tso}, {Tsukada}, {Tsuna},
  {Tuyenbayev}, {Ueno}, {Ugolini}, {Unnikrishnan}, {Urban}, {Usman},
  {Vahlbruch}, {Vajente}, {Valdes}, {Vallisneri}, {van Bakel}, {van Beuzekom},
  {van den Brand}, {Van Den Broeck}, {Vander-Hyde}, {van der Schaaf}, {van
  Heijningen}, {van Veggel}, {Vardaro}, {Varma}, {Vass}, {Vas{\'u}th},
  {Vecchio}, {Vedovato}, {Veitch}, {Veitch}, {Venkateswara}, {Venugopalan},
  {Verkindt}, {Vetrano}, {Vicer{\'e}}, {Viets}, {Vinciguerra}, {Vine}, {Vinet},
  {Vitale}, {Vo}, {Vocca}, {Vorvick}, {Vyatchanin}, {Wade}, {Wade}, {Wade},
  {Walet}, {Walker}, {Wallace}, {Walsh}, {Wang}, {Wang}, {Wang}, {Wang},
  {Wang}, {Ward}, {Warner}, {Was}, {Watchi}, {Weaver}, {Wei}, {Weinert},
  {Weinstein}, {Weiss}, {Wen}, {Wessel}, {We{\ss}els}, {Westerweck},
  {Westphal}, {Wette}, {Whelan}, {Whitcomb}, {Whiting}, {Whittle}, {Wilken},
  {Williams}, {Williams}, {Williamson}, {Willis}, {Willke}, {Wimmer},
  {Winkler}, {Wipf}, {Wittel}, {Woan}, {Woehler}, {Wofford}, {Wong}, {Worden},
  {Wright}, {Wu}, {Wysocki}, {Xiao}, {Yamamoto}, {Yancey}, {Yang}, {Yap},
  {Yazback}, {Yu}, {Yu}, {Yvert}, {Zadro{\.Z}ny}, {Zanolin}, {Zelenova},
  {Zendri}, {Zevin}, {Zhang}, {Zhang}, {Zhang}, {Zhang}, {Zhao}, {Zhou},
  {Zhou}, {Zhu}, {Zhu}, {Zimmerman}, {Zucker}, {Zweizig}, {LIGO Scientific
  Collaboration}, \& {Virgo Collaboration}}]{Abbott17:gw}
{Abbott}, B.~P., {Abbott}, R., {Abbott}, T.~D., {et~al.} 2017{\natexlab{a}},
  \prl, 119, 161101, \dodoi{10.1103/PhysRevLett.119.161101}

\bibitem[{{Abbott} {et~al.}(2017{\natexlab{b}}){Abbott}, {Abbott}, {Abbott},
  {Acernese}, {Ackley}, {Adams}, {Adams}, {Addesso}, {Adhikari}, {Adya},
  {Affeldt}, {Afrough}, {Agarwal}, {Agathos}, {Agatsuma}, {Aggarwal}, {Aguiar},
  {Aiello}, {Ain}, {Ajith}, {Allen}, {Allen}, {Allocca}, {Altin}, {Amato},
  {Ananyeva}, {Anderson}, {Anderson}, {Angelova}, {Antier}, {Appert}, {Arai},
  {Araya}, {Areeda}, {Arnaud}, {Arun}, {Ascenzi}, {Ashton}, {Ast}, {Aston},
  {Astone}, {Atallah}, {Aufmuth}, {Aulbert}, {AultONeal}, {Austin},
  {Avila-Alvarez}, {Babak}, {Bacon}, {Bader}, {Bae}, {Baker}, {Baldaccini},
  {Ballardin}, {Ballmer}, {Banagiri}, {Barayoga}, {Barclay}, {Barish},
  {Barker}, {Barkett}, {Barone}, {Barr}, {Barsotti}, {Barsuglia}, {Barta},
  {Barthelmy}, {Bartlett}, {Bartos}, {Bassiri}, {Basti}, {Batch}, {Bawaj},
  {Bayley}, {Bazzan}, {B{\'e}csy}, {Beer}, {Bejger}, {Belahcene}, {Bell},
  {Berger}, {Bergmann}, {Bero}, {Berry}, {Bersanetti}, {Bertolini},
  {Betzwieser}, {Bhagwat}, {Bhandare}, {Bilenko}, {Billingsley}, {Billman},
  {Birch}, {Birney}, {Birnholtz}, {Biscans}, {Biscoveanu}, {Bisht}, {Bitossi},
  {Biwer}, {Bizouard}, {Blackburn}, {Blackman}, {Blair}, {Blair}, {Blair},
  {Bloemen}, {Bock}, {Bode}, {Boer}, {Bogaert}, {Bohe}, {Bondu}, {Bonilla},
  {Bonnand}, {Boom}, {Bork}, {Boschi}, {Bose}, {Bossie}, {Bouffanais}, {Bozzi},
  {Bradaschia}, {Brady}, {Branchesi}, {Brau}, {Briant}, {Brillet}, {Brinkmann},
  {Brisson}, {Brockill}, {Broida}, {Brooks}, {Brown}, {Brown}, {Brunett},
  {Buchanan}, {Buikema}, {Bulik}, {Bulten}, {Buonanno}, {Buskulic}, {Buy},
  {Byer}, {Cabero}, {Cadonati}, {Cagnoli}, {Cahillane}, {Calder{\'o}n
  Bustillo}, {Callister}, {Calloni}, {Camp}, {Canepa}, {Canizares}, {Cannon},
  {Cao}, {Cao}, {Capano}, {Capocasa}, {Carbognani}, {Caride}, {Carney},
  {Casanueva Diaz}, {Casentini}, {Caudill}, {Cavagli{\`a}}, {Cavalier},
  {Cavalieri}, {Cella}, {Cepeda}, {Cerd{\'a}-Dur{\'a}n}, {Cerretani},
  {Cesarini}, {Chamberlin}, {Chan}, {Chao}, {Charlton}, {Chase},
  {Chassande-Mottin}, {Chatterjee}, {Chatziioannou}, {Cheeseboro}, {Chen},
  {Chen}, {Chen}, {Cheng}, {Chia}, {Chincarini}, {Chiummo}, {Chmiel}, {Cho},
  {Cho}, {Chow}, {Christensen}, {Chu}, {Chua}, {Chua}, {Chung}, {Chung},
  {Ciani}, {Ciolfi}, {Cirelli}, {Cirone}, {Clara}, {Clark}, {Clearwater},
  {Cleva}, {Cocchieri}, {Coccia}, {Cohadon}, {Cohen}, {Colla}, {Collette},
  {Cominsky}, {Constancio}, {Conti}, {Cooper}, {Corban}, {Corbitt},
  {Cordero-Carri{\'o}n}, {Corley}, {Cornish}, {Corsi}, {Cortese}, {Costa},
  {Coughlin}, {Coughlin}, {Coulon}, {Countryman}, {Couvares}, {Covas}, {Cowan},
  {Coward}, {Cowart}, {Coyne}, {Coyne}, {Creighton}, {Creighton}, {Cripe},
  {Crowder}, {Cullen}, {Cumming}, {Cunningham}, {Cuoco}, {Dal Canton},
  {D{\'a}lya}, {Danilishin}, {D'Antonio}, {Danzmann}, {Dasgupta}, {Da Silva
  Costa}, {Dattilo}, {Dave}, {Davier}, {Davis}, {Daw}, {Day}, {De}, {DeBra},
  {Degallaix}, {De Laurentis}, {Del{\'e}glise}, {Del Pozzo}, {Demos}, {Denker},
  {Dent}, {De Pietri}, {Dergachev}, {De Rosa}, {DeRosa}, {De Rossi}, {DeSalvo},
  {de Varona}, {Devenson}, {Dhurandhar}, {D{\'\i}az}, {Di Fiore}, {Di
  Giovanni}, {Di Girolamo}, {Di Lieto}, {Di Pace}, {Di Palma}, {Di Renzo},
  {Doctor}, {Dolique}, {Donovan}, {Dooley}, {Doravari}, {Dorrington},
  {Douglas}, {Dovale {\'A}lvarez}, {Downes}, {Drago}, {Dreissigacker},
  {Driggers}, {Du}, {Ducrot}, {Dupej}, {Dwyer}, {Edo}, {Edwards}, {Effler},
  {Ehrens}, {Eichholz}, {Eikenberry}, {Eisenstein}, {Essick}, {Estevez},
  {Etienne}, {Etzel}, {Evans}, {Evans}, {Factourovich}, {Fafone}, {Fair},
  {Fairhurst}, {Fan}, {Farinon}, {Farr}, {Farr}, {Fauchon-Jones}, {Favata},
  {Fays}, {Fee}, {Fehrmann}, {Feicht}, {Fejer}, {Fernandez-Galiana},
  {Ferrante}, {Ferreira}, {Ferrini}, {Fidecaro}, {Finstad}, {Fiori},
  {Fiorucci}, {Fishbach}, {Fisher}, {Fitz-Axen}, {Flaminio}, {Fletcher},
  {Fong}, {Font}, {Forsyth}, {Forsyth}, {Fournier}, {Frasca}, {Frasconi},
  {Frei}, {Freise}, {Frey}, {Frey}, {Fries}, {Fritschel}, {Frolov}, {Fulda},
  {Fyffe}, {Gabbard}, {Gadre}, {Gaebel}, {Gair}, {Gammaitoni}, {Ganija},
  {Gaonkar}, {Garcia-Quiros}, {Garufi}, {Gateley}, {Gaudio}, {Gaur},
  {Gayathri}, {Gehrels}, {Gemme}, {Genin}, {Gennai}, {George}, {George},
  {Gergely}, {Germain}, {Ghonge}, {Ghosh}, {Ghosh}, {Ghosh}, {Giaime},
  {Giardina}, {Giazotto}, {Gill}, {Glover}, {Goetz}, {Goetz}, {Gomes},
  {Goncharov}, {Gonz{\'a}lez}, {Gonzalez Castro}, {Gopakumar}, {Gorodetsky},
  {Gossan}, {Gosselin}, {Gouaty}, {Grado}, {Graef}, {Granata}, {Grant}, {Gras},
  {Gray}, {Greco}, {Green}, {Gretarsson}, {Griswold}, {Groot}, {Grote},
  {Grunewald}, {Gruning}, {Guidi}, {Guo}, {Gupta}, {Gupta}, {Gushwa},
  {Gustafson}, {Gustafson}, {Halim}, {Hall}, {Hall}, {Hamilton}, {Hammond},
  {Haney}, {Hanke}, {Hanks}, {Hanna}, {Hannam}, {Hannuksela}, {Hanson},
  {Hardwick}, {Harms}, {Harry}, {Harry}, {Hart}, {Haster}, {Haughian}, {Healy},
  {Heidmann}, {Heintze}, {Heitmann}, {Hello}, {Hemming}, {Hendry}, {Heng},
  {Hennig}, {Heptonstall}, {Heurs}, {Hild}, {Hinderer}, {Hoak}, {Hofman},
  {Holt}, {Holz}, {Hopkins}, {Horst}, {Hough}, {Houston}, {Howell}, {Hreibi},
  {Hu}, {Huerta}, {Huet}, {Hughey}, {Husa}, {Huttner}, {Huynh-Dinh}, {Indik},
  {Inta}, {Intini}, {Isa}, {Isac}, {Isi}, {Iyer}, {Izumi}, {Jacqmin}, {Jani},
  {Jaranowski}, {Jawahar}, {Jim{\'e}nez-Forteza}, {Johnson}, {Jones}, {Jones},
  {Jonker}, {Ju}, {Junker}, {Kalaghatgi}, {Kalogera}, {Kamai}, {Kandhasamy},
  {Kang}, {Kanner}, {Kapadia}, {Karki}, {Karvinen}, {Kasprzack}, {Katolik},
  {Katsavounidis}, {Katzman}, {Kaufer}, {Kawabe}, {K{\'e}f{\'e}lian}, {Keitel},
  {Kemball}, {Kennedy}, {Kent}, {Key}, {Khalili}, {Khan}, {Khan}, {Khan},
  {Khazanov}, {Kijbunchoo}, {Kim}, {Kim}, {Kim}, {Kim}, {Kim}, {Kim},
  {Kimbrell}, {King}, {King}, {Kinley-Hanlon}, {Kirchhoff}, {Kissel},
  {Kleybolte}, {Klimenko}, {Knowles}, {Koch}, {Koehlenbeck}, {Koley},
  {Kondrashov}, {Kontos}, {Korobko}, {Korth}, {Kowalska}, {Kozak},
  {Kr{\"a}mer}, {Kringel}, {Krishnan}, {Kr{\'o}lak}, {Kuehn}, {Kumar}, {Kumar},
  {Kumar}, {Kuo}, {Kutynia}, {Kwang}, {Lackey}, {Lai}, {Landry}, {Lang},
  {Lange}, {Lantz}, {Lanza}, {Larson}, {Lartaux-Vollard}, {Lasky}, {Laxen},
  {Lazzarini}, {Lazzaro}, {Leaci}, {Leavey}, {Lee}, {Lee}, {Lee}, {Lee}, {Lee},
  {Lehmann}, {Lenon}, {Leonardi}, {Leroy}, {Letendre}, {Levin}, {Li}, {Linker},
  {Littenberg}, {Liu}, {Lo}, {Lockerbie}, {London}, {Lord}, {Lorenzini},
  {Loriette}, {Lormand}, {Losurdo}, {Lough}, {Lousto}, {Lovelace}, {L{\"u}ck},
  {Lumaca}, {Lundgren}, {Lynch}, {Ma}, {Macas}, {Macfoy}, {Machenschalk},
  {MacInnis}, {Macleod}, {Maga{\~n}a Hernandez}, {Maga{\~n}a-Sandoval},
  {Maga{\~n}a Zertuche}, {Magee}, {Majorana}, {Maksimovic}, {Man}, {Mandic},
  {Mangano}, {Mansell}, {Manske}, {Mantovani}, {Marchesoni}, {Marion},
  {M{\'a}rka}, {M{\'a}rka}, {Markakis}, {Markosyan}, {Markowitz}, {Maros},
  {Marquina}, {Marsh}, {Martelli}, {Martellini}, {Martin}, {Martin},
  {Martynov}, {Mason}, {Massera}, {Masserot}, {Massinger}, {Masso-Reid},
  {Mastrogiovanni}, {Matas}, {Matichard}, {Matone}, {Mavalvala}, {Mazumder},
  {McCarthy}, {McClelland}, {McCormick}, {McCuller}, {McGuire}, {McIntyre},
  {McIver}, {McManus}, {McNeill}, {McRae}, {McWilliams}, {Meacher}, {Meadors},
  {Mehmet}, {Meidam}, {Mejuto-Villa}, {Melatos}, {Mendell}, {Mercer}, {Merilh},
  {Merzougui}, {Meshkov}, {Messenger}, {Messick}, {Metzdorff}, {Meyers},
  {Miao}, {Michel}, {Middleton}, {Mikhailov}, {Milano}, {Miller}, {Miller},
  {Miller}, {Millhouse}, {Milovich-Goff}, {Minazzoli}, {Minenkov}, {Ming},
  {Mishra}, {Mitra}, {Mitrofanov}, {Mitselmakher}, {Mittleman}, {Moffa},
  {Moggi}, {Mogushi}, {Mohan}, {Mohapatra}, {Montani}, {Moore}, {Moraru},
  {Moreno}, {Morriss}, {Mours}, {Mow-Lowry}, {Mueller}, {Muir}, {Mukherjee},
  {Mukherjee}, {Mukherjee}, {Mukund}, {Mullavey}, {Munch}, {Mu{\~n}iz},
  {Muratore}, {Murray}, {Napier}, {Nardecchia}, {Naticchioni}, {Nayak},
  {Neilson}, {Nelemans}, {Nelson}, {Nery}, {Neunzert}, {Nevin}, {Newport},
  {Newton}, {Ng}, {Nguyen}, {Nguyen}, {Nichols}, {Nielsen}, {Nissanke}, {Nitz},
  {Noack}, {Nocera}, {Nolting}, {North}, {Nuttall}, {Oberling}, {O'Dea},
  {Ogin}, {Oh}, {Oh}, {Ohme}, {Okada}, {Oliver}, {Oppermann}, {Oram},
  {O'Reilly}, {Ormiston}, {Ortega}, {O'Shaughnessy}, {Ossokine}, {Ottaway},
  {Overmier}, {Owen}, {Pace}, {Page}, {Page}, {Pai}, {Pai}, {Palamos},
  {Palashov}, {Palomba}, {Pal-Singh}, {Pan}, {Pan}, {Pang}, {Pang}, {Pankow},
  {Pannarale}, {Pant}, {Paoletti}, {Paoli}, {Papa}, {Parida}, {Parker},
  {Pascucci}, {Pasqualetti}, {Passaquieti}, {Passuello}, {Patil}, {Patricelli},
  {Pearlstone}, {Pedraza}, {Pedurand}, {Pekowsky}, {Pele}, {Penn}, {Perez},
  {Perreca}, {Perri}, {Pfeiffer}, {Phelps}, {Piccinni}, {Pichot},
  {Piergiovanni}, {Pierro}, {Pillant}, {Pinard}, {Pinto}, {Pirello}, {Pitkin},
  {Poe}, {Poggiani}, {Popolizio}, {Porter}, {Post}, {Powell}, {Prasad},
  {Pratt}, {Pratten}, {Predoi}, {Prestegard}, {Price}, {Prijatelj}, {Principe},
  {Privitera}, {Prodi}, {Prokhorov}, {Puncken}, {Punturo}, {Puppo},
  {P{\"u}rrer}, {Qi}, {Quetschke}, {Quintero}, {Quitzow-James}, {Raab},
  {Rabeling}, {Radkins}, {Raffai}, {Raja}, {Rajan}, {Rajbhandari}, {Rakhmanov},
  {Ramirez}, {Ramos-Buades}, {Rapagnani}, {Raymond}, {Razzano}, {Read},
  {Regimbau}, {Rei}, {Reid}, {Reitze}, {Ren}, {Reyes}, {Ricci}, {Ricker},
  {Rieger}, {Riles}, {Rizzo}, {Robertson}, {Robie}, {Robinet}, {Rocchi},
  {Rolland}, {Rollins}, {Roma}, {Romano}, {Romel}, {Romie}, {Rosi{\'n}ska},
  {Ross}, {Rowan}, {R{\"u}diger}, {Ruggi}, {Rutins}, {Ryan}, {Sachdev},
  {Sadecki}, {Sadeghian}, {Sakellariadou}, {Salconi}, {Saleem}, {Salemi},
  {Samajdar}, {Sammut}, {Sampson}, {Sanchez}, {Sanchez}, {Sanchis-Gual},
  {Sandberg}, {Sanders}, {Sassolas}, {Sathyaprakash}, {Saulson}, {Sauter},
  {Savage}, {Sawadsky}, {Schale}, {Scheel}, {Scheuer}, {Schmidt}, {Schmidt},
  {Schnabel}, {Schofield}, {Sch{\"o}nbeck}, {Schreiber}, {Schuette}, {Schulte},
  {Schutz}, {Schwalbe}, {Scott}, {Scott}, {Seidel}, {Sellers}, {Sengupta},
  {Sentenac}, {Sequino}, {Sergeev}, {Shaddock}, {Shaffer}, {Shah}, {Shahriar},
  {Shaner}, {Shao}, {Shapiro}, {Shawhan}, {Sheperd}, {Shoemaker}, {Shoemaker},
  {Siellez}, {Siemens}, {Sieniawska}, {Sigg}, {Silva}, {Singer}, {Singh},
  {Singhal}, {Sintes}, {Slagmolen}, {Smith}, {Smith}, {Smith}, {Somala}, {Son},
  {Sonnenberg}, {Sorazu}, {Sorrentino}, {Souradeep}, {Spencer}, {Srivastava},
  {Staats}, {Staley}, {Steinke}, {Steinlechner}, {Steinlechner}, {Steinmeyer},
  {Stevenson}, {Stone}, {Stops}, {Strain}, {Stratta}, {Strigin}, {Strunk},
  {Sturani}, {Stuver}, {Summerscales}, {Sun}, {Sunil}, {Suresh}, {Sutton},
  {Swinkels}, {Szczepa{\'n}czyk}, {Tacca}, {Tait}, {Talbot}, {Talukder},
  {Tanner}, {T{\'a}pai}, {Taracchini}, {Tasson}, {Taylor}, {Taylor}, {Tewari},
  {Theeg}, {Thies}, {Thomas}, {Thomas}, {Thomas}, {Thorne}, {Thorne}, {Thrane},
  {Tiwari}, {Tiwari}, {Tokmakov}, {Toland}, {Tonelli}, {Tornasi},
  {Torres-Forn{\'e}}, {Torrie}, {T{\"o}yr{\"a}}, {Travasso}, {Traylor},
  {Trinastic}, {Tringali}, {Trozzo}, {Tsang}, {Tse}, {Tso}, {Tsukada}, {Tsuna},
  {Tuyenbayev}, {Ueno}, {Ugolini}, {Unnikrishnan}, {Urban}, {Usman},
  {Vahlbruch}, {Vajente}, {Valdes}, {van Bakel}, {van Beuzekom}, {van den
  Brand}, {Van Den Broeck}, {Vander-Hyde}, {van der Schaaf}, {van Heijningen},
  {van Veggel}, {Vardaro}, {Varma}, {Vass}, {Vas{\'u}th}, {Vecchio},
  {Vedovato}, {Veitch}, {Veitch}, {Venkateswara}, {Venugopalan}, {Verkindt},
  {Vetrano}, {Vicer{\'e}}, {Viets}, {Vinciguerra}, {Vine}, {Vinet}, {Vitale},
  {Vo}, {Vocca}, {Vorvick}, {Vyatchanin}, {Wade}, {Wade}, {Wade}, {Walet},
  {Walker}, {Wallace}, {Walsh}, {Wang}, {Wang}, {Wang}, {Wang}, {Wang}, {Ward},
  {Warner}, {Was}, {Watchi}, {Weaver}, {Wei}, {Weinert}, {Weinstein}, {Weiss},
  {Wen}, {Wessel}, {Wessels}, {Westerweck}, {Westphal}, {Wette}, {Whelan},
  {Whitcomb}, {Whiting}, {Whittle}, {Wilken}, {Williams}, {Williams},
  {Williamson}, {Willis}, {Willke}, {Wimmer}, {Winkler}, {Wipf}, {Wittel},
  {Woan}, {Woehler}, {Wofford}, {Wong}, {Worden}, {Wright}, {Wu}, {Wysocki},
  {Xiao}, {Yamamoto}, {Yancey}, {Yang}, {Yap}, {Yazback}, {Yu}, {Yu}, {Yvert},
  {Zadro{\.z}ny}, {Zanolin}, {Zelenova}, {Zendri}, {Zevin}, {Zhang}, {Zhang},
  {Zhang}, {Zhang}, {Zhao}, {Zhou}, {Zhou}, {Zhu}, {Zhu}, {Zimmerman},
  {Zucker}, {Zweizig}, {LIGO Scientific Collaboration}, {Virgo Collaboration},
  {Wilson-Hodge}, {Bissaldi}, {Blackburn}, {Briggs}, {Burns}, {Cleveland},
  {Connaughton}, {Gibby}, {Giles}, {Goldstein}, {Hamburg}, {Jenke}, {Hui},
  {Kippen}, {Kocevski}, {McBreen}, {Meegan}, {Paciesas}, {Poolakkil}, {Preece},
  {Racusin}, {Roberts}, {Stanbro}, {Veres}, {von Kienlin}, {GBM}, {Savchenko},
  {Ferrigno}, {Kuulkers}, {Bazzano}, {Bozzo}, {Brandt}, {Chenevez},
  {Courvoisier}, {Diehl}, {Domingo}, {Hanlon}, {Jourdain}, {Laurent}, {Lebrun},
  {Lutovinov}, {Martin-Carrillo}, {Mereghetti}, {Natalucci}, {Rodi}, {Roques},
  {Sunyaev}, {Ubertini}, {INTEGRAL}, {Aartsen}, {Ackermann}, {Adams},
  {Aguilar}, {Ahlers}, {Ahrens}, {Samarai}, {Altmann}, {Andeen}, {Anderson},
  {Ansseau}, {Anton}, {Arg{\"u}elles}, {Auffenberg}, {Axani}, {Bagherpour},
  {Bai}, {Barron}, {Barwick}, {Baum}, {Bay}, {Beatty}, {Becker Tjus},
  {Bernardini}, {Besson}, {Binder}, {Bindig}, {Blaufuss}, {Blot}, {Bohm},
  {B{\"o}rner}, {Bos}, {Bose}, {B{\"o}ser}, {Botner}, {Bourbeau}, {Bourbeau},
  {Bradascio}, {Braun}, {Brayeur}, {Brenzke}, {Bretz}, {Bron},
  {Brostean-Kaiser}, {Burgman}, {Carver}, {Casey}, {Casier}, {Cheung},
  {Chirkin}, {Christov}, {Clark}, {Classen}, {Coenders}, {Collin}, {Conrad},
  {Cowen}, {Cross}, {Day}, {de Andr{\'e}}, {De Clercq}, {DeLaunay},
  {Dembinski}, {De Ridder}, {Desiati}, {de Vries}, {de Wasseige}, {de With},
  {DeYoung}, {D{\'\i}az-V{\'e}lez}, {di Lorenzo}, {Dujmovic}, {Dumm},
  {Dunkman}, {Dvorak}, {Eberhardt}, {Ehrhardt}, {Eichmann}, {Eller}, {Evenson},
  {Fahey}, {Fazely}, {Felde}, {Filimonov}, {Finley}, {Flis}, {Franckowiak},
  {Friedman}, {Fuchs}, {Gaisser}, {Gallagher}, {Gerhardt}, {Ghorbani}, {Giang},
  {Glauch}, {Gl{\"u}senkamp}, {Goldschmidt}, {Gonzalez}, {Grant}, {Griffith},
  {Haack}, {Hallgren}, {Halzen}, {Hanson}, {Hebecker}, {Heereman}, {Helbing},
  {Hellauer}, {Hickford}, {Hignight}, {Hill}, {Hoffman}, {Hoffmann},
  {Hokanson-Fasig}, {Hoshina}, {Huang}, {Huber}, {Hultqvist}, {H{\"u}nnefeld},
  {In}, {Ishihara}, {Jacobi}, {Japaridze}, {Jeong}, {Jero}, {Jones},
  {Kalaczynski}, {Kang}, {Kappes}, {Karg}, {Karle}, {Kauer}, {Keivani},
  {Kelley}, {Kheirandish}, {Kim}, {Kim}, {Kintscher}, {Kiryluk}, {Kittler},
  {Klein}, {Kohnen}, {Koirala}, {Kolanoski}, {K{\"o}pke}, {Kopper}, {Kopper},
  {Koschinsky}, {Koskinen}, {Kowalski}, {Krings}, {Kroll}, {Kr{\"u}ckl},
  {Kunnen}, {Kunwar}, {Kurahashi}, {Kuwabara}, {Kyriacou}, {Labare},
  {Lanfranchi}, {Larson}, {Lauber}, {Lesiak-Bzdak}, {Leuermann}, {Liu}, {Lu},
  {L{\"u}nemann}, {Luszczak}, {Madsen}, {Maggi}, {Mahn}, {Mancina}, {Maruyama},
  {Mase}, {Maunu}, {McNally}, {Meagher}, {Medici}, {Meier}, {Menne}, {Merino},
  {Meures}, {Miarecki}, {Micallef}, {Moment{\'e}}, {Montaruli}, {Moore},
  {Moulai}, {Nahnhauer}, {Nakarmi}, {Naumann}, {Neer}, {Niederhausen},
  {Nowicki}, {Nygren}, {Obertacke Pollmann}, {Olivas}, {O'Murchadha},
  {Palczewski}, {Pandya}, {Pankova}, {Peiffer}, {Pepper}, {P{\'e}rez de los
  Heros}, {Pieloth}, {Pinat}, {Price}, {Przybylski}, {Raab}, {R{\"a}del},
  {Rameez}, {Rawlins}, {Rea}, {Reimann}, {Relethford}, {Relich}, {Resconi},
  {Rhode}, {Richman}, {Robertson}, {Rongen}, {Rott}, {Ruhe}, {Ryckbosch},
  {Rysewyk}, {S{\"a}lzer}, {Sanchez Herrera}, {Sandrock}, {Sandroos},
  {Santander}, {Sarkar}, {Sarkar}, {Satalecka}, {Schlunder}, {Schmidt},
  {Schneider}, {Schoenen}, {Sch{\"o}neberg}, {Schumacher}, {Seckel},
  {Seunarine}, {Soedingrekso}, {Soldin}, {Song}, {Spiczak}, {Spiering},
  {Stachurska}, {Stamatikos}, {Stanev}, {Stasik}, {Stettner}, {Steuer},
  {Stezelberger}, {Stokstad}, {St{\"o}ssl}, {Strotjohann}, {Stuttard},
  {Sullivan}, {Sutherland}, {Taboada}, {Tatar}, {Tenholt}, {Ter-Antonyan},
  {Terliuk}, {Te{\v{s}}i{\'c}}, {Tilav}, {Toale}, {Tobin}, {Toscano}, {Tosi},
  {Tselengidou}, {Tung}, {Turcati}, {Turley}, {Ty}, {Unger}, {Usner},
  {Vandenbroucke}, {Van Driessche}, {van Eijndhoven}, {Vanheule}, {van Santen},
  {Vehring}, {Vogel}, {Vraeghe}, {Walck}, {Wallace}, {Wallraff}, {Wandler},
  {Wandkowsky}, {Waza}, {Weaver}, {Weiss}, {Wendt}, {Werthebach}, {Whelan},
  {Wiebe}, {Wiebusch}, {Wille}, {Williams}, {Wills}, {Wolf}, {Wood}, {Woolsey},
  {Woschnagg}, {Xu}, {Xu}, {Xu}, {Yanez}, {Yodh}, {Yoshida}, {Yuan}, {Zoll},
  {IceCube Collaboration}, {Balasubramanian}, {Mate}, {Bhalerao},
  {Bhattacharya}, {Vibhute}, {Dewangan}, {Rao}, {Vadawale}, {AstroSat Cadmium
  Zinc Telluride Imager Team}, {Svinkin}, {Hurley}, {Aptekar}, {Frederiks},
  {Golenetskii}, {Kozlova}, {Lysenko}, {Oleynik}, {Tsvetkova}, {Ulanov},
  {Cline}, {IPN Collaboration}, {Li}, {Xiong}, {Zhang}, {Lu}, {Song}, {Cao},
  {Chang}, {Chen}, {Chen}, {Chen}, {Chen}, {Chen}, {Chen}, {Cui}, {Cui},
  {Deng}, {Dong}, {Du}, {Fu}, {Gao}, {Gao}, {Gao}, {Ge}, {Gu}, {Guan}, {Guo},
  {Han}, {Hu}, {Huang}, {Huo}, {Jia}, {Jiang}, {Jiang}, {Jin}, {Jin}, {Li},
  {Li}, {Li}, {Li}, {Li}, {Li}, {Li}, {Li}, {Li}, {Li}, {Li}, {Liang}, {Liao},
  {Liu}, {Liu}, {Liu}, {Liu}, {Liu}, {Liu}, {Liu}, {Lu}, {Lu}, {Luo}, {Ma},
  {Meng}, {Nang}, {Nie}, {Ou}, {Qu}, {Sai}, {Sun}, {Tan}, {Tao}, {Tao}, {Tuo},
  {Wang}, {Wang}, {Wang}, {Wang}, {Wang}, {Wen}, {Wu}, {Wu}, {Xiao}, {Xu},
  {Xu}, {Yan}, {Yang}, {Yang}, {Yang}, {Zhang}, {Zhang}, {Zhang}, {Zhang},
  {Zhang}, {Zhang}, {Zhang}, {Zhang}, {Zhang}, {Zhang}, {Zhang}, {Zhang},
  {Zhang}, {Zhang}, {Zhang}, {Zhang}, {Zhang}, {Zhang}, {Zhao}, {Zhao}, {Zhao},
  {Zheng}, {Zhu}, {Zhu}, {Zou}, {Insight-HXMT Collaboration}, {Albert},
  {Andr{\'e}}, {Anghinolfi}, {Ardid}, {Aubert}, {Aublin}, {Avgitas}, {Baret},
  {Barrios-Mart{\'\i}}, {Basa}, {Belhorma}, {Bertin}, {Biagi}, {Bormuth},
  {Bourret}, {Bouwhuis}, {Br{\^a}nza{\textcommabelow s}}, {Bruijn}, {Brunner},
  {Busto}, {Capone}, {Caramete}, {Carr}, {Celli}, {Cherkaoui El Moursli},
  {Chiarusi}, {Circella}, {Coelho}, {Coleiro}, {Coniglione}, {Costantini},
  {Coyle}, {Creusot}, {D{\'\i}az}, {Deschamps}, {De Bonis}, {Distefano}, {Di
  Palma}, {Domi}, {Donzaud}, {Dornic}, {Drouhin}, {Eberl}, {El Bojaddaini}, {El
  Khayati}, {Els{\"a}sser}, {Enzenh{\"o}fer}, {Ettahiri}, {Fassi}, {Felis},
  {Fusco}, {Gay}, {Giordano}, {Glotin}, {Gr{\'e}goire}, {Ruiz}, {Graf},
  {Hallmann}, {van Haren}, {Heijboer}, {Hello}, {Hern{\'a}ndez-Rey},
  {H{\"o}ssl}, {Hofest{\"a}dt}, {Hugon}, {Illuminati}, {James}, {de Jong},
  {Jongen}, {Kadler}, {Kalekin}, {Katz}, {Kiessling}, {Kouchner}, {Kreter},
  {Kreykenbohm}, {Kulikovskiy}, {Lachaud}, {Lahmann}, {Lef{\`e}vre}, {Leonora},
  {Lotze}, {Loucatos}, {Marcelin}, {Margiotta}, {Marinelli},
  {Mart{\'\i}nez-Mora}, {Mele}, {Melis}, {Michael}, {Migliozzi}, {Moussa},
  {Navas}, {Nezri}, {Organokov}, {P{\u{a}}v{\u{a}}la{\textcommabelow s}},
  {Pellegrino}, {Perrina}, {Piattelli}, {Popa}, {Pradier}, {Quinn}, {Racca},
  {Riccobene}, {S{\'a}nchez-Losa}, {Salda{\~n}a}, {Salvadori}, {Samtleben},
  {Sanguineti}, {Sapienza}, {Sieger}, {Spurio}, {Stolarczyk}, {Taiuti},
  {Tayalati}, {Trovato}, {Turpin}, {T{\"o}nnis}, {Vallage}, {Van Elewyck},
  {Versari}, {Vivolo}, {Vizzoca}, {Wilms}, {Zornoza}, {Z{\'u}{\~n}iga},
  {ANTARES Collaboration}, {Beardmore}, {Breeveld}, {Burrows}, {Cenko},
  {Cusumano}, {D'A{\`\i}}, {de Pasquale}, {Emery}, {Evans}, {Giommi},
  {Gronwall}, {Kennea}, {Krimm}, {Kuin}, {Lien}, {Marshall}, {Melandri},
  {Nousek}, {Oates}, {Osborne}, {Pagani}, {Page}, {Palmer}, {Perri}, {Siegel},
  {Sbarufatti}, {Tagliaferri}, {Tohuvavohu}, {Swift Collaboration}, {Tavani},
  {Verrecchia}, {Bulgarelli}, {Evangelista}, {Pacciani}, {Feroci}, {Pittori},
  {Giuliani}, {Del Monte}, {Donnarumma}, {Argan}, {Trois}, {Ursi}, {Cardillo},
  {Piano}, {Longo}, {Lucarelli}, {Munar-Adrover}, {Fuschino}, {Labanti},
  {Marisaldi}, {Minervini}, {Fioretti}, {Parmiggiani}, {Gianotti}, {Trifoglio},
  {Di Persio}, {Antonelli}, {Barbiellini}, {Caraveo}, {Cattaneo}, {Costa},
  {Colafrancesco}, {D'Amico}, {Ferrari}, {Morselli}, {Paoletti}, {Picozza},
  {Pilia}, {Rappoldi}, {Soffitta}, {Vercellone}, {AGILE Team}, {Foley},
  {Coulter}, {Kilpatrick}, {Drout}, {Piro}, {Shappee}, {Siebert}, {Simon},
  {Ulloa}, {Kasen}, {Madore}, {Murguia-Berthier}, {Pan}, {Prochaska},
  {Ramirez-Ruiz}, {Rest}, {Rojas-Bravo}, {1M2H Team}, {Berger},
  {Soares-Santos}, {Annis}, {Alexander}, {Allam}, {Balbinot}, {Blanchard},
  {Brout}, {Butler}, {Chornock}, {Cook}, {Cowperthwaite}, {Diehl},
  {Drlica-Wagner}, {Drout}, {Durret}, {Eftekhari}, {Finley}, {Fong}, {Frieman},
  {Fryer}, {Garc{\'\i}a-Bellido}, {Gruendl}, {Hartley}, {Herner}, {Kessler},
  {Lin}, {Lopes}, {Louren{\c{c}}o}, {Margutti}, {Marshall}, {Matheson},
  {Medina}, {Metzger}, {Mu{\~n}oz}, {Muir}, {Nicholl}, {Nugent}, {Palmese},
  {Paz-Chinch{\'o}n}, {Quataert}, {Sako}, {Sauseda}, {Schlegel}, {Scolnic},
  {Secco}, {Smith}, {Sobreira}, {Villar}, {Vivas}, {Wester}, {Williams},
  {Yanny}, {Zenteno}, {Zhang}, {Abbott}, {Banerji}, {Bechtol},
  {Benoit-L{\'e}vy}, {Bertin}, {Brooks}, {Buckley-Geer}, {Burke}, {Capozzi},
  {Carnero Rosell}, {Carrasco Kind}, {Castander}, {Crocce}, {Cunha},
  {D'Andrea}, {da Costa}, {Davis}, {DePoy}, {Desai}, {Dietrich}, {Eifler},
  {Fernandez}, {Flaugher}, {Fosalba}, {Gaztanaga}, {Gerdes}, {Giannantonio},
  {Goldstein}, {Gruen}, {Gschwend}, {Gutierrez}, {Honscheid}, {James},
  {Jeltema}, {Johnson}, {Johnson}, {Kent}, {Krause}, {Kron}, {Kuehn}, {Lahav},
  {Lima}, {Maia}, {March}, {Martini}, {McMahon}, {Menanteau}, {Miller},
  {Miquel}, {Mohr}, {Nichol}, {Ogando}, {Plazas}, {Romer}, {Roodman}, {Rykoff},
  {Sanchez}, {Scarpine}, {Schindler}, {Schubnell}, {Sevilla-Noarbe}, {Sheldon},
  {Smith}, {Smith}, {Stebbins}, {Suchyta}, {Swanson}, {Tarle}, {Thomas},
  {Troxel}, {Tucker}, {Vikram}, {Walker}, {Wechsler}, {Weller}, {Carlin},
  {Gill}, {Li}, {Marriner}, {Neilsen}, {Dark Energy Camera GW-EM
  Collaboration}, {DES Collaboration}, {Haislip}, {Kouprianov}, {Reichart},
  {Sand}, {Tartaglia}, {Valenti}, {Yang}, {DLT40 Collaboration}, {Benetti},
  {Brocato}, {Campana}, {Cappellaro}, {Covino}, {D'Avanzo}, {D'Elia}, {Getman},
  {Ghirlanda}, {Ghisellini}, {Limatola}, {Nicastro}, {Palazzi}, {Pian},
  {Piranomonte}, {Possenti}, {Rossi}, {Salafia}, {Tomasella}, {Amati},
  {Antonelli}, {Bernardini}, {Bufano}, {Capaccioli}, {Casella}, {Dadina}, {De
  Cesare}, {Di Paola}, {Giuffrida}, {Giunta}, {Israel}, {Lisi}, {Maiorano},
  {Mapelli}, {Masetti}, {Pescalli}, {Pulone}, {Salvaterra}, {Schipani},
  {Spera}, {Stamerra}, {Stella}, {Testa}, {Turatto}, {Vergani}, {Aresu},
  {Bachetti}, {Buffa}, {Burgay}, {Buttu}, {Caria}, {Carretti}, {Casasola},
  {Castangia}, {Carboni}, {Casu}, {Concu}, {Corongiu}, {Deiana}, {Egron},
  {Fara}, {Gaudiomonte}, {Gusai}, {Ladu}, {Loru}, {Leurini}, {Marongiu},
  {Melis}, {Melis}, {Migoni}, {Milia}, {Navarrini}, {Orlati}, {Ortu}, {Palmas},
  {Pellizzoni}, {Perrodin}, {Pisanu}, {Poppi}, {Righini}, {Saba}, {Serra},
  {Serrau}, {Stagni}, {Surcis}, {Vacca}, {Vargiu}, {Hunt}, {Jin}, {Klose},
  {Kouveliotou}, {Mazzali}, {M{\o}ller}, {Nava}, {Piran}, {Selsing}, {Vergani},
  {Wiersema}, {Toma}, {Higgins}, {Mundell}, {di Serego Alighieri}, {G{\'o}tz},
  {Gao}, {Gomboc}, {Kaper}, {Kobayashi}, {Kopac}, {Mao}, {Starling}, {Steele},
  {van der Horst}, {GRAWITA: GRAvitational Wave Inaf TeAm}, {Acero}, {Atwood},
  {Baldini}, {Barbiellini}, {Bastieri}, {Berenji}, {Bellazzini}, {Bissaldi},
  {Blandford}, {Bloom}, {Bonino}, {Bottacini}, {Bregeon}, {Buehler}, {Buson},
  {Cameron}, {Caputo}, {Caraveo}, {Cavazzuti}, {Chekhtman}, {Cheung}, {Chiang},
  {Ciprini}, {Cohen-Tanugi}, {Cominsky}, {Costantin}, {Cuoco}, {D'Ammando}, {de
  Palma}, {Digel}, {Di Lalla}, {Di Mauro}, {Di Venere}, {Dubois}, {Fegan},
  {Focke}, {Franckowiak}, {Fukazawa}, {Funk}, {Fusco}, {Gargano}, {Gasparrini},
  {Giglietto}, {Giordano}, {Giroletti}, {Glanzman}, {Green}, {Grondin},
  {Guillemot}, {Guiriec}, {Harding}, {Horan}, {J{\'o}hannesson}, {Kamae},
  {Kensei}, {Kuss}, {La Mura}, {Latronico}, {Lemoine-Goumard}, {Longo},
  {Loparco}, {Lovellette}, {Lubrano}, {Magill}, {Maldera}, {Manfreda},
  {Mazziotta}, {McEnery}, {Meyer}, {Michelson}, {Mirabal}, {Monzani},
  {Moretti}, {Morselli}, {Moskalenko}, {Negro}, {Nuss}, {Ojha}, {Omodei},
  {Orienti}, {Orlando}, {Palatiello}, {Paliya}, {Paneque}, {Pesce-Rollins},
  {Piron}, {Porter}, {Principe}, {Rain{\`o}}, {Rando}, {Razzano}, {Razzaque},
  {Reimer}, {Reimer}, {Reposeur}, {Rochester}, {Saz Parkinson}, {Sgr{\`o}},
  {Siskind}, {Spada}, {Spandre}, {Suson}, {Takahashi}, {Tanaka}, {Thayer},
  {Thayer}, {Thompson}, {Tibaldo}, {Torres}, {Torresi}, {Troja}, {Venters},
  {Vianello}, {Zaharijas}, {Fermi Large Area Telescope Collaboration},
  {Allison}, {Bannister}, {Dobie}, {Kaplan}, {Lenc}, {Lynch}, {Murphy},
  {Sadler}, {Australia Telescope Compact Array}, {Hotan}, {James}, {Oslowski},
  {Raja}, {Shannon}, {Whiting}, {Australian SKA Pathfinder}, {Arcavi},
  {Howell}, {McCully}, {Hosseinzadeh}, {Hiramatsu}, {Poznanski}, {Barnes},
  {Zaltzman}, {Vasylyev}, {Maoz}, {Las Cumbres Observatory Group}, {Cooke},
  {Bailes}, {Wolf}, {Deller}, {Lidman}, {Wang}, {Gendre}, {Andreoni}, {Ackley},
  {Pritchard}, {Bessell}, {Chang}, {M{\"o}ller}, {Onken}, {Scalzo},
  {Ridden-Harper}, {Sharp}, {Tucker}, {Farrell}, {Elmer}, {Johnston},
  {Venkatraman Krishnan}, {Keane}, {Green}, {Jameson}, {Hu}, {Ma}, {Sun}, {Wu},
  {Wang}, {Shang}, {Hu}, {Ashley}, {Yuan}, {Li}, {Tao}, {Zhu}, {Zhang},
  {Suntzeff}, {Zhou}, {Yang}, {Orange}, {Morris}, {Cucchiara}, {Giblin},
  {Klotz}, {Staff}, {Thierry}, {Schmidt}, {OzGrav}, {(Deeper}, {Wider},
  {program}, {AST3}, {CAASTRO Collaborations}, {Tanvir}, {Levan}, {Cano}, {de
  Ugarte-Postigo}, {Gonz{\'a}lez-Fern{\'a}ndez}, {Greiner}, {Hjorth}, {Irwin},
  {Kr{\"u}hler}, {Mandel}, {Milvang-Jensen}, {O'Brien}, {Rol}, {Rosetti},
  {Rosswog}, {Rowlinson}, {Steeghs}, {Th{\"o}ne}, {Ulaczyk}, {Watson}, {Bruun},
  {Cutter}, {Figuera Jaimes}, {Fujii}, {Fruchter}, {Gompertz}, {Jakobsson},
  {Hodosan}, {J{\`e}rgensen}, {Kangas}, {Kann}, {Rabus}, {Schr{\o}der},
  {Stanway}, {Wijers}, {VINROUGE Collaboration}, {Lipunov}, {Gorbovskoy},
  {Kornilov}, {Tyurina}, {Balanutsa}, {Kuznetsov}, {Vlasenko}, {Podesta},
  {Lopez}, {Podesta}, {Levato}, {Saffe}, {Mallamaci}, {Budnev}, {Gress},
  {Kuvshinov}, {Gorbunov}, {Vladimirov}, {Zimnukhov}, {Gabovich}, {Yurkov},
  {Sergienko}, {Rebolo}, {Serra-Ricart}, {Tlatov}, {Ishmuhametova}, {MASTER
  Collaboration}, {Abe}, {Aoki}, {Aoki}, {Asakura}, {Baar}, {Barway}, {Bond},
  {Doi}, {Finet}, {Fujiyoshi}, {Furusawa}, {Honda}, {Itoh}, {Kanda},
  {Kawabata}, {Kawabata}, {Kim}, {Koshida}, {Kuroda}, {Lee}, {Liu},
  {Matsubayashi}, {Miyazaki}, {Morihana}, {Morokuma}, {Motohara}, {Murata},
  {Nagai}, {Nagashima}, {Nagayama}, {Nakaoka}, {Nakata}, {Ohsawa}, {Ohshima},
  {Ohta}, {Okita}, {Saito}, {Saito}, {Sako}, {Sekiguchi}, {Sumi}, {Tajitsu},
  {Takahashi}, {Takayama}, {Tamura}, {Tanaka}, {Tanaka}, {Terai}, {Tominaga},
  {Tristram}, {Uemura}, {Utsumi}, {Yamaguchi}, {Yasuda}, {Yoshida}, {Zenko},
  {J-GEM}, {Adams}, {Anupama}, {Bally}, {Barway}, {Bellm}, {Blagorodnova},
  {Cannella}, {Chandra}, {Chatterjee}, {Clarke}, {Cobb}, {Cook}, {Copperwheat},
  {De}, {Emery}, {Feindt}, {Foster}, {Fox}, {Frail}, {Fremling}, {Frohmaier},
  {Garcia}, {Ghosh}, {Giacintucci}, {Goobar}, {Gottlieb}, {Grefenstette},
  {Hallinan}, {Harrison}, {Heida}, {Helou}, {Ho}, {Horesh}, {Hotokezaka}, {Ip},
  {Itoh}, {Jacobs}, {Jencson}, {Kasen}, {Kasliwal}, {Kassim}, {Kim}, {Kiran},
  {Kuin}, {Kulkarni}, {Kupfer}, {Lau}, {Madsen}, {Mazzali}, {Miller},
  {Miyasaka}, {Mooley}, {Myers}, {Nakar}, {Ngeow}, {Nugent}, {Ofek},
  {Palliyaguru}, {Pavana}, {Perley}, {Peters}, {Pike}, {Piran}, {Qi}, {Quimby},
  {Rana}, {Rosswog}, {Rusu}, {Sadler}, {Van Sistine}, {Sollerman}, {Xu}, {Yan},
  {Yatsu}, {Yu}, {Zhang}, {Zhao}, {GROWTH}, {JAGWAR}, {Caltech-NRAO},
  {TTU-NRAO}, {NuSTAR Collaborations}, {Chambers}, {Huber}, {Schultz},
  {Bulger}, {Flewelling}, {Magnier}, {Lowe}, {Wainscoat}, {Waters}, {Willman},
  {Pan-STARRS}, {Ebisawa}, {Hanyu}, {Harita}, {Hashimoto}, {Hidaka}, {Hori},
  {Ishikawa}, {Isobe}, {Iwakiri}, {Kawai}, {Kawai}, {Kawamuro}, {Kawase},
  {Kitaoka}, {Makishima}, {Matsuoka}, {Mihara}, {Morita}, {Morita}, {Nakahira},
  {Nakajima}, {Nakamura}, {Negoro}, {Oda}, {Sakamaki}, {Sasaki}, {Serino},
  {Shidatsu}, {Shimomukai}, {Sugawara}, {Sugita}, {Sugizaki}, {Tachibana},
  {Takao}, {Tanimoto}, {Tomida}, {Tsuboi}, {Tsunemi}, {Ueda}, {Ueno}, {Yamada},
  {Yamaoka}, {Yamauchi}, {Yatabe}, {Yoneyama}, {Yoshii}, {MAXI Team}, {Coward},
  {Crisp}, {Macpherson}, {Andreoni}, {Laugier}, {Noysena}, {Klotz}, {Gendre},
  {Thierry}, {Turpin}, {Consortium}, {Im}, {Choi}, {Kim}, {Yoon}, {Lim}, {Lee},
  {Lee}, {Kim}, {Ko}, {Joe}, {Kwon}, {Kim}, {Lim}, {Choi}, {KU Collaboration},
  {Fynbo}, {Malesani}, {Xu}, {Optical Telescope}, {Smartt}, {Jerkstrand},
  {Kankare}, {Sim}, {Fraser}, {Inserra}, {Maguire}, {Leloudas}, {Magee},
  {Shingles}, {Smith}, {Young}, {Kotak}, {Gal-Yam}, {Lyman}, {Homan},
  {Agliozzo}, {Anderson}, {Angus}, {Ashall}, {Barbarino}, {Bauer}, {Berton},
  {Botticella}, {Bulla}, {Cannizzaro}, {Cartier}, {Cikota}, {Clark}, {De Cia},
  {Della Valle}, {Dennefeld}, {Dessart}, {Dimitriadis}, {Elias-Rosa}, {Firth},
  {Fl{\"o}rs}, {Frohmaier}, {Galbany}, {Gonz{\'a}lez-Gait{\'a}n}, {Gromadzki},
  {Guti{\'e}rrez}, {Hamanowicz}, {Harmanen}, {Heintz}, {Hernandez}, {Hodgkin},
  {Hook}, {Izzo}, {James}, {Jonker}, {Kerzendorf}, {Kostrzewa-Rutkowska},
  {Kromer}, {Kuncarayakti}, {Lawrence}, {Manulis}, {Mattila}, {McBrien},
  {M{\"u}ller}, {Nordin}, {O'Neill}, {Onori}, {Palmerio}, {Pastorello},
  {Patat}, {Pignata}, {Podsiadlowski}, {Razza}, {Reynolds}, {Roy}, {Ruiter},
  {Rybicki}, {Salmon}, {Pumo}, {Prentice}, {Seitenzahl}, {Smith}, {Sollerman},
  {Sullivan}, {Szegedi}, {Taddia}, {Taubenberger}, {Terreran}, {Van Soelen},
  {Vos}, {Walton}, {Wright}, {Wyrzykowski}, {Yaron}, {pre=''(''>ePESSTO},
  {Chen}, {Kr{\"u}hler}, {Schady}, {Wiseman}, {Greiner}, {Rau}, {Schweyer},
  {Klose}, {Nicuesa Guelbenzu}, {GROND}, {Palliyaguru}, {Tech University},
  {Shara}, {Williams}, {Vaisanen}, {Potter}, {Romero Colmenero}, {Crawford},
  {Buckley}, {Mao}, {SALT Group}, {D{\'\i}az}, {Macri}, {Garc{\'\i}a Lambas},
  {Mendes de Oliveira}, {Nilo Castell{\'o}n}, {Ribeiro}, {S{\'a}nchez},
  {Schoenell}, {Abramo}, {Akras}, {Alcaniz}, {Artola}, {Beroiz}, {Bonoli},
  {Cabral}, {Camuccio}, {Chavushyan}, {Coelho}, {Colazo}, {Costa-Duarte},
  {Cuevas Larenas}, {Dom{\'\i}nguez Romero}, {Dultzin}, {Fern{\'a}ndez},
  {Garc{\'\i}a}, {Girardini}, {Gon{\c{c}}alves}, {Gon{\c{c}}alves}, {Gurovich},
  {Jim{\'e}nez-Teja}, {Kanaan}, {Lares}, {Lopes de Oliveira}, {L{\'o}pez-Cruz},
  {Melia}, {Molino}, {Padilla}, {Pe{\~n}uela}, {Placco}, {Qui{\~n}ones},
  {Ram{\'\i}rez Rivera}, {Renzi}, {Riguccini}, {R{\'\i}os-L{\'o}pez},
  {Rodriguez}, {Sampedro}, {Schneiter}, {Sodr{\'e}}, {Starck}, {Torres-Flores},
  {Tornatore}, {Zadro{\.z}ny}, {Castillo}, {TOROS: Transient Robotic
  Observatory of South Collaboration}, {Castro-Tirado}, {Tello}, {Hu}, {Zhang},
  {Cunniffe}, {Castell{\'o}n}, {Hiriart}, {Caballero-Garc{\'\i}a},
  {Jel{\'\i}nek}, {Kub{\'a}nek}, {P{\'e}rez del Pulgar}, {Park}, {Jeong},
  {Castro Cer{\'o}n}, {Pandey}, {Yock}, {Querel}, {Fan}, {Wang}, {BOOTES
  Collaboration}, {Beardsley}, {Brown}, {Crosse}, {Emrich}, {Franzen},
  {Gaensler}, {Horsley}, {Johnston-Hollitt}, {Kenney}, {Morales}, {Pallot},
  {Sokolowski}, {Steele}, {Tingay}, {Trott}, {Walker}, {Wayth}, {Williams},
  {Wu}, {Murchison Widefield Array}, {Yoshida}, {Sakamoto}, {Kawakubo},
  {Yamaoka}, {Takahashi}, {Asaoka}, {Ozawa}, {Torii}, {Shimizu}, {Tamura},
  {Ishizaki}, {Cherry}, {Ricciarini}, {Penacchioni}, {Marrocchesi}, {CALET
  Collaboration}, {Pozanenko}, {Volnova}, {Mazaeva}, {Minaev}, {Krugov},
  {Kusakin}, {Reva}, {Moskvitin}, {Rumyantsev}, {Inasaridze}, {Klunko},
  {Tungalag}, {Schmalz}, {Burhonov}, {IKI-GW Follow-up Collaboration},
  {Abdalla}, {Abramowski}, {Aharonian}, {Ait Benkhali}, {Ang{\"u}ner},
  {Arakawa}, {Arrieta}, {Aubert}, {Backes}, {Balzer}, {Barnard}, {Becherini},
  {Becker Tjus}, {Berge}, {Bernhard}, {Bernl{\"o}hr}, {Blackwell},
  {B{\"o}ttcher}, {Boisson}, {Bolmont}, {Bonnefoy}, {Bordas}, {Bregeon},
  {Brun}, {Brun}, {Bryan}, {B{\"u}chele}, {Bulik}, {Capasso}, {Caroff},
  {Carosi}, {Casanova}, {Cerruti}, {Chakraborty}, {Chaves}, {Chen},
  {Chevalier}, {Colafrancesco}, {Condon}, {Conrad}, {Davids}, {Decock}, {Deil},
  {Devin}, {deWilt}, {Dirson}, {Djannati-Ata{\"\i}}, {Donath}, {O'C. Drury},
  {Dutson}, {Dyks}, {Edwards}, {Egberts}, {Emery}, {Ernenwein}, {Eschbach},
  {Farnier}, {Fegan}, {Fernandes}, {Fiasson}, {Fontaine}, {Funk},
  {F{\"u}ssling}, {Gabici}, {Gallant}, {Garrigoux}, {Gat{\'e}}, {Giavitto},
  {Giebels}, {Glawion}, {Glicenstein}, {Gottschall}, {Grondin}, {Hahn},
  {Haupt}, {Hawkes}, {Heinzelmann}, {Henri}, {Hermann}, {Hinton}, {Hofmann},
  {Hoischen}, {Holch}, {Holler}, {Horns}, {Ivascenko}, {Iwasaki},
  {Jacholkowska}, {Jamrozy}, {Jankowsky}, {Jankowsky}, {Jingo}, {Jouvin},
  {Jung-Richardt}, {Kastendieck}, {Katarzy{\'n}ski}, {Katsuragawa},
  {Kerszberg}, {Khangulyan}, {Kh{\'e}lifi}, {King}, {Klepser}, {Klochkov},
  {Klu{\'z}niak}, {Komin}, {Kosack}, {Krakau}, {Kraus}, {Kr{\"u}ger}, {Laffon},
  {Lamanna}, {Lau}, {Lees}, {Lefaucheur}, {Lemi{\`e}re}, {Lemoine-Goumard},
  {Lenain}, {Leser}, {Lohse}, {Lorentz}, {Liu}, {Lypova}, {Malyshev},
  {Marandon}, {Marcowith}, {Mariaud}, {Marx}, {Maurin}, {Maxted}, {Mayer},
  {Meintjes}, {Meyer}, {Mitchell}, {Moderski}, {Mohamed}, {Mohrmann},
  {Mor{\r{a}}}, {Moulin}, {Murach}, {Nakashima}, {de Naurois}, {Ndiyavala},
  {Niederwanger}, {Niemiec}, {Oakes}, {O'Brien}, {Odaka}, {Ohm}, {Ostrowski},
  {Oya}, {Padovani}, {Panter}, {Parsons}, {Pekeur}, {Pelletier}, {Perennes},
  {Petrucci}, {Peyaud}, {Piel}, {Pita}, {Poireau}, {Poon}, {Prokhorov},
  {Prokoph}, {P{\"u}hlhofer}, {Punch}, {Quirrenbach}, {Raab}, {Rauth},
  {Reimer}, {Reimer}, {Renaud}, {de los Reyes}, {Rieger}, {Rinchiuso},
  {Romoli}, {Rowell}, {Rudak}, {Rulten}, {Sahakian}, {Saito}, {Sanchez},
  {Santangelo}, {Sasaki}, {Schlickeiser}, {Sch{\"u}ssler}, {Schulz},
  {Schwanke}, {Schwemmer}, {Seglar-Arroyo}, {Settimo}, {Seyffert}, {Shafi},
  {Shilon}, {Shiningayamwe}, {Simoni}, {Sol}, {Spanier}, {Spir-Jacob},
  {Stawarz}, {Steenkamp}, {Stegmann}, {Steppa}, {Sushch}, {Takahashi},
  {Tavernet}, {Tavernier}, {Taylor}, {Terrier}, {Tibaldo}, {Tiziani},
  {Tluczykont}, {Trichard}, {Tsirou}, {Tsuji}, {Tuffs}, {Uchiyama}, {van der
  Walt}, {van Eldik}, {van Rensburg}, {van Soelen}, {Vasileiadis}, {Veh},
  {Venter}, {Viana}, {Vincent}, {Vink}, {Voisin}, {V{\"o}lk}, {Vuillaume},
  {Wadiasingh}, {Wagner}, {Wagner}, {Wagner}, {White}, {Wierzcholska},
  {Willmann}, {W{\"o}rnlein}, {Wouters}, {Yang}, {Zaborov}, {Zacharias},
  {Zanin}, {Zdziarski}, {Zech}, {Zefi}, {Ziegler}, {Zorn}, {{\.Z}ywucka},
  {H.~E.~S.~S. Collaboration}, {Fender}, {Broderick}, {Rowlinson}, {Wijers},
  {Stewart}, {ter Veen}, {Shulevski}, {LOFAR Collaboration}, {Kavic},
  {Simonetti}, {League}, {Tsai}, {Obenberger}, {Nathaniel}, {Taylor}, {Dowell},
  {Liebling}, {Estes}, {Lippert}, {Sharma}, {Vincent}, {Farella}, {Wavelength
  Array}, {Abeysekara}, {Albert}, {Alfaro}, {Alvarez}, {Arceo},
  {Arteaga-Vel{\'a}zquez}, {Avila Rojas}, {Ayala Solares}, {Barber}, {Becerra
  Gonzalez}, {Becerril}, {Belmont-Moreno}, {BenZvi}, {Berley}, {Bernal},
  {Braun}, {Brisbois}, {Caballero-Mora}, {Capistr{\'a}n}, {Carrami{\~n}ana},
  {Casanova}, {Castillo}, {Cotti}, {Cotzomi}, {Couti{\~n}o de Le{\'o}n}, {De
  Le{\'o}n}, {De la Fuente}, {Diaz Hernandez}, {Dichiara}, {Dingus},
  {DuVernois}, {D{\'\i}az-V{\'e}lez}, {Ellsworth}, {Engel},
  {Enr{\'\i}quez-Rivera}, {Fiorino}, {Fleischhack}, {Fraija},
  {Garc{\'\i}a-Gonz{\'a}lez}, {Garfias}, {Gerhardt}, {Gonz{\~o}lez Mu{\~n}oz},
  {Gonz{\'a}lez}, {Goodman}, {Hampel-Arias}, {Harding}, {Hernandez},
  {Hernandez-Almada}, {Hona}, {H{\"u}ntemeyer}, {Iriarte}, {Jardin-Blicq},
  {Joshi}, {Kaufmann}, {Kieda}, {Lara}, {Lauer}, {Lennarz}, {Le{\'o}n Vargas},
  {Linnemann}, {Longinotti}, {Raya}, {Luna-Garc{\'\i}a}, {L{\'o}pez-Coto},
  {Malone}, {Marinelli}, {Martinez}, {Martinez-Castellanos},
  {Mart{\'\i}nez-Castro}, {Mart{\'\i}nez-Huerta}, {Matthews},
  {Miranda-Romagnoli}, {Moreno}, {Mostaf{\'a}}, {Nellen}, {Newbold}, {Nisa},
  {Noriega-Papaqui}, {Pelayo}, {Pretz}, {P{\'e}rez-P{\'e}rez}, {Ren}, {Rho},
  {Rivi{\`e}re}, {Rosa-Gonz{\'a}lez}, {Rosenberg}, {Ruiz-Velasco}, {Salazar},
  {Salesa Greus}, {Sandoval}, {Schneider}, {Schoorlemmer}, {Sinnis}, {Smith},
  {Springer}, {Surajbali}, {Tibolla}, {Tollefson}, {Torres}, {Ukwatta},
  {Weisgarber}, {Westerhoff}, {Wisher}, {Wood}, {Yapici}, {Yodh}, {Younk},
  {Zhou}, {{\'A}lvarez}, {HAWC Collaboration}, {Aab}, {Abreu}, {Aglietta},
  {Albuquerque}, {Albury}, {Allekotte}, {Almela}, {Alvarez Castillo},
  {Alvarez-Mu{\~n}iz}, {Anastasi}, {Anchordoqui}, {Andrada}, {Andringa},
  {Aramo}, {Arsene}, {Asorey}, {Assis}, {Avila}, {Badescu}, {Balaceanu},
  {Barbato}, {Barreira Luz}, {Becker}, {Bellido}, {Berat}, {Bertaina},
  {Bertou}, {Biermann}, {Biteau}, {Blaess}, {Blanco}, {Blazek}, {Bleve},
  {Boh{\'a}{\v{c}}ov{\'a}}, {Bonifazi}, {Borodai}, {Botti}, {Brack}, {Brancus},
  {Bretz}, {Bridgeman}, {Briechle}, {Buchholz}, {Bueno}, {Buitink}, {Buscemi},
  {Caballero-Mora}, {Caccianiga}, {Cancio}, {Canfora}, {Caruso}, {Castellina},
  {Catalani}, {Cataldi}, {Cazon}, {Chavez}, {Chinellato}, {Chudoba}, {Clay},
  {Cobos Cerutti}, {Colalillo}, {Coleman}, {Collica}, {Coluccia},
  {Concei{\c{c}}{\~a}o}, {Consolati}, {Contreras}, {Cooper}, {Coutu},
  {Covault}, {Cronin}, {D'Amico}, {Daniel}, {Dasso}, {Daumiller}, {Dawson},
  {Day}, {de Almeida}, {de Jong}, {De Mauro}, {de Mello Neto}, {De Mitri}, {de
  Oliveira}, {de Souza}, {Debatin}, {Deligny}, {D{\'\i}az Castro}, {Diogo},
  {Dobrigkeit}, {D'Olivo}, {Dorosti}, {Dos Anjos}, {Dova}, {Dundovic}, {Ebr},
  {Engel}, {Erdmann}, {Erfani}, {Escobar}, {Espadanal}, {Etchegoyen}, {Falcke},
  {Farmer}, {Farrar}, {Fauth}, {Fazzini}, {Feldbusch}, {Fenu}, {Fick},
  {Figueira}, {Filip{\v{c}}i{\v{c}}}, {Freire}, {Fujii}, {Fuster},
  {Ga{\"\i}or}, {Garc{\'\i}a}, {Gat{\'e}}, {Gemmeke}, {Gherghel-Lascu}, {Ghia},
  {Giaccari}, {Giammarchi}, {Giller}, {G{\l}as}, {Glaser}, {Golup}, {G{\'o}mez
  Berisso}, {G{\'o}mez Vitale}, {Gonz{\'a}lez}, {Gorgi}, {Gottowik}, {Grillo},
  {Grubb}, {Guarino}, {Guedes}, {Halliday}, {Hampel}, {Hansen}, {Harari},
  {Harrison}, {Harvey}, {Haungs}, {Hebbeker}, {Heck}, {Heimann}, {Herve},
  {Hill}, {Hojvat}, {Holt}, {Homola}, {H{\"o}randel}, {Horvath},
  {Hrabovsk{\'y}}, {Huege}, {Hulsman}, {Insolia}, {Isar}, {Jandt}, {Johnsen},
  {Josebachuili}, {Jurysek}, {K{\"a}{\"a}p{\"a}}, {Kampert}, {Keilhauer},
  {Kemmerich}, {Kemp}, {Kieckhafer}, {Klages}, {Kleifges}, {Kleinfeller},
  {Krause}, {Krohm}, {Kuempel}, {Kukec Mezek}, {Kunka}, {Kuotb Awad}, {Lago},
  {LaHurd}, {Lang}, {Lauscher}, {Legumina}, {Leigui de Oliveira},
  {Letessier-Selvon}, {Lhenry-Yvon}, {Link}, {Lo Presti}, {Lopes}, {L{\'o}pez},
  {L{\'o}pez Casado}, {Lorek}, {Luce}, {Lucero}, {Malacari}, {Mallamaci},
  {Mandat}, {Mantsch}, {Mariazzi}, {Maris}, {Marsella}, {Martello}, {Martinez},
  {Mart{\'\i}nez Bravo}, {Mas{\'\i}as Meza}, {Mathes}, {Mathys}, {Matthews},
  {Matthiae}, {Mayotte}, {Mazur}, {Medina}, {Medina-Tanco}, {Melo},
  {Menshikov}, {Merenda}, {Michal}, {Micheletti}, {Middendorf}, {Miramonti},
  {Mitrica}, {Mockler}, {Mollerach}, {Montanet}, {Morello}, {Morlino},
  {M{\"u}ller}, {M{\"u}ller}, {Muller}, {M{\"u}ller}, {Mussa}, {Naranjo},
  {Nguyen}, {Niculescu-Oglinzanu}, {Niechciol}, {Niemietz}, {Niggemann},
  {Nitz}, {Nosek}, {Novotny}, {No{\v{z}}ka}, {N{\'u}{\~n}ez}, {Oikonomou},
  {Olinto}, {Palatka}, {Pallotta}, {Papenbreer}, {Parente}, {Parra}, {Paul},
  {Pech}, {Pedreira}, {P{\c{e}}kala}, {Pe{\~n}a-Rodriguez}, {Pereira},
  {Perlin}, {Perrone}, {Peters}, {Petrera}, {Phuntsok}, {Pierog}, {Pimenta},
  {Pirronello}, {Platino}, {Plum}, {Poh}, {Porowski}, {Prado}, {Privitera},
  {Prouza}, {Quel}, {Querchfeld}, {Quinn}, {Ramos-Pollan}, {Rautenberg},
  {Ravignani}, {Ridky}, {Riehn}, {Risse}, {Ristori}, {Rizi}, {Rodrigues de
  Carvalho}, {Rodriguez Fernandez}, {Rodriguez Rojo}, {Roncoroni}, {Roth},
  {Roulet}, {Rovero}, {Ruehl}, {Saffi}, {Saftoiu}, {Salamida}, {Salazar},
  {Saleh}, {Salina}, {S{\'a}nchez}, {Sanchez-Lucas}, {Santos}, {Santos},
  {Sarazin}, {Sarmento}, {Sarmiento-Cano}, {Sato}, {Schauer}, {Scherini},
  {Schieler}, {Schimp}, {Schmidt}, {Scholten}, {Schov{\'a}nek}, {Schr{\"o}der},
  {Schr{\"o}der}, {Schulz}, {Schumacher}, {Sciutto}, {Segreto}, {Shadkam},
  {Shellard}, {Sigl}, {Silli}, {{\v{S}}m{\'\i}da}, {Snow}, {Sommers},
  {Sonntag}, {Soriano}, {Squartini}, {Stanca}, {Stani{\v{c}}}, {Stasielak},
  {Stassi}, {Stolpovskiy}, {Strafella}, {Streich}, {Suarez},
  {Suarez-Dur{\'a}n}, {Sudholz}, {Suomij{\"a}rvi}, {Supanitsky},
  {{\v{S}}up{\'\i}k}, {Swain}, {Szadkowski}, {Taboada}, {Taborda},
  {Timmermans}, {Todero Peixoto}, {Tomankova}, {Tom{\'e}}, {Torralba Elipe},
  {Travnicek}, {Trini}, {Tueros}, {Ulrich}, {Unger}, {Urban}, {Vald{\'e}s
  Galicia}, {Vali{\~n}o}, {Valore}, {van Aar}, {van Bodegom}, {van den Berg},
  {van Vliet}, {Varela}, {Vargas C{\'a}rdenas}, {V{\'a}zquez}, {Veberi{\v{c}}},
  {Ventura}, {Vergara Quispe}, {Verzi}, {Vicha}, {Villase{\~n}or}, {Vorobiov},
  {Wahlberg}, {Wainberg}, {Walz}, {Watson}, {Weber}, {Weindl}, {Wiede{\'n}ski},
  {Wiencke}, {Wilczy{\'n}ski}, {Wirtz}, {Wittkowski}, {Wundheiler}, {Yang},
  {Yushkov}, {Zas}, {Zavrtanik}, {Zavrtanik}, {Zepeda}, {Zimmermann},
  {Ziolkowski}, {Zong}, {Zuccarello}, {Pierre Auger Collaboration}, {Kim},
  {Schulze}, {Bauer}, {Corral-Santana}, {de Gregorio-Monsalvo},
  {Gonz{\'a}lez-L{\'o}pez}, {Hartmann}, {Ishwara-Chandra}, {Mart{\'\i}n},
  {Mehner}, {Misra}, {Micha{\l}owski}, {Resmi}, {ALMA Collaboration}, {Paragi},
  {Agudo}, {An}, {Beswick}, {Casadio}, {Frey}, {Jonker}, {Kettenis}, {Marcote},
  {Moldon}, {Szomoru}, {van Langevelde}, {Yang}, {Euro VLBI Team}, {Cwiek},
  {Cwiok}, {Czyrkowski}, {Dabrowski}, {Kasprowicz}, {Mankiewicz}, {Nawrocki},
  {Opiela}, {Piotrowski}, {Wrochna}, {Zaremba}, {{\.Z}arnecki}, {Pi of the Sky
  Collaboration}, {Haggard}, {Nynka}, {Ruan}, {Chandra Team at McGill
  University}, {Bland}, {Booler}, {Devillepoix}, {de Gois}, {Hancock}, {Howie},
  {Paxman}, {Sansom}, {Towner}, {Desert Fireball Network}, {Tonry}, {Coughlin},
  {Stubbs}, {Denneau}, {Heinze}, {Stalder}, {Weiland}, {ATLAS}, {Eatough},
  {Kramer}, {Kraus}, {Time Resolution Universe Survey}, {Troja}, {Piro},
  {Becerra Gonz{\'a}lez}, {Butler}, {Fox}, {Khandrika}, {Kutyrev}, {Lee},
  {Ricci}, {Ryan}, {S{\'a}nchez-Ram{\'\i}rez}, {Veilleux}, {Watson},
  {Wieringa}, {Burgess}, {van Eerten}, {Fontes}, {Fryer}, {Korobkin},
  {Wollaeger}, {RIMAS}, {RATIR}, {Camilo}, {Foley}, {Goedhart}, {Makhathini},
  {Oozeer}, {Smirnov}, {Fender}, {Woudt}, \& {South
  Africa/MeerKAT}}]{Abbott17:mma}
---. 2017{\natexlab{b}}, \apjl, 848, L12, \dodoi{10.3847/2041-8213/aa91c9}

\bibitem[{{Abbott} {et~al.}(2018){Abbott}, {Abbott}, {Abbott}, {Abernathy},
  {Acernese}, {Ackley}, {Adams}, {Adams}, {Addesso}, {Adhikari}, {Adya},
  {Affeldt}, {Agathos}, {Agatsuma}, {Aggarwal}, {Aguiar}, {Aiello}, {Ain},
  {Ajith}, {Akutsu}, {Allen}, {Allocca}, {Altin}, {Ananyeva}, {Anderson},
  {Anderson}, {Ando}, {Appert}, {Arai}, {Araya}, {Araya}, {Areeda}, {Arnaud},
  {Arun}, {Asada}, {Ascenzi}, {Ashton}, {Aso}, {Ast}, {Aston}, {Astone},
  {Atsuta}, {Aufmuth}, {Aulbert}, {Avila-Alvarez}, {Awai}, {Babak}, {Bacon},
  {Bader}, {Baiotti}, {Baker}, {Baldaccini}, {Ballardin}, {Ballmer},
  {Barayoga}, {Barclay}, {Barish}, {Barker}, {Barone}, {Barr}, {Barsotti},
  {Barsuglia}, {Barta}, {Bartlett}, {Barton}, {Bartos}, {Bassiri}, {Basti},
  {Batch}, {Baune}, {Bavigadda}, {Bazzan}, {B{\'e}csy}, {Beer}, {Bejger},
  {Belahcene}, {Belgin}, {Bell}, {Berger}, {Bergmann}, {Berry}, {Bersanetti},
  {Bertolini}, {Betzwieser}, {Bhagwat}, {Bhandare}, {Bilenko}, {Billingsley},
  {Billman}, {Birch}, {Birney}, {Birnholtz}, {Biscans}, {Bisht}, {Bitossi},
  {Biwer}, {Bizouard}, {Blackburn}, {Blackman}, {Blair}, {Blair}, {Blair},
  {Bloemen}, {Bock}, {Boer}, {Bogaert}, {Bohe}, {Bondu}, {Bonnand}, {Boom},
  {Bork}, {Boschi}, {Bose}, {Bouffanais}, {Bozzi}, {Bradaschia}, {Brady},
  {Braginsky}, {Branchesi}, {Brau}, {Briant}, {Brillet}, {Brinkmann},
  {Brisson}, {Brockill}, {Broida}, {Brooks}, {Brown}, {Brown}, {Brown},
  {Brunett}, {Buchanan}, {Buikema}, {Bulik}, {Bulten}, {Buonanno}, {Buskulic},
  {Buy}, {Byer}, {Cabero}, {Cadonati}, {Cagnoli}, {Cahillane}, {Calder{\'o}n
  Bustillo}, {Callister}, {Calloni}, {Camp}, {Cannon}, {Cao}, {Cao}, {Capano},
  {Capocasa}, {Carbognani}, {Caride}, {Casanueva Diaz}, {Casentini}, {Caudill},
  {Cavagli{\`a}}, {Cavalier}, {Cavalieri}, {Cella}, {Cepeda}, {Cerboni
  Baiardi}, {Cerretani}, {Cesarini}, {Chamberlin}, {Chan}, {Chao}, {Charlton},
  {Chassande-Mottin}, {Cheeseboro}, {Chen}, {Chen}, {Cheng}, {Chincarini},
  {Chiummo}, {Chmiel}, {Cho}, {Cho}, {Chow}, {Christensen}, {Chu}, {Chua},
  {Chua}, {Chung}, {Ciani}, {Clara}, {Clark}, {Cleva}, {Cocchieri}, {Coccia},
  {Cohadon}, {Colla}, {Collette}, {Cominsky}, {Constancio}, {Conti}, {Cooper},
  {Corbitt}, {Cornish}, {Corsi}, {Cortese}, {Costa}, {Coughlin}, {Coughlin},
  {Coulon}, {Countryman}, {Couvares}, {Covas}, {Cowan}, {Coward}, {Cowart},
  {Coyne}, {Coyne}, {Creighton}, {Creighton}, {Cripe}, {Crowder}, {Cullen},
  {Cumming}, {Cunningham}, {Cuoco}, {Dal Canton}, {Danilishin}, {D'Antonio},
  {Danzmann}, {Dasgupta}, {da Silva Costa}, {Dattilo}, {Dave}, {Davier},
  {Davies}, {Davis}, {Daw}, {Day}, {Day}, {de}, {Debra}, {Debreczeni},
  {Degallaix}, {de Laurentis}, {Del{\'e}glise}, {Del Pozzo}, {Denker}, {Dent},
  {Dergachev}, {De Rosa}, {Derosa}, {Desalvo}, {Devine}, {Dhurandhar},
  {D{\'\i}az}, {di Fiore}, {di Giovanni}, {di Girolamo}, {di Lieto}, {di Pace},
  {di Palma}, {di Virgilio}, {Doctor}, {Doi}, {Dolique}, {Donovan}, {Dooley},
  {Doravari}, {Dorrington}, {Douglas}, {Dovale {\'A}lvarez}, {Downes}, {Drago},
  {Drever}, {Driggers}, {Du}, {Ducrot}, {Dwyer}, {Eda}, {Edo}, {Edwards},
  {Effler}, {Eggenstein}, {Ehrens}, {Eichholz}, {Eikenberry}, {Eisenstein},
  {Essick}, {Etienne}, {Etzel}, {Evans}, {Evans}, {Everett}, {Factourovich},
  {Fafone}, {Fair}, {Fairhurst}, {Fan}, {Farinon}, {Farr}, {Farr},
  {Fauchon-Jones}, {Favata}, {Fays}, {Fehrmann}, {Fejer}, {Fern{\'a}ndez
  Galiana}, {Ferrante}, {Ferreira}, {Ferrini}, {Fidecaro}, {Fiori}, {Fiorucci},
  {Fisher}, {Flaminio}, {Fletcher}, {Fong}, {Forsyth}, {Fournier}, {Frasca},
  {Frasconi}, {Frei}, {Freise}, {Frey}, {Frey}, {Fries}, {Fritschel}, {Frolov},
  {Fujii}, {Fujimoto}, {Fulda}, {Fyffe}, {Gabbard}, {Gadre}, {Gaebel}, {Gair},
  {Gammaitoni}, {Gaonkar}, {Garufi}, {Gaur}, {Gayathri}, {Gehrels}, {Gemme},
  {Genin}, {Gennai}, {George}, {Gergely}, {Germain}, {Ghonge}, {Ghosh},
  {Ghosh}, {Ghosh}, {Giaime}, {Giardina}, {Giazotto}, {Gill}, {Glaefke},
  {Goetz}, {Goetz}, {Gondan}, {Gonz{\'a}lez}, {Gonzalez Castro}, {Gopakumar},
  {Gorodetsky}, {Gossan}, {Gosselin}, {Gouaty}, {Grado}, {Graef}, {Granata},
  {Grant}, {Gras}, {Gray}, {Greco}, {Green}, {Groot}, {Grote}, {Grunewald},
  {Guidi}, {Guo}, {Gupta}, {Gupta}, {Gushwa}, {Gustafson}, {Gustafson},
  {Hacker}, {Hagiwara}, {Hall}, {Hall}, {Hammond}, {Haney}, {Hanke}, {Hanks},
  {Hanna}, {Hannam}, {Hanson}, {Hardwick}, {Harms}, {Harry}, {Harry}, {Hart},
  {Hartman}, {Haster}, {Haughian}, {Hayama}, {Healy}, {Heidmann}, {Heintze},
  {Heitmann}, {Hello}, {Hemming}, {Hendry}, {Heng}, {Hennig}, {Henry},
  {Heptonstall}, {Heurs}, {Hild}, {Hirose}, {Hoak}, {Hofman}, {Holt}, {Holz},
  {Hopkins}, {Hough}, {Houston}, {Howell}, {Hu}, {Huerta}, {Huet}, {Hughey},
  {Husa}, {Huttner}, {Huynh-Dinh}, {Indik}, {Ingram}, {Inta}, {Ioka}, {Isa},
  {Isac}, {Isi}, {Isogai}, {Itoh}, {Iyer}, {Izumi}, {Jacqmin}, {Jani},
  {Jaranowski}, {Jawahar}, {Jim{\'e}nez-Forteza}, {Johnson}, {Jones}, {Jones},
  {Jonker}, {Ju}, {Junker}, {Kagawa}, {Kajita}, {Kakizaki}, {Kalaghatgi},
  {Kalogera}, {Kamiizumi}, {Kanda}, {Kandhasamy}, {Kanemura}, {Kaneyama},
  {Kang}, {Kanner}, {Karki}, {Karvinen}, {Kasprzack}, {Kataoka},
  {Katsavounidis}, {Katzman}, {Kaufer}, {Kaur}, {Kawabe}, {Kawai}, {Kawamura},
  {K{\'e}f{\'e}lian}, {Keitel}, {Kelley}, {Kennedy}, {Key}, {Khalili}, {Khan},
  {Khan}, {Khan}, {Khazanov}, {Kijbunchoo}, {Kim}, {Kim}, {Kim}, {Kim}, {Kim},
  {Kim}, {Kimbrell}, {Kimura}, {King}, {King}, {Kirchhoff}, {Kissel}, {Klein},
  {Kleybolte}, {Klimenko}, {Koch}, {Koehlenbeck}, {Kojima}, {Kokeyama},
  {Koley}, {Komori}, {Kondrashov}, {Kontos}, {Korobko}, {Korth}, {Kotake},
  {Kowalska}, {Kozak}, {Kr{\"a}mer}, {Kringel}, {Krishnan}, {Kr{\'o}lak},
  {Kuehn}, {Kumar}, {Kumar}, {Kumar}, {Kuo}, {Kuroda}, {Kutynia}, {Kuwahara},
  {Lackey}, {Landry}, {Lang}, {Lange}, {Lantz}, {Lanza}, {Lartaux-Vollard},
  {Lasky}, {Laxen}, {Lazzarini}, {Lazzaro}, {Leaci}, {Leavey}, {Lebigot},
  {Lee}, {Lee}, {Lee}, {Lee}, {Lee}, {Lehmann}, {Lenon}, {Leonardi}, {Leong},
  {Leroy}, {Letendre}, {Levin}, {Li}, {Libson}, {Littenberg}, {Liu},
  {Lockerbie}, {Lombardi}, {London}, {Lord}, {Lorenzini}, {Loriette},
  {Lormand}, {Losurdo}, {Lough}, {Lousto}, {Lovelace}, {L{\"u}ck}, {Lundgren},
  {Lynch}, {Ma}, {Macfoy}, {Machenschalk}, {Macinnis}, {MacLeod},
  {Maga{\~n}a-Sandoval}, {Majorana}, {Maksimovic}, {Malvezzi}, {Man}, {Mandic},
  {Mangano}, {Mano}, {Mansell}, {Manske}, {Mantovani}, {Marchesoni}, {Marchio},
  {Marion}, {M{\'a}rka}, {M{\'a}rka}, {Markosyan}, {Maros}, {Martelli},
  {Martellini}, {Martin}, {Martynov}, {Mason}, {Masserot}, {Massinger},
  {Masso-Reid}, {Mastrogiovanni}, {Matichard}, {Matone}, {Matsumoto},
  {Matsushima}, {Mavalvala}, {Mazumder}, {McCarthy}, {McClelland}, {McCormick},
  {McGrath}, {McGuire}, {McIntyre}, {McIver}, {McManus}, {McRae}, {McWilliams},
  {Meacher}, {Meadors}, {Meidam}, {Melatos}, {Mendell}, {Mendoza-Gandara},
  {Mercer}, {Merilh}, {Merzougui}, {Meshkov}, {Messenger}, {Messick},
  {Metzdorff}, {Meyers}, {Mezzani}, {Miao}, {Michel}, {Michimura}, {Middleton},
  {Mikhailov}, {Milano}, {Miller}, {Miller}, {Miller}, {Miller}, {Millhouse},
  {Minenkov}, {Ming}, {Mirshekari}, {Mishra}, {Mitrofanov}, {Mitselmakher},
  {Mittleman}, {Miyakawa}, {Miyamoto}, {Miyamoto}, {Miyoki}, {Moggi}, {Mohan},
  {Mohapatra}, {Montani}, {Moore}, {Moore}, {Moraru}, {Moreno}, {Morii},
  {Morisaki}, {Moriwaki}, {Morriss}, {Mours}, {Mow-Lowry}, {Mueller}, {Muir},
  {Mukherjee}, {Mukherjee}, {Mukherjee}, {Mukund}, {Mullavey}, {Munch},
  {Muniz}, {Murray}, {Mytidis}, {Nagano}, {Nakamura}, {Nakamura}, {Nakano},
  {Nakano}, {Nakano}, {Nakao}, {Napier}, {Nardecchia}, {Narikawa},
  {Naticchioni}, {Nelemans}, {Nelson}, {Neri}, {Nery}, {Neunzert}, {Newport},
  {Newton}, {Nguyen}, {Ni}, {Nielsen}, {Nissanke}, {Nitz}, {Noack}, {Nocera},
  {Nolting}, {Normandin}, {Nuttall}, {Oberling}, {Ochsner}, {Oelker}, {Ogin},
  {Oh}, {Oh}, {Ohashi}, {Ohishi}, {Ohkawa}, {Ohme}, {Okutomi}, {Oliver}, {Ono},
  {Ono}, {Oohara}, {Oppermann}, {Oram}, {O'Reilly}, {O'Shaughnessy}, {Ottaway},
  {Overmier}, {Owen}, {Pace}, {Page}, {Pai}, {Pai}, {Palamos}, {Palashov},
  {Palomba}, {Pal-Singh}, {Pan}, {Pankow}, {Pannarale}, {Pant}, {Paoletti},
  {Paoli}, {Papa}, {Paris}, {Parker}, {Pascucci}, {Pasqualetti}, {Passaquieti},
  {Passuello}, {Patricelli}, {Pearlstone}, {Pedraza}, {Pedurand}, {Pekowsky},
  {Pele}, {Pe{\~n}a Arellano}, {Penn}, {Perez}, {Perreca}, {Perri}, {Pfeiffer},
  {Phelps}, {Piccinni}, {Pichot}, {Piergiovanni}, {Pierro}, {Pillant},
  {Pinard}, {Pinto}, {Pitkin}, {Poe}, {Poggiani}, {Popolizio}, {Post},
  {Powell}, {Prasad}, {Pratt}, {Predoi}, {Prestegard}, {Prijatelj}, {Principe},
  {Privitera}, {Prodi}, {Prokhorov}, {Puncken}, {Punturo}, {Puppo},
  {P{\"u}rrer}, {Qi}, {Qin}, {Qiu}, {Quetschke}, {Quintero}, {Quitzow-James},
  {Raab}, {Rabeling}, {Radkins}, {Raffai}, {Raja}, {Rajan}, {Rakhmanov},
  {Rapagnani}, {Raymond}, {Razzano}, {Re}, {Read}, {Regimbau}, {Rei}, {Reid},
  {Reitze}, {Rew}, {Reyes}, {Rhoades}, {Ricci}, {Riles}, {Rizzo}, {Robertson},
  {Robie}, {Robinet}, {Rocchi}, {Rolland}, {Rollins}, {Roma}, {Romano},
  {Romie}, {Rosi{\'n}ska}, {Rowan}, {R{\"u}diger}, {Ruggi}, {Ryan}, {Sachdev},
  {Sadecki}, {Sadeghian}, {Sago}, {Saijo}, {Saito}, {Sakai}, {Sakellariadou},
  {Salconi}, {Saleem}, {Salemi}, {Samajdar}, {Sammut}, {Sampson}, {Sanchez},
  {Sandberg}, {Sanders}, {Sasaki}, {Sassolas}, {Sathyaprakash}, {Sato}, {Sato},
  {Saulson}, {Sauter}, {Savage}, {Sawadsky}, {Schale}, {Scheuer}, {Schmidt},
  {Schmidt}, {Schmidt}, {Schnabel}, {Schofield}, {Sch{\"o}nbeck}, {Schreiber},
  {Schuette}, {Schutz}, {Schwalbe}, {Scott}, {Scott}, {Sekiguchi}, {Sekiguchi},
  {Sellers}, {Sengupta}, {Sentenac}, {Sequino}, {Sergeev}, {Setyawati},
  {Shaddock}, {Shaffer}, {Shahriar}, {Shapiro}, {Shawhan}, {Sheperd},
  {Shibata}, {Shikano}, {Shimoda}, {Shoda}, {Shoemaker}, {Shoemaker},
  {Siellez}, {Siemens}, {Sieniawska}, {Sigg}, {Silva}, {Singer}, {Singer},
  {Singh}, {Singh}, {Singhal}, {Sintes}, {Slagmolen}, {Smith}, {Smith},
  {Smith}, {Somiya}, {Son}, {Sorazu}, {Sorrentino}, {Souradeep}, {Spencer},
  {Srivastava}, {Staley}, {Steinke}, {Steinlechner}, {Steinlechner},
  {Steinmeyer}, {Stephens}, {Stevenson}, {Stone}, {Strain}, {Straniero},
  {Stratta}, {Strigin}, {Sturani}, {Stuver}, {Sugimoto}, {Summerscales}, {Sun},
  {Sunil}, {Sutton}, {Suzuki}, {Swinkels}, {Szczepa{\'n}czyk}, {Tacca},
  {Tagoshi}, {Takada}, {Takahashi}, {Takahashi}, {Takamori}, {Talukder},
  {Tanaka}, {Tanaka}, {Tanaka}, {Tanner}, {T{\'a}pai}, {Taracchini}, {Tatsumi},
  {Taylor}, {Telada}, {Theeg}, {Thomas}, {Thomas}, {Thomas}, {Thorne},
  {Thrane}, {Tippens}, {Tiwari}, {Tiwari}, {Tokmakov}, {Toland}, {Tomaru},
  {Tomlinson}, {Tonelli}, {Tornasi}, {Torrie}, {T{\"o}yr{\"a}}, {Travasso},
  {Traylor}, {Trifir{\`o}}, {Trinastic}, {Tringali}, {Trozzo}, {Tse}, {Tso},
  {Tsubono}, {Tsuzuki}, {Turconi}, {Tuyenbayev}, {Uchiyama}, {Uehara}, {Ueki},
  {Ueno}, {Ugolini}, {Unnikrishnan}, {Urban}, {Ushiba}, {Usman}, {Vahlbruch},
  {Vajente}, {Valdes}, {van Bakel}, {van Beuzekom}, {van den Brand}, {van den
  Broeck}, {Vander-Hyde}, {van der Schaaf}, {van Heijningen}, {van Putten},
  {van Veggel}, {Vardaro}, {Varma}, {Vass}, {Vas{\'u}th}, {Vecchio},
  {Vedovato}, {Veitch}, {Veitch}, {Venkateswara}, {Venugopalan}, {Verkindt},
  {Vetrano}, {Vicer{\'e}}, {Viets}, {Vinciguerra}, {Vine}, {Vinet}, {Vitale},
  {Vo}, {Vocca}, {Vorvick}, {Voss}, {Vousden}, {Vyatchanin}, {Wade}, {Wade},
  {Wade}, {Wakamatsu}, {Walker}, {Wallace}, {Walsh}, {Wang}, {Wang}, {Wang},
  {Wang}, {Ward}, {Warner}, {Was}, {Watchi}, {Weaver}, {Wei}, {Weinert},
  {Weinstein}, {Weiss}, {Wen}, {We{\ss}els}, {Westphal}, {Wette}, {Whelan},
  {Whiting}, {Whittle}, {Williams}, {Williams}, {Williamson}, {Willis},
  {Willke}, {Wimmer}, {Winkler}, {Wipf}, {Wittel}, {Woan}, {Woehler}, {Worden},
  {Wright}, {Wu}, {Wu}, {Yam}, {Yamamoto}, {Yamamoto}, {Yamamoto}, {Yancey},
  {Yano}, {Yap}, {Yokoyama}, {Yokozawa}, {Yoon}, {Yu}, {Yu}, {Yuzurihara},
  {Yvert}, {Zadro{\.z}ny}, {Zangrando}, {Zanolin}, {Zeidler}, {Zendri},
  {Zevin}, {Zhang}, {Zhang}, {Zhang}, {Zhang}, {Zhao}, {Zhou}, {Zhou}, {Zhu},
  {Zhu}, {Zucker}, {Zweizig}, {Kagra Collaboration}, \& {VIRGO
  Collaboration}}]{LIGO20}
---. 2018, Living Reviews in Relativity, 21, 3,
  \dodoi{10.1007/s41114-018-0012-9}

\bibitem[{{Alexander} {et~al.}(2017){Alexander}, {Berger}, {Fong}, {Williams},
  {Guidorzi}, {Margutti}, {Metzger}, {Annis}, {Blanchard}, {Brout}, {Brown},
  {Chen}, {Chornock}, {Cowperthwaite}, {Drout}, {Eftekhari}, {Frieman}, {Holz},
  {Nicholl}, {Rest}, {Sako}, {Soares-Santos}, \& {Villar}}]{Alexander17}
{Alexander}, K.~D., {Berger}, E., {Fong}, W., {et~al.} 2017, \apjl, 848, L21,
  \dodoi{10.3847/2041-8213/aa905d}

\bibitem[{{Alexander} {et~al.}(2018){Alexander}, {Margutti}, {Blanchard},
  {Fong}, {Berger}, {Hajela}, {Eftekhari}, {Chornock}, {Cowperthwaite},
  {Giannios}, {Guidorzi}, {Kathirgamaraju}, {MacFadyen}, {Metzger}, {Nicholl},
  {Sironi}, {Villar}, {Williams}, {Xie}, \& {Zrake}}]{Alexander18}
{Alexander}, K.~D., {Margutti}, R., {Blanchard}, P.~K., {et~al.} 2018, \apjl,
  863, L18, \dodoi{10.3847/2041-8213/aad637}

\bibitem[{{Andreoni} {et~al.}(2017){Andreoni}, {Ackley}, {Cooke}, {Acharyya},
  {Allison}, {Anderson}, {Ashley}, {Baade}, {Bailes}, {Bannister}, {Beardsley},
  {Bessell}, {Bian}, {Bland}, {Boer}, {Booler}, {Brandeker}, {Brown},
  {Buckley}, {Chang}, {Coward}, {Crawford}, {Crisp}, {Crosse}, {Cucchiara},
  {Cup{\'a}k}, {de Gois}, {Deller}, {Devillepoix}, {Dobie}, {Elmer}, {Emrich},
  {Farah}, {Farrell}, {Franzen}, {Gaensler}, {Galloway}, {Gendre}, {Giblin},
  {Goobar}, {Green}, {Hancock}, {Hartig}, {Howell}, {Horsley}, {Hotan},
  {Howie}, {Hu}, {Hu}, {James}, {Johnston}, {Johnston-Hollitt}, {Kaplan},
  {Kasliwal}, {Keane}, {Kenney}, {Klotz}, {Lau}, {Laugier}, {Lenc}, {Li},
  {Liang}, {Lidman}, {Luvaul}, {Lynch}, {Ma}, {Macpherson}, {Mao},
  {McClelland}, {McCully}, {M{\"o}ller}, {Morales}, {Morris}, {Murphy},
  {Noysena}, {Onken}, {Orange}, {Os{\l}owski}, {Pallot}, {Paxman}, {Potter},
  {Pritchard}, {Raja}, {Ridden-Harper}, {Romero-Colmenero}, {Sadler}, {Sansom},
  {Scalzo}, {Schmidt}, {Scott}, {Seghouani}, {Shang}, {Shannon}, {Shao},
  {Shara}, {Sharp}, {Sokolowski}, {Sollerman}, {Staff}, {Steele}, {Sun},
  {Suntzeff}, {Tao}, {Tingay}, {Towner}, {Thierry}, {Trott}, {Tucker},
  {V{\"a}is{\"a}nen}, {Krishnan}, {Walker}, {Wang}, {Wang}, {Wayth}, {Whiting},
  {Williams}, {Williams}, {Wolf}, {Wu}, {Wu}, {Yang}, {Yuan}, {Zhang}, {Zhou},
  \& {Zovaro}}]{Andreoni17}
{Andreoni}, I., {Ackley}, K., {Cooke}, J., {et~al.} 2017, \pasa, 34, e069,
  \dodoi{10.1017/pasa.2017.65}

\bibitem[{{Arcavi} {et~al.}(2017){Arcavi}, {Hosseinzadeh}, {Howell}, {McCully},
  {Poznanski}, {Kasen}, {Barnes}, {Zaltzman}, {Vasylyev}, {Maoz}, \&
  {Valenti}}]{Arcavi17}
{Arcavi}, I., {Hosseinzadeh}, G., {Howell}, D.~A., {et~al.} 2017, \nat, 551,
  64, \dodoi{10.1038/nature24291}

\bibitem[{{Arnett}(1982)}]{Arnett82}
{Arnett}, W.~D. 1982, \apj, 253, 785, \dodoi{10.1086/159681}

\bibitem[{{Astropy Collaboration} {et~al.}(2013){Astropy Collaboration},
  {Robitaille}, {Tollerud}, {Greenfield}, {Droettboom}, {Bray}, {Aldcroft},
  {Davis}, {Ginsburg}, {Price-Whelan}, {Kerzendorf}, {Conley}, {Crighton},
  {Barbary}, {Muna}, {Ferguson}, {Grollier}, {Parikh}, {Nair}, {Unther},
  {Deil}, {Woillez}, {Conseil}, {Kramer}, {Turner}, {Singer}, {Fox}, {Weaver},
  {Zabalza}, {Edwards}, {Azalee Bostroem}, {Burke}, {Casey}, {Crawford},
  {Dencheva}, {Ely}, {Jenness}, {Labrie}, {Lim}, {Pierfederici}, {Pontzen},
  {Ptak}, {Refsdal}, {Servillat}, \& {Streicher}}]{astropy}
{Astropy Collaboration}, {Robitaille}, T.~P., {Tollerud}, E.~J., {et~al.} 2013,
  \aap, 558, A33, \dodoi{10.1051/0004-6361/201322068}

\bibitem[{{Baillot d'Etivaux} {et~al.}(2019){Baillot d'Etivaux}, {Guillot},
  {Margueron}, {Webb}, {Catelan}, \& {Reisenegger}}]{Baillot19}
{Baillot d'Etivaux}, N., {Guillot}, S., {Margueron}, J., {et~al.} 2019, \apj,
  887, 48, \dodoi{10.3847/1538-4357/ab4f6c}

\bibitem[{{Balasubramanian} {et~al.}(2021){Balasubramanian}, {Corsi}, {Mooley},
  {Brightman}, {Hallinan}, {Hotokezaka}, {Kaplan}, {Lazzati}, \&
  {Murphy}}]{Balasubramanian21}
{Balasubramanian}, A., {Corsi}, A., {Mooley}, K.~P., {et~al.} 2021, \apjl, 914,
  L20, \dodoi{10.3847/2041-8213/abfd38}

\bibitem[{{Barnes} \& {Kasen}(2013)}]{Barnes13}
{Barnes}, J., \& {Kasen}, D. 2013, \apj, 775, 18,
  \dodoi{10.1088/0004-637X/775/1/18}

\bibitem[{{Barnes} {et~al.}(2016){Barnes}, {Kasen}, {Wu}, \&
  {Mart{\'\i}nez-Pinedo}}]{Barnes16}
{Barnes}, J., {Kasen}, D., {Wu}, M.-R., \& {Mart{\'\i}nez-Pinedo}, G. 2016,
  \apj, 829, 110, \dodoi{10.3847/0004-637X/829/2/110}

\bibitem[{{Becker}(2015)}]{hotpants}
{Becker}, A. 2015, {HOTPANTS: High Order Transform of PSF ANd Template
  Subtraction}.
\newblock \doeprint{1504.004}

\bibitem[{{Belczynski} {et~al.}(2018){Belczynski}, {Askar}, {Arca-Sedda},
  {Chruslinska}, {Donnari}, {Giersz}, {Benacquista}, {Spurzem}, {Jin},
  {Wiktorowicz}, \& {Belloni}}]{Belczynski18}
{Belczynski}, K., {Askar}, A., {Arca-Sedda}, M., {et~al.} 2018, \aap, 615, A91,
  \dodoi{10.1051/0004-6361/201732428}

\bibitem[{{Bertin} \& {Arnouts}(1996)}]{sextractor}
{Bertin}, E., \& {Arnouts}, S. 1996, \aaps, 117, 393,
  \dodoi{10.1051/aas:1996164}

\bibitem[{{B{\'\i}lek} {et~al.}(2015){B{\'\i}lek}, {Ebrov{\'a}}, {Jungwiert},
  {J{\'\i}lkov{\'a}}, \& {Barto{\v{s}}kov{\'a}}}]{Bilek15}
{B{\'\i}lek}, M., {Ebrov{\'a}}, I., {Jungwiert}, B., {J{\'\i}lkov{\'a}}, L., \&
  {Barto{\v{s}}kov{\'a}}, K. 2015, Canadian Journal of Physics, 93, 203,
  \dodoi{10.1139/cjp-2014-0170}

\bibitem[{{Blakeslee} {et~al.}(2012){Blakeslee}, {Cho}, {Peng}, {Ferrarese},
  {Jord{\'a}n}, \& {Martel}}]{Blakeslee12}
{Blakeslee}, J.~P., {Cho}, H., {Peng}, E.~W., {et~al.} 2012, \apj, 746, 88,
  \dodoi{10.1088/0004-637X/746/1/88}

\bibitem[{{Blanchard} {et~al.}(2017){Blanchard}, {Berger}, {Fong}, {Nicholl},
  {Leja}, {Conroy}, {Alexander}, {Margutti}, {Williams}, {Doctor}, {Chornock},
  {Villar}, {Cowperthwaite}, {Annis}, {Brout}, {Brown}, {Chen}, {Eftekhari},
  {Frieman}, {Holz}, {Metzger}, {Rest}, {Sako}, \&
  {Soares-Santos}}]{Blanchard17}
{Blanchard}, P.~K., {Berger}, E., {Fong}, W., {et~al.} 2017, \apjl, 848, L22,
  \dodoi{10.3847/2041-8213/aa9055}

\bibitem[{{Bradley} {et~al.}(2020){Bradley}, {Sip{\H{o}}cz}, {Robitaille},
  {Tollerud}, {Vin{\'\i}cius}, {Deil}, {Barbary}, {Wilson}, {Busko},
  {G{\"u}nther}, {Cara}, {Conseil}, {Bostroem}, {Droettboom}, {Bray}, {Andersen
  Bratholm}, {Lim}, {Barentsen}, {Craig}, {Pascual}, {Perren}, {Greco},
  {Donath}, {De Val-Borro}, {Kerzendorf}, {Bach}, {Weaver}, {D'Eugenio},
  {Souchereau}, \& {Ferreira}}]{photutils}
{Bradley}, L., {Sip{\H{o}}cz}, B., {Robitaille}, T., {et~al.} 2020,
  {astropy/photutils: 1.0.1}, 1.0.1,  Zenodo, \dodoi{10.5281/zenodo.596036}

\bibitem[{{Brodie} \& {Strader}(2006)}]{BrodieStrader06}
{Brodie}, J.~P., \& {Strader}, J. 2006, \araa, 44, 193,
  \dodoi{10.1146/annurev.astro.44.051905.092441}

\bibitem[{{Calzetti} {et~al.}(2000){Calzetti}, {Armus}, {Bohlin}, {Kinney},
  {Koornneef}, \& {Storchi-Bergmann}}]{Calzetti00}
{Calzetti}, D., {Armus}, L., {Bohlin}, R.~C., {et~al.} 2000, \apj, 533, 682,
  \dodoi{10.1086/308692}

\bibitem[{{Cantiello} {et~al.}(2018){Cantiello}, {Jensen}, {Blakeslee},
  {Berger}, {Levan}, {Tanvir}, {Raimondo}, {Brocato}, {Alexander}, {Blanchard},
  {Branchesi}, {Cano}, {Chornock}, {Covino}, {Cowperthwaite}, {D'Avanzo},
  {Eftekhari}, {Fong}, {Fruchter}, {Grado}, {Hjorth}, {Holz}, {Lyman},
  {Mandel}, {Margutti}, {Nicholl}, {Villar}, \& {Williams}}]{Cantiello18}
{Cantiello}, M., {Jensen}, J.~B., {Blakeslee}, J.~P., {et~al.} 2018, \apjl,
  854, L31, \dodoi{10.3847/2041-8213/aaad64}

\bibitem[{{Chabrier}(2003)}]{Chabrier03}
{Chabrier}, G. 2003, \pasp, 115, 763, \dodoi{10.1086/376392}

\bibitem[{{Chornock} {et~al.}(2017){Chornock}, {Berger}, {Kasen},
  {Cowperthwaite}, {Nicholl}, {Villar}, {Alexander}, {Blanchard}, {Eftekhari},
  {Fong}, {Margutti}, {Williams}, {Annis}, {Brout}, {Brown}, {Chen}, {Drout},
  {Farr}, {Foley}, {Frieman}, {Fryer}, {Herner}, {Holz}, {Kessler}, {Matheson},
  {Metzger}, {Quataert}, {Rest}, {Sako}, {Scolnic}, {Smith}, \&
  {Soares-Santos}}]{Chornock17}
{Chornock}, R., {Berger}, E., {Kasen}, D., {et~al.} 2017, \apjl, 848, L19,
  \dodoi{10.3847/2041-8213/aa905c}

\bibitem[{{Colbert} {et~al.}(2001){Colbert}, {Mulchaey}, \&
  {Zabludoff}}]{Colbert2001}
{Colbert}, J.~W., {Mulchaey}, J.~S., \& {Zabludoff}, A.~I. 2001, \aj, 121, 808,
  \dodoi{10.1086/318758}

\bibitem[{{Conroy}(2013)}]{Conroy13}
{Conroy}, C. 2013, \araa, 51, 393, \dodoi{10.1146/annurev-astro-082812-141017}

\bibitem[{{Conroy} \& {Gunn}(2010)}]{Conroy10}
{Conroy}, C., \& {Gunn}, J.~E. 2010, \apj, 712, 833,
  \dodoi{10.1088/0004-637X/712/2/833}

\bibitem[{{Conroy} {et~al.}(2009){Conroy}, {Gunn}, \& {White}}]{Conroy09}
{Conroy}, C., {Gunn}, J.~E., \& {White}, M. 2009, \apj, 699, 486,
  \dodoi{10.1088/0004-637X/699/1/486}

\bibitem[{{Coulter} {et~al.}(2017){Coulter}, {Foley}, {Kilpatrick}, {Drout},
  {Piro}, {Shappee}, {Siebert}, {Simon}, {Ulloa}, {Kasen}, {Madore},
  {Murguia-Berthier}, {Pan}, {Prochaska}, {Ramirez-Ruiz}, {Rest}, \&
  {Rojas-Bravo}}]{Coulter17}
{Coulter}, D.~A., {Foley}, R.~J., {Kilpatrick}, C.~D., {et~al.} 2017, Science,
  358, 1556, \dodoi{10.1126/science.aap9811}

\bibitem[{{Cowperthwaite} {et~al.}(2017){Cowperthwaite}, {Berger}, {Villar},
  {Metzger}, {Nicholl}, {Chornock}, {Blanchard}, {Fong}, {Margutti},
  {Soares-Santos}, {Alexander}, {Allam}, {Annis}, {Brout}, {Brown}, {Butler},
  {Chen}, {Diehl}, {Doctor}, {Drout}, {Eftekhari}, {Farr}, {Finley}, {Foley},
  {Frieman}, {Fryer}, {Garc{\'\i}a-Bellido}, {Gill}, {Guillochon}, {Herner},
  {Holz}, {Kasen}, {Kessler}, {Marriner}, {Matheson}, {Neilsen}, {Quataert},
  {Palmese}, {Rest}, {Sako}, {Scolnic}, {Smith}, {Tucker}, {Williams},
  {Balbinot}, {Carlin}, {Cook}, {Durret}, {Li}, {Lopes}, {Louren{\c{c}}o},
  {Marshall}, {Medina}, {Muir}, {Mu{\~n}oz}, {Sauseda}, {Schlegel}, {Secco},
  {Vivas}, {Wester}, {Zenteno}, {Zhang}, {Abbott}, {Banerji}, {Bechtol},
  {Benoit-L{\'e}vy}, {Bertin}, {Buckley-Geer}, {Burke}, {Capozzi}, {Carnero
  Rosell}, {Carrasco Kind}, {Castander}, {Crocce}, {Cunha}, {D'Andrea}, {da
  Costa}, {Davis}, {DePoy}, {Desai}, {Dietrich}, {Drlica-Wagner}, {Eifler},
  {Evrard}, {Fernandez}, {Flaugher}, {Fosalba}, {Gaztanaga}, {Gerdes},
  {Giannantonio}, {Goldstein}, {Gruen}, {Gruendl}, {Gutierrez}, {Honscheid},
  {Jain}, {James}, {Jeltema}, {Johnson}, {Johnson}, {Kent}, {Krause}, {Kron},
  {Kuehn}, {Nuropatkin}, {Lahav}, {Lima}, {Lin}, {Maia}, {March}, {Martini},
  {McMahon}, {Menanteau}, {Miller}, {Miquel}, {Mohr}, {Neilsen}, {Nichol},
  {Ogando}, {Plazas}, {Roe}, {Romer}, {Roodman}, {Rykoff}, {Sanchez},
  {Scarpine}, {Schindler}, {Schubnell}, {Sevilla-Noarbe}, {Smith}, {Smith},
  {Sobreira}, {Suchyta}, {Swanson}, {Tarle}, {Thomas}, {Thomas}, {Troxel},
  {Vikram}, {Walker}, {Wechsler}, {Weller}, {Yanny}, \&
  {Zuntz}}]{Cowperthwaite17}
{Cowperthwaite}, P.~S., {Berger}, E., {Villar}, V.~A., {et~al.} 2017, \apjl,
  848, L17, \dodoi{10.3847/2041-8213/aa8fc7}

\bibitem[{{D{\'\i}az} {et~al.}(2017){D{\'\i}az}, {Macri}, {Garcia Lambas},
  {Mendes de Oliveira}, {Nilo Castell{\'o}n}, {Ribeiro}, {S{\'a}nchez},
  {Schoenell}, {Abramo}, {Akras}, {Alcaniz}, {Artola}, {Beroiz}, {Bonoli},
  {Cabral}, {Camuccio}, {Castillo}, {Chavushyan}, {Coelho}, {Colazo},
  {Costa-Duarte}, {Cuevas Larenas}, {DePoy}, {Dom{\'\i}nguez Romero},
  {Dultzin}, {Fern{\'a}ndez}, {Garc{\'\i}a}, {Girardini}, {Gon{\c{c}}alves},
  {Gon{\c{c}}alves}, {Gurovich}, {Jim{\'e}nez-Teja}, {Kanaan}, {Lares}, {Lopes
  de Oliveira}, {L{\'o}pez-Cruz}, {Marshall}, {Melia}, {Molino}, {Padilla},
  {Pe{\~n}uela}, {Placco}, {Qui{\~n}ones}, {Ram{\'\i}rez Rivera}, {Renzi},
  {Riguccini}, {R{\'\i}os-L{\'o}pez}, {Rodriguez}, {Sampedro}, {Schneiter},
  {Sodr{\'e}}, {Starck}, {Torres-Flores}, {Tornatore}, \&
  {Zadro{\.z}ny}}]{Diaz17}
{D{\'\i}az}, M.~C., {Macri}, L.~M., {Garcia Lambas}, D., {et~al.} 2017, \apjl,
  848, L29, \dodoi{10.3847/2041-8213/aa9060}

\bibitem[{{Dolphin}(2016)}]{dolphot}
{Dolphin}, A. 2016, {DOLPHOT: Stellar photometry}.
\newblock \doeprint{1608.013}

\bibitem[{{Drout} {et~al.}(2009){Drout}, {Massey}, {Meynet}, {Tokarz}, \&
  {Caldwell}}]{Drout09}
{Drout}, M.~R., {Massey}, P., {Meynet}, G., {Tokarz}, S., \& {Caldwell}, N.
  2009, \apj, 703, 441, \dodoi{10.1088/0004-637X/703/1/441}

\bibitem[{{Drout} {et~al.}(2017){Drout}, {Piro}, {Shappee}, {Kilpatrick},
  {Simon}, {Contreras}, {Coulter}, {Foley}, {Siebert}, {Morrell}, {Boutsia},
  {Di Mille}, {Holoien}, {Kasen}, {Kollmeier}, {Madore}, {Monson},
  {Murguia-Berthier}, {Pan}, {Prochaska}, {Ramirez-Ruiz}, {Rest}, {Adams},
  {Alatalo}, {Ba{\~n}ados}, {Baughman}, {Beers}, {Bernstein}, {Bitsakis},
  {Campillay}, {Hansen}, {Higgs}, {Ji}, {Maravelias}, {Marshall}, {Moni Bidin},
  {Prieto}, {Rasmussen}, {Rojas-Bravo}, {Strom}, {Ulloa},
  {Vargas-Gonz{\'a}lez}, {Wan}, \& {Whitten}}]{Drout17}
{Drout}, M.~R., {Piro}, A.~L., {Shappee}, B.~J., {et~al.} 2017, Science, 358,
  1570, \dodoi{10.1126/science.aaq0049}

\bibitem[{{Dupraz} \& {Combes}(1986)}]{Dupraz86}
{Dupraz}, C., \& {Combes}, F. 1986, \aap, 166, 53

\bibitem[{{Ebrov{\'a}} {et~al.}(2020){Ebrov{\'a}}, {B{\'\i}lek},
  {Y{\i}ld{\i}z}, \& {Eli{\'a}{\v{s}}ek}}]{Ebrova20}
{Ebrov{\'a}}, I., {B{\'\i}lek}, M., {Y{\i}ld{\i}z}, M.~K., \&
  {Eli{\'a}{\v{s}}ek}, J. 2020, \aap, 634, A73,
  \dodoi{10.1051/0004-6361/201935219}

\bibitem[{{Evans} {et~al.}(2017){Evans}, {Cenko}, {Kennea}, {Emery}, {Kuin},
  {Korobkin}, {Wollaeger}, {Fryer}, {Madsen}, {Harrison}, {Xu}, {Nakar},
  {Hotokezaka}, {Lien}, {Campana}, {Oates}, {Troja}, {Breeveld}, {Marshall},
  {Barthelmy}, {Beardmore}, {Burrows}, {Cusumano}, {D'A{\`\i}}, {D'Avanzo},
  {D'Elia}, {de Pasquale}, {Even}, {Fontes}, {Forster}, {Garcia}, {Giommi},
  {Grefenstette}, {Gronwall}, {Hartmann}, {Heida}, {Hungerford}, {Kasliwal},
  {Krimm}, {Levan}, {Malesani}, {Melandri}, {Miyasaka}, {Nousek}, {O'Brien},
  {Osborne}, {Pagani}, {Page}, {Palmer}, {Perri}, {Pike}, {Racusin}, {Rosswog},
  {Siegel}, {Sakamoto}, {Sbarufatti}, {Tagliaferri}, {Tanvir}, \&
  {Tohuvavohu}}]{Evans17}
{Evans}, P.~A., {Cenko}, S.~B., {Kennea}, J.~A., {et~al.} 2017, Science, 358,
  1565, \dodoi{10.1126/science.aap9580}

\bibitem[{{Fong} {et~al.}(2019){Fong}, {Blanchard}, {Alexander}, {Strader},
  {Margutti}, {Hajela}, {Villar}, {Wu}, {Ye}, {Berger}, {Chornock},
  {Coppejans}, {Cowperthwaite}, {Eftekhari}, {Giannios}, {Guidorzi},
  {Kathirgamaraju}, {Laskar}, {Macfadyen}, {Metzger}, {Nicholl}, {Paterson},
  {Terreran}, {Sand}, {Sironi}, {Williams}, {Xie}, \& {Zrake}}]{Fong19}
{Fong}, W., {Blanchard}, P.~K., {Alexander}, K.~D., {et~al.} 2019, \apjl, 883,
  L1, \dodoi{10.3847/2041-8213/ab3d9e}

\bibitem[{{Fong} {et~al.}(2021{\natexlab{a}}){Fong}, {Laskar}, {Rastinejad},
  {Escorial}, {Schroeder}, {Barnes}, {Kilpatrick}, {Paterson}, {Berger},
  {Metzger}, {Dong}, {Nugent}, {Strausbaugh}, {Blanchard}, {Goyal},
  {Cucchiara}, {Terreran}, {Alexander}, {Eftekhari}, {Fryer}, {Margalit},
  {Margutti}, \& {Nicholl}}]{Fong21a}
{Fong}, W., {Laskar}, T., {Rastinejad}, J., {et~al.} 2021{\natexlab{a}}, \apj,
  906, 127, \dodoi{10.3847/1538-4357/abc74a}

\bibitem[{{Fong} {et~al.}(2021{\natexlab{b}}){Fong}, {Dong}, {Leja},
  {Bhandari}, {Day}, {Deller}, {Kumar}, {Prochaska}, {Scott}, {Bannister},
  {Eftekhari}, {Gordon}, {Heintz}, {James}, {Kilpatrick}, {Mahony}, {Rouco
  Escorial}, {Ryder}, {Shannon}, \& {Tejos}}]{Fong21}
{Fong}, W.-f., {Dong}, Y., {Leja}, J., {et~al.} 2021{\natexlab{b}}, arXiv
  e-prints, arXiv:2106.11993.
\newblock \doarXiv{2106.11993}

\bibitem[{{Fruchter} \& {Hook}(1997)}]{Fruchter97}
{Fruchter}, A.~S., \& {Hook}, R.~N. 1997, in The Hubble Space Telescope and the
  High Redshift Universe, ed. N.~R. {Tanvir}, A.~{Aragon-Salamanca}, \& J.~V.
  {Wall}, 137

\bibitem[{{Fruchter} \& {Hook}(2002)}]{Fruchter02}
{Fruchter}, A.~S., \& {Hook}, R.~N. 2002, \pasp, 114, 144,
  \dodoi{10.1086/338393}

\bibitem[{{Garcia}(1993)}]{Garcia93}
{Garcia}, A.~M. 1993, \aaps, 100, 47

\bibitem[{{Goldstein} {et~al.}(2017){Goldstein}, {Veres}, {Burns}, {Briggs},
  {Hamburg}, {Kocevski}, {Wilson-Hodge}, {Preece}, {Poolakkil}, {Roberts},
  {Hui}, {Connaughton}, {Racusin}, {von Kienlin}, {Dal Canton}, {Christensen},
  {Littenberg}, {Siellez}, {Blackburn}, {Broida}, {Bissaldi}, {Cleveland},
  {Gibby}, {Giles}, {Kippen}, {McBreen}, {McEnery}, {Meegan}, {Paciesas}, \&
  {Stanbro}}]{GW170817:fermi}
{Goldstein}, A., {Veres}, P., {Burns}, E., {et~al.} 2017, \apjl, 848, L14,
  \dodoi{10.3847/2041-8213/aa8f41}

\bibitem[{{Graham} {et~al.}(1983){Graham}, {Meikle}, {Selby}, {Allen}, {Evans},
  {Pearce}, {Bode}, {Longmore}, \& {Williams}}]{Graham83}
{Graham}, J.~R., {Meikle}, W.~P.~S., {Selby}, M.~J., {et~al.} 1983, \nat, 304,
  709, \dodoi{10.1038/304709a0}

\bibitem[{{Granot} {et~al.}(2018){Granot}, {Gill}, {Guetta}, \& {De
  Colle}}]{Granot18}
{Granot}, J., {Gill}, R., {Guetta}, D., \& {De Colle}, F. 2018, \mnras, 481,
  1597, \dodoi{10.1093/mnras/sty2308}

\bibitem[{{Haggard} {et~al.}(2017){Haggard}, {Nynka}, {Ruan}, {Kalogera},
  {Cenko}, {Evans}, \& {Kennea}}]{Haggard17}
{Haggard}, D., {Nynka}, M., {Ruan}, J.~J., {et~al.} 2017, \apjl, 848, L25,
  \dodoi{10.3847/2041-8213/aa8ede}

\bibitem[{{Hajela} {et~al.}(2020){Hajela}, {Margutti}, {Alexander}, {Laskar},
  {Giannios}, \& {Villar}}]{Hajela20}
{Hajela}, A., {Margutti}, R., {Alexander}, K.~D., {et~al.} 2020, GRB
  Coordinates Network, 29019, 1

\bibitem[{{Hajela} {et~al.}(2019){Hajela}, {Margutti}, {Alexander},
  {Kathirgamaraju}, {Baldeschi}, {Guidorzi}, {Giannios}, {Fong}, {Wu},
  {MacFadyen}, {Paggi}, {Berger}, {Blanchard}, {Chornock}, {Coppejans},
  {Cowperthwaite}, {Eftekhari}, {Gomez}, {Hosseinzadeh}, {Laskar}, {Metzger},
  {Nicholl}, {Paterson}, {Radice}, {Sironi}, {Terreran}, {Villar}, {Williams},
  {Xie}, \& {Zrake}}]{Hajela19}
---. 2019, \apjl, 886, L17, \dodoi{10.3847/2041-8213/ab5226}

\bibitem[{{Hajela} {et~al.}(2021){Hajela}, {Margutti}, {Bright}, {Alexander},
  {Metzger}, {Nedora}, {Kathirgamaraju}, {Margalit}, {Radice}, {Berger},
  {MacFadyen}, {Giannios}, {Chornock}, {Heywood}, {Sironi}, {Gottlieb},
  {Coppejans}, {Laskar}, {Cendes}, {Barniol Duran}, {Eftekhari}, {Fong},
  {McDowell}, {Nicholl}, {Xie}, {Zrake}, {Bernuzzi}, {Broekgaarden},
  {Kilpatrick}, {Terreran}, {Villar}, {Blanchard}, {Gomez}, {Hosseinzadeh},
  {Matthews}, \& {Rastinejad}}]{Hajela21}
{Hajela}, A., {Margutti}, R., {Bright}, J.~S., {et~al.} 2021, arXiv e-prints,
  arXiv:2104.02070.
\newblock \doarXiv{2104.02070}

\bibitem[{{Hallinan} {et~al.}(2017){Hallinan}, {Corsi}, {Mooley}, {Hotokezaka},
  {Nakar}, {Kasliwal}, {Kaplan}, {Frail}, {Myers}, {Murphy}, {De}, {Dobie},
  {Allison}, {Bannister}, {Bhalerao}, {Chandra}, {Clarke}, {Giacintucci}, {Ho},
  {Horesh}, {Kassim}, {Kulkarni}, {Lenc}, {Lockman}, {Lynch}, {Nichols},
  {Nissanke}, {Palliyaguru}, {Peters}, {Piran}, {Rana}, {Sadler}, \&
  {Singer}}]{Hallinan17}
{Hallinan}, G., {Corsi}, A., {Mooley}, K.~P., {et~al.} 2017, Science, 358,
  1579, \dodoi{10.1126/science.aap9855}

\bibitem[{{Hernquist} \& {Quinn}(1988)}]{Hernquist88}
{Hernquist}, L., \& {Quinn}, P.~J. 1988, \apj, 331, 682, \dodoi{10.1086/166592}

\bibitem[{{Hoffman} {et~al.}(2021){Hoffman}, {Hack}, {Avila}, {Martlin},
  {Bajaj}, \& {Cohen}}]{drizzlepac}
{Hoffman}, S., {Hack}, W., {Avila}, R.~J., {et~al.} 2021, {The DrizzlePac
  Handbook} (Baltimore: STScI)

\bibitem[{{Jiang} {et~al.}(2021){Jiang}, {Wang}, {Hu}, {Sun}, {Dou}, \&
  {Xiao}}]{Jiang21}
{Jiang}, N., {Wang}, T., {Hu}, X., {et~al.} 2021, \apj, 911, 31,
  \dodoi{10.3847/1538-4357/abe772}

\bibitem[{{Johnson} {et~al.}(2021){Johnson}, {Leja}, {Conroy}, \&
  {Speagle}}]{Johnson21}
{Johnson}, B.~D., {Leja}, J., {Conroy}, C., \& {Speagle}, J.~S. 2021, \apjs,
  254, 22, \dodoi{10.3847/1538-4365/abef67}

\bibitem[{{Kasen} {et~al.}(2013){Kasen}, {Badnell}, \& {Barnes}}]{Kasen13}
{Kasen}, D., {Badnell}, N.~R., \& {Barnes}, J. 2013, \apj, 774, 25,
  \dodoi{10.1088/0004-637X/774/1/25}

\bibitem[{{Kasen} {et~al.}(2017){Kasen}, {Metzger}, {Barnes}, {Quataert}, \&
  {Ramirez-Ruiz}}]{Kasen17}
{Kasen}, D., {Metzger}, B., {Barnes}, J., {Quataert}, E., \& {Ramirez-Ruiz}, E.
  2017, \nat, 551, 80, \dodoi{10.1038/nature24453}

\bibitem[{{Kasliwal} {et~al.}(2017){Kasliwal}, {Nakar}, {Singer}, {Kaplan},
  {Cook}, {Van Sistine}, {Lau}, {Fremling}, {Gottlieb}, {Jencson}, {Adams},
  {Feindt}, {Hotokezaka}, {Ghosh}, {Perley}, {Yu}, {Piran}, {Allison},
  {Anupama}, {Balasubramanian}, {Bannister}, {Bally}, {Barnes}, {Barway},
  {Bellm}, {Bhalerao}, {Bhattacharya}, {Blagorodnova}, {Bloom}, {Brady},
  {Cannella}, {Chatterjee}, {Cenko}, {Cobb}, {Copperwheat}, {Corsi}, {De},
  {Dobie}, {Emery}, {Evans}, {Fox}, {Frail}, {Frohmaier}, {Goobar}, {Hallinan},
  {Harrison}, {Helou}, {Hinderer}, {Ho}, {Horesh}, {Ip}, {Itoh}, {Kasen},
  {Kim}, {Kuin}, {Kupfer}, {Lynch}, {Madsen}, {Mazzali}, {Miller}, {Mooley},
  {Murphy}, {Ngeow}, {Nichols}, {Nissanke}, {Nugent}, {Ofek}, {Qi}, {Quimby},
  {Rosswog}, {Rusu}, {Sadler}, {Schmidt}, {Sollerman}, {Steele}, {Williamson},
  {Xu}, {Yan}, {Yatsu}, {Zhang}, \& {Zhao}}]{Kasliwal17}
{Kasliwal}, M.~M., {Nakar}, E., {Singer}, L.~P., {et~al.} 2017, Science, 358,
  1559, \dodoi{10.1126/science.aap9455}

\bibitem[{{Kilpatrick} {et~al.}(2017){Kilpatrick}, {Foley}, {Kasen},
  {Murguia-Berthier}, {Ramirez-Ruiz}, {Coulter}, {Drout}, {Piro}, {Shappee},
  {Boutsia}, {Contreras}, {Di Mille}, {Madore}, {Morrell}, {Pan}, {Prochaska},
  {Rest}, {Rojas-Bravo}, {Siebert}, {Simon}, \& {Ulloa}}]{Kilpatrick17}
{Kilpatrick}, C.~D., {Foley}, R.~J., {Kasen}, D., {et~al.} 2017, Science, 358,
  1583, \dodoi{10.1126/science.aaq0073}

\bibitem[{{Kilpatrick} {et~al.}(2021{\natexlab{a}}){Kilpatrick}, {Drout},
  {Auchettl}, {Dimitriadis}, {Foley}, {Jones}, {DeMarchi}, {French}, {Gall},
  {Hjorth}, {Jacobson-Gal{\'a}n}, {Margutti}, {Piro}, {Ramirez-Ruiz}, {Rest},
  \& {Rojas-Bravo}}]{Kilpatrick21}
{Kilpatrick}, C.~D., {Drout}, M.~R., {Auchettl}, K., {et~al.}
  2021{\natexlab{a}}, \mnras, 504, 2073, \dodoi{10.1093/mnras/stab838}

\bibitem[{{Kilpatrick} {et~al.}(2021{\natexlab{b}}){Kilpatrick}, {Fong},
  {Hajela}, {Alexander}, {Berger}, {Blanchard}, {Chornock}, {Margutti},
  {Paterson}, \& {Rastinejad}}]{Kilpatrick21a}
{Kilpatrick}, C.~D., {Fong}, W., {Hajela}, A., {et~al.} 2021{\natexlab{b}}, GRB
  Coordinates Network, 29263, 1

\bibitem[{{Kimble} {et~al.}(2008){Kimble}, {MacKenty}, {O'Connell}, \&
  {Townsend}}]{WFC3}
{Kimble}, R.~A., {MacKenty}, J.~W., {O'Connell}, R.~W., \& {Townsend}, J.~A.
  2008, in Society of Photo-Optical Instrumentation Engineers (SPIE) Conference
  Series, Vol. 7010, Space Telescopes and Instrumentation 2008: Optical,
  Infrared, and Millimeter, ed. J.~{Oschmann}, Jacobus~M., M.~W.~M. {de
  Graauw}, \& H.~A. {MacEwen}, 70101E, \dodoi{10.1117/12.789581}

\bibitem[{{Kriek} \& {Conroy}(2013)}]{Kriek13}
{Kriek}, M., \& {Conroy}, C. 2013, \apjl, 775, L16,
  \dodoi{10.1088/2041-8205/775/1/L16}

\bibitem[{{Lamb} {et~al.}(2019){Lamb}, {Lyman}, {Levan}, {Tanvir}, {Kangas},
  {Fruchter}, {Gompertz}, {Hjorth}, {Mandel}, {Oates}, {Steeghs}, \&
  {Wiersema}}]{Lamb19}
{Lamb}, G.~P., {Lyman}, J.~D., {Levan}, A.~J., {et~al.} 2019, \apjl, 870, L15,
  \dodoi{10.3847/2041-8213/aaf96b}

\bibitem[{{Larsen} {et~al.}(2001){Larsen}, {Brodie}, {Huchra}, {Forbes}, \&
  {Grillmair}}]{Larsen01}
{Larsen}, S.~S., {Brodie}, J.~P., {Huchra}, J.~P., {Forbes}, D.~A., \&
  {Grillmair}, C.~J. 2001, \aj, 121, 2974, \dodoi{10.1086/321081}

\bibitem[{{Lattimer} \& {Schramm}(1976)}]{Lattimer76}
{Lattimer}, J.~M., \& {Schramm}, D.~N. 1976, \apj, 210, 549,
  \dodoi{10.1086/154860}

\bibitem[{{Lazzati} {et~al.}(2018){Lazzati}, {Perna}, {Morsony},
  {Lopez-Camara}, {Cantiello}, {Ciolfi}, {Giacomazzo}, \&
  {Workman}}]{Lazzati18}
{Lazzati}, D., {Perna}, R., {Morsony}, B.~J., {et~al.} 2018, \prl, 120, 241103,
  \dodoi{10.1103/PhysRevLett.120.241103}

\bibitem[{{Lee} {et~al.}(2018){Lee}, {Kang}, \& {Im}}]{Lee18}
{Lee}, M.~G., {Kang}, J., \& {Im}, M. 2018, \apjl, 859, L6,
  \dodoi{10.3847/2041-8213/aac2e9}

\bibitem[{{Leja} {et~al.}(2019){Leja}, {Carnall}, {Johnson}, {Conroy}, \&
  {Speagle}}]{Leja19a}
{Leja}, J., {Carnall}, A.~C., {Johnson}, B.~D., {Conroy}, C., \& {Speagle},
  J.~S. 2019, \apj, 876, 3, \dodoi{10.3847/1538-4357/ab133c}

\bibitem[{{Leja} {et~al.}(2017){Leja}, {Johnson}, {Conroy}, {van Dokkum}, \&
  {Byler}}]{Leja17}
{Leja}, J., {Johnson}, B.~D., {Conroy}, C., {van Dokkum}, P.~G., \& {Byler}, N.
  2017, \apj, 837, 170, \dodoi{10.3847/1538-4357/aa5ffe}

\bibitem[{{Levan} {et~al.}(2017){Levan}, {Lyman}, {Tanvir}, {Hjorth}, {Mandel},
  {Stanway}, {Steeghs}, {Fruchter}, {Troja}, {Schr{\o}der}, {Wiersema},
  {Bruun}, {Cano}, {Cenko}, {de Ugarte Postigo}, {Evans}, {Fairhurst}, {Fox},
  {Fynbo}, {Gompertz}, {Greiner}, {Im}, {Izzo}, {Jakobsson}, {Kangas},
  {Khandrika}, {Lien}, {Malesani}, {O'Brien}, {Osborne}, {Palazzi}, {Pian},
  {Perley}, {Rosswog}, {Ryan}, {Schulze}, {Sutton}, {Th{\"o}ne}, {Watson}, \&
  {Wijers}}]{Levan17}
{Levan}, A.~J., {Lyman}, J.~D., {Tanvir}, N.~R., {et~al.} 2017, \apjl, 848,
  L28, \dodoi{10.3847/2041-8213/aa905f}

\bibitem[{{Li} \& {Paczy{\'n}ski}(1998)}]{Li98}
{Li}, L.-X., \& {Paczy{\'n}ski}, B. 1998, \apjl, 507, L59,
  \dodoi{10.1086/311680}

\bibitem[{{Li} {et~al.}(2015){Li}, {de Grijs}, {Anders}, \& {Li}}]{Li15}
{Li}, S., {de Grijs}, R., {Anders}, P., \& {Li}, C. 2015, \apjs, 216, 6,
  \dodoi{10.1088/0067-0049/216/1/6}

\bibitem[{{LIGO Scientific Collaboration} {et~al.}(2015){LIGO Scientific
  Collaboration}, {Aasi}, {Abbott}, {Abbott}, {Abbott}, {Abernathy}, {Ackley},
  {Adams}, {Adams}, {Addesso}, {Adhikari}, {Adya}, {Affeldt}, {Aggarwal},
  {Aguiar}, {Ain}, {Ajith}, {Alemic}, {Allen}, {Amariutei}, {Anderson},
  {Anderson}, {Arai}, {Araya}, {Arceneaux}, {Areeda}, {Ashton}, {Ast}, {Aston},
  {Aufmuth}, {Aulbert}, {Aylott}, {Babak}, {Baker}, {Ballmer}, {Barayoga},
  {Barbet}, {Barclay}, {Barish}, {Barker}, {Barr}, {Barsotti}, {Bartlett},
  {Barton}, {Bartos}, {Bassiri}, {Batch}, {Baune}, {Behnke}, {Bell}, {Bell},
  {Benacquista}, {Bergman}, {Bergmann}, {Berry}, {Betzwieser}, {Bhagwat},
  {Bhandare}, {Bilenko}, {Billingsley}, {Birch}, {Biscans}, {Biwer},
  {Blackburn}, {Blackburn}, {Blair}, {Blair}, {Bock}, {Bodiya}, {Bojtos},
  {Bond}, {Bork}, {Born}, {Bose}, {Brady}, {Braginsky}, {Brau}, {Bridges},
  {Brinkmann}, {Brooks}, {Brown}, {Brown}, {Brown}, {Buchman}, {Buikema},
  {Buonanno}, {Cadonati}, {Calder{\'o}n Bustillo}, {Camp}, {Cannon}, {Cao},
  {Capano}, {Caride}, {Caudill}, {Cavagli{\`a}}, {Cepeda}, {Chakraborty},
  {Chalermsongsak}, {Chamberlin}, {Chao}, {Charlton}, {Chen}, {Cho}, {Cho},
  {Chow}, {Christensen}, {Chu}, {Chung}, {Ciani}, {Clara}, {Clark}, {Collette},
  {Cominsky}, {Constancio}, {Cook}, {Corbitt}, {Cornish}, {Corsi}, {Costa},
  {Coughlin}, {Countryman}, {Couvares}, {Coward}, {Cowart}, {Coyne}, {Coyne},
  {Craig}, {Creighton}, {Creighton}, {Cripe}, {Crowder}, {Cumming},
  {Cunningham}, {Cutler}, {Dahl}, {Dal Canton}, {Damjanic}, {Danilishin},
  {Danzmann}, {Dartez}, {Dave}, {Daveloza}, {Davies}, {Daw}, {DeBra}, {Del
  Pozzo}, {Denker}, {Dent}, {Dergachev}, {DeRosa}, {DeSalvo}, {Dhurandhar},
  {D́{\i}az}, {Di Palma}, {Dojcinoski}, {Dominguez}, {Donovan}, {Dooley},
  {Doravari}, {Douglas}, {Downes}, {Driggers}, {Du}, {Dwyer}, {Eberle}, {Edo},
  {Edwards}, {Edwards}, {Effler}, {Eggenstein}, {Ehrens}, {Eichholz},
  {Eikenberry}, {Essick}, {Etzel}, {Evans}, {Evans}, {Factourovich},
  {Fairhurst}, {Fan}, {Fang}, {Farr}, {Farr}, {Favata}, {Fays}, {Fehrmann},
  {Fejer}, {Feldbaum}, {Ferreira}, {Fisher}, {Frei}, {Freise}, {Frey},
  {Fricke}, {Fritschel}, {Frolov}, {Fuentes-Tapia}, {Fulda}, {Fyffe}, {Gair},
  {Gaonkar}, {Gehrels}, {Gergely}, {Giaime}, {Giardina}, {Gleason}, {Goetz},
  {Goetz}, {Gondan}, {Gonz{\'a}lez}, {Gordon}, {Gorodetsky}, {Gossan},
  {Go{\ss}ler}, {Gr{\"a}f}, {Graff}, {Grant}, {Gras}, {Gray}, {Greenhalgh},
  {Gretarsson}, {Grote}, {Grunewald}, {Guido}, {Guo}, {Gushwa}, {Gustafson},
  {Gustafson}, {Hacker}, {Hall}, {Hammond}, {Hanke}, {Hanks}, {Hanna},
  {Hannam}, {Hanson}, {Hardwick}, {Harry}, {Harry}, {Hart}, {Hartman},
  {Haster}, {Haughian}, {Hee}, {Heintze}, {Heinzel}, {Hendry}, {Heng},
  {Heptonstall}, {Heurs}, {Hewitson}, {Hild}, {Hoak}, {Hodge}, {Hollitt},
  {Holt}, {Hopkins}, {Hosken}, {Hough}, {Houston}, {Howell}, {Hu}, {Huerta},
  {Hughey}, {Husa}, {Huttner}, {Huynh}, {Huynh-Dinh}, {Idrisy}, {Indik},
  {Ingram}, {Inta}, {Islas}, {Isler}, {Isogai}, {Iyer}, {Izumi}, {Jacobson},
  {Jang}, {Jawahar}, {Ji}, {Jim{\'e}nez-Forteza}, {Johnson}, {Jones}, {Jones},
  {Ju}, {Haris}, {Kalogera}, {Kandhasamy}, {Kang}, {Kanner}, {Katsavounidis},
  {Katzman}, {Kaufer}, {Kaufer}, {Kaur}, {Kawabe}, {Kawazoe}, {Keiser},
  {Keitel}, {Kelley}, {Kells}, {Keppel}, {Key}, {Khalaidovski}, {Khalili},
  {Khazanov}, {Kim}, {Kim}, {Kim}, {Kim}, {Kim}, {King}, {King}, {Kinzel},
  {Kissel}, {Klimenko}, {Kline}, {Koehlenbeck}, {Kokeyama}, {Kondrashov},
  {Korobko}, {Korth}, {Kozak}, {Kringel}, {Krishnan}, {Krueger}, {Kuehn},
  {Kumar}, {Kumar}, {Kuo}, {Landry}, {Lantz}, {Larson}, {Lasky}, {Lazzarini},
  {Lazzaro}, {Le}, {Leaci}, {Leavey}, {Lebigot}, {Lee}, {Lee}, {Lee}, {Leong},
  {Levin}, {Levine}, {Lewis}, {Li}, {Libbrecht}, {Libson}, {Lin}, {Littenberg},
  {Lockerbie}, {Lockett}, {Logue}, {Lombardi}, {Lormand}, {Lough}, {Lubinski},
  {L{\"u}ck}, {Lundgren}, {Lynch}, {Ma}, {Macarthur}, {MacDonald},
  {Machenschalk}, {MacInnis}, {Macleod}, {Maga{\~n}a-Sandoval}, {Magee},
  {Mageswaran}, {Maglione}, {Mailand}, {Mandel}, {Mandic}, {Mangano},
  {Mansell}, {M{\'a}rka}, {M{\'a}rka}, {Markosyan}, {Maros}, {Martin},
  {Martin}, {Martynov}, {Marx}, {Mason}, {Massinger}, {Matichard}, {Matone},
  {Mavalvala}, {Mazumder}, {Mazzolo}, {McCarthy}, {McClelland}, {McCormick},
  {McGuire}, {McIntyre}, {McIver}, {McLin}, {McWilliams}, {Meadors},
  {Meinders}, {Melatos}, {Mendell}, {Mercer}, {Meshkov}, {Messenger}, {Meyers},
  {Miao}, {Middleton}, {Mikhailov}, {Miller}, {Miller}, {Millhouse}, {Ming},
  {Mirshekari}, {Mishra}, {Mitra}, {Mitrofanov}, {Mitselmakher}, {Mittleman},
  {Moe}, {Mohanty}, {Mohapatra}, {Moore}, {Moraru}, {Moreno}, {Morriss},
  {Mossavi}, {Mow-Lowry}, {Mueller}, {Mueller}, {Mukherjee}, {Mullavey},
  {Munch}, {Murphy}, {Murray}, {Mytidis}, {Nash}, {Nayak}, {Necula}, {Nedkova},
  {Newton}, {Nguyen}, {Nielsen}, {Nissanke}, {Nitz}, {Nolting}, {Normandin},
  {Nuttall}, {Ochsner}, {O'Dell}, {Oelker}, {Ogin}, {Oh}, {Oh}, {Ohme},
  {Oppermann}, {Oram}, {O'Reilly}, {Ortega}, {O'Shaughnessy}, {Osthelder},
  {Ott}, {Ottaway}, {Ottens}, {Overmier}, {Owen}, {Padilla}, {Pai}, {Pai},
  {Palashov}, {Pal-Singh}, {Pan}, {Pankow}, {Pannarale}, {Pant}, {Papa},
  {Paris}, {Patrick}, {Pedraza}, {Pekowsky}, {Pele}, {Penn}, {Perreca},
  {Phelps}, {Pierro}, {Pinto}, {Pitkin}, {Poeld}, {Post}, {Poteomkin},
  {Powell}, {Prasad}, {Predoi}, {Premachandra}, {Prestegard}, {Price},
  {Principe}, {Privitera}, {Prix}, {Prokhorov}, {Puncken}, {P{\"u}rrer}, {Qin},
  {Quetschke}, {Quintero}, {Quiroga}, {Quitzow-James}, {Raab}, {Rabeling},
  {Radkins}, {Raffai}, {Raja}, {Rajalakshmi}, {Rakhmanov}, {Ramirez},
  {Raymond}, {Reed}, {Reid}, {Reitze}, {Reula}, {Riles}, {Robertson}, {Robie},
  {Rollins}, {Roma}, {Romano}, {Romanov}, {Romie}, {Rowan}, {R{\"u}diger},
  {Ryan}, {Sachdev}, {Sadecki}, {Sadeghian}, {Saleem}, {Salemi}, {Sammut},
  {Sandberg}, {Sanders}, {Sannibale}, {Santiago-Prieto}, {Sathyaprakash},
  {Saulson}, {Savage}, {Sawadsky}, {Scheuer}, {Schilling}, {Schmidt},
  {Schnabel}, {Schofield}, {Schreiber}, {Schuette}, {Schutz}, {Scott}, {Scott},
  {Sellers}, {Sengupta}, {Sergeev}, {Serna}, {Sevigny}, {Shaddock}, {Shahriar},
  {Shaltev}, {Shao}, {Shapiro}, {Shawhan}, {Shoemaker}, {Sidery}, {Siemens},
  {Sigg}, {Silva}, {Simakov}, {Singer}, {Singer}, {Singh}, {Sintes},
  {Slagmolen}, {Smith}, {Smith}, {Smith}, {Smith-Lefebvre}, {Son}, {Sorazu},
  {Souradeep}, {Staley}, {Stebbins}, {Steinke}, {Steinlechner}, {Steinlechner},
  {Steinmeyer}, {Stephens}, {Steplewski}, {Stevenson}, {Stone}, {Strain},
  {Strigin}, {Sturani}, {Stuver}, {Summerscales}, {Sutton}, {Szczepanczyk},
  {Szeifert}, {Talukder}, {Tanner}, {T{\'a}pai}, {Tarabrin}, {Taracchini},
  {Taylor}, {Tellez}, {Theeg}, {Thirugnanasambandam}, {Thomas}, {Thomas},
  {Thorne}, {Thorne}, {Thrane}, {Tiwari}, {Tomlinson}, {Torres}, {Torrie},
  {Traylor}, {Tse}, {Tshilumba}, {Ugolini}, {Unnikrishnan}, {Urban}, {Usman},
  {Vahlbruch}, {Vajente}, {Valdes}, {Vallisneri}, {van Veggel}, {Vass},
  {Vaulin}, {Vecchio}, {Veitch}, {Veitch}, {Venkateswara}, {Vincent-Finley},
  {Vitale}, {Vo}, {Vorvick}, {Vousden}, {Vyatchanin}, {Wade}, {Wade}, {Wade},
  {Walker}, {Wallace}, {Walsh}, {Wang}, {Wang}, {Wang}, {Ward}, {Warner},
  {Was}, {Weaver}, {Weinert}, {Weinstein}, {Weiss}, {Welborn}, {Wen},
  {Wessels}, {Westphal}, {Wette}, {Whelan}, {Whitcomb}, {White}, {Whiting},
  {Wilkinson}, {Williams}, {Williams}, {Williamson}, {Willis}, {Willke},
  {Wimmer}, {Winkler}, {Wipf}, {Wittel}, {Woan}, {Worden}, {Xie}, {Yablon},
  {Yakushin}, {Yam}, {Yamamoto}, {Yancey}, {Yang}, {Zanolin}, {Zhang}, {Zhang},
  {Zhang}, {Zhang}, {Zhao}, {Zhou}, {Zhu}, {Zucker}, {Zuraw}, \&
  {Zweizig}}]{LIGO15}
{LIGO Scientific Collaboration}, {Aasi}, J., {Abbott}, B.~P., {et~al.} 2015,
  Classical and Quantum Gravity, 32, 074001,
  \dodoi{10.1088/0264-9381/32/7/074001}

\bibitem[{{Lindegren} {et~al.}(2018){Lindegren}, {Hern{\'a}ndez}, {Bombrun},
  {Klioner}, {Bastian}, {Ramos-Lerate}, {de Torres}, {Steidelm{\"u}ller},
  {Stephenson}, {Hobbs}, {Lammers}, {Biermann}, {Geyer}, {Hilger}, {Michalik},
  {Stampa}, {McMillan}, {Casta{\~n}eda}, {Clotet}, {Comoretto}, {Davidson},
  {Fabricius}, {Gracia}, {Hambly}, {Hutton}, {Mora}, {Portell}, {van Leeuwen},
  {Abbas}, {Abreu}, {Altmann}, {Andrei}, {Anglada}, {Balaguer-N{\'u}{\~n}ez},
  {Barache}, {Becciani}, {Bertone}, {Bianchi}, {Bouquillon}, {Bourda},
  {Br{\"u}semeister}, {Bucciarelli}, {Busonero}, {Buzzi}, {Cancelliere},
  {Carlucci}, {Charlot}, {Cheek}, {Crosta}, {Crowley}, {de Bruijne}, {de
  Felice}, {Drimmel}, {Esquej}, {Fienga}, {Fraile}, {Gai}, {Garralda},
  {Gonz{\'a}lez-Vidal}, {Guerra}, {Hauser}, {Hofmann}, {Holl}, {Jordan},
  {Lattanzi}, {Lenhardt}, {Liao}, {Licata}, {Lister}, {L{\"o}ffler},
  {Marchant}, {Martin-Fleitas}, {Messineo}, {Mignard}, {Morbidelli}, {Poggio},
  {Riva}, {Rowell}, {Salguero}, {Sarasso}, {Sciacca}, {Siddiqui}, {Smart},
  {Spagna}, {Steele}, {Taris}, {Torra}, {van Elteren}, {van Reeven}, \&
  {Vecchiato}}]{Lindegren18}
{Lindegren}, L., {Hern{\'a}ndez}, J., {Bombrun}, A., {et~al.} 2018, \aap, 616,
  A2, \dodoi{10.1051/0004-6361/201832727}

\bibitem[{{Lippuner} {et~al.}(2017){Lippuner}, {Fern{\'a}ndez}, {Roberts},
  {Foucart}, {Kasen}, {Metzger}, \& {Ott}}]{Lippuner17}
{Lippuner}, J., {Fern{\'a}ndez}, R., {Roberts}, L.~F., {et~al.} 2017, \mnras,
  472, 904, \dodoi{10.1093/mnras/stx1987}

\bibitem[{{Lipunov} {et~al.}(2017){Lipunov}, {Gorbovskoy}, {Kornilov}, {.
  Tyurina}, {Balanutsa}, {Kuznetsov}, {Vlasenko}, {Kuvshinov}, {Gorbunov},
  {Buckley}, {Krylov}, {Podesta}, {Lopez}, {Podesta}, {Levato}, {Saffe},
  {Mallamachi}, {Potter}, {Budnev}, {Gress}, {Ishmuhametova}, {Vladimirov},
  {Zimnukhov}, {Yurkov}, {Sergienko}, {Gabovich}, {Rebolo}, {Serra-Ricart},
  {Israelyan}, {Chazov}, {Wang}, {Tlatov}, \& {Panchenko}}]{Lipunov17}
{Lipunov}, V.~M., {Gorbovskoy}, E., {Kornilov}, V.~G., {et~al.} 2017, \apjl,
  850, L1, \dodoi{10.3847/2041-8213/aa92c0}

\bibitem[{{Lu} {et~al.}(2021){Lu}, {McKee}, \& {Mooley}}]{Lu21}
{Lu}, W., {McKee}, C.~F., \& {Mooley}, K.~P. 2021, \mnras,
  \dodoi{10.1093/mnras/stab2388}

\bibitem[{{Lyman} {et~al.}(2018){Lyman}, {Lamb}, {Levan}, {Mandel}, {Tanvir},
  {Kobayashi}, {Gompertz}, {Hjorth}, {Fruchter}, {Kangas}, {Steeghs}, {Steele},
  {Cano}, {Copperwheat}, {Evans}, {Fynbo}, {Gall}, {Im}, {Izzo}, {Jakobsson},
  {Milvang-Jensen}, {O'Brien}, {Osborne}, {Palazzi}, {Perley}, {Pian},
  {Rosswog}, {Rowlinson}, {Schulze}, {Stanway}, {Sutton}, {Th{\"o}ne}, {de
  Ugarte Postigo}, {Watson}, {Wiersema}, \& {Wijers}}]{Lyman18}
{Lyman}, J.~D., {Lamb}, G.~P., {Levan}, A.~J., {et~al.} 2018, Nature Astronomy,
  2, 751, \dodoi{10.1038/s41550-018-0511-3}

\bibitem[{{Malin} \& {Carter}(1983)}]{MalinCarter83}
{Malin}, D.~F., \& {Carter}, D. 1983, \apj, 274, 534, \dodoi{10.1086/161467}

\bibitem[{{Margutti} \& {Chornock}(2020)}]{Margutti20}
{Margutti}, R., \& {Chornock}, R. 2020, arXiv e-prints, arXiv:2012.04810.
\newblock \doarXiv{2012.04810}

\bibitem[{{Margutti} {et~al.}(2017){Margutti}, {Berger}, {Fong}, {Guidorzi},
  {Alexander}, {Metzger}, {Blanchard}, {Cowperthwaite}, {Chornock},
  {Eftekhari}, {Nicholl}, {Villar}, {Williams}, {Annis}, {Brown}, {Chen},
  {Doctor}, {Frieman}, {Holz}, {Sako}, \& {Soares-Santos}}]{Margutti17}
{Margutti}, R., {Berger}, E., {Fong}, W., {et~al.} 2017, \apjl, 848, L20,
  \dodoi{10.3847/2041-8213/aa9057}

\bibitem[{{Margutti} {et~al.}(2018){Margutti}, {Alexander}, {Xie}, {Sironi},
  {Metzger}, {Kathirgamaraju}, {Fong}, {Blanchard}, {Berger}, {MacFadyen},
  {Giannios}, {Guidorzi}, {Hajela}, {Chornock}, {Cowperthwaite}, {Eftekhari},
  {Nicholl}, {Villar}, {Williams}, \& {Zrake}}]{Margutti18}
{Margutti}, R., {Alexander}, K.~D., {Xie}, X., {et~al.} 2018, \apjl, 856, L18,
  \dodoi{10.3847/2041-8213/aab2ad}

\bibitem[{{McCully} {et~al.}(2017){McCully}, {Hiramatsu}, {Howell},
  {Hosseinzadeh}, {Arcavi}, {Kasen}, {Barnes}, {Shara}, {Williams},
  {V{\"a}is{\"a}nen}, {Potter}, {Romero-Colmenero}, {Crawford}, {Buckley},
  {Cooke}, {Andreoni}, {Pritchard}, {Mao}, {Gromadzki}, \& {Burke}}]{McCully17}
{McCully}, C., {Hiramatsu}, D., {Howell}, D.~A., {et~al.} 2017, \apjl, 848,
  L32, \dodoi{10.3847/2041-8213/aa9111}

\bibitem[{{McMaster} {et~al.}(2008){McMaster}, {Biretta}, {Brammer}, {Burrows},
  {Casertano}, {Chiaberge}, {Clampin}, {Dixon}, \& {et al.}}]{McMaster08}
{McMaster}, M., {Biretta}, J., {Brammer}, G., {et~al.} 2008, {Wide Field and
  Planetary Camera 2 Instrument Handbook v. 10.0} (Baltimore: STScI)

\bibitem[{{Metzger} {et~al.}(2017){Metzger}, {Berger}, \&
  {Margalit}}]{Metzger17}
{Metzger}, B.~D., {Berger}, E., \& {Margalit}, B. 2017, \apj, 841, 14,
  \dodoi{10.3847/1538-4357/aa633d}

\bibitem[{{Metzger} {et~al.}(2010){Metzger}, {Mart{\'\i}nez-Pinedo}, {Darbha},
  {Quataert}, {Arcones}, {Kasen}, {Thomas}, {Nugent}, {Panov}, \&
  {Zinner}}]{Metzger10}
{Metzger}, B.~D., {Mart{\'\i}nez-Pinedo}, G., {Darbha}, S., {et~al.} 2010,
  \mnras, 406, 2650, \dodoi{10.1111/j.1365-2966.2010.16864.x}

\bibitem[{{Miller} {et~al.}(2019){Miller}, {Ryan}, {Dolence}, {Burrows},
  {Fontes}, {Fryer}, {Korobkin}, {Lippuner}, {Mumpower}, \&
  {Wollaeger}}]{Miller19}
{Miller}, J.~M., {Ryan}, B.~R., {Dolence}, J.~C., {et~al.} 2019, \prd, 100,
  023008, \dodoi{10.1103/PhysRevD.100.023008}

\bibitem[{{Mooley} {et~al.}(2018){Mooley}, {Deller}, {Gottlieb}, {Nakar},
  {Hallinan}, {Bourke}, {Frail}, {Horesh}, {Corsi}, \& {Hotokezaka}}]{Mooley18}
{Mooley}, K.~P., {Deller}, A.~T., {Gottlieb}, O., {et~al.} 2018, \nat, 561,
  355, \dodoi{10.1038/s41586-018-0486-3}

\bibitem[{{Nicholl} {et~al.}(2017){Nicholl}, {Berger}, {Kasen}, {Metzger},
  {Elias}, {Brice{\~n}o}, {Alexander}, {Blanchard}, {Chornock},
  {Cowperthwaite}, {Eftekhari}, {Fong}, {Margutti}, {Villar}, {Williams},
  {Brown}, {Annis}, {Bahramian}, {Brout}, {Brown}, {Chen}, {Clemens},
  {Dennihy}, {Dunlap}, {Holz}, {Marchesini}, {Massaro}, {Moskowitz},
  {Pelisoli}, {Rest}, {Ricci}, {Sako}, {Soares-Santos}, \&
  {Strader}}]{Nicholl17}
{Nicholl}, M., {Berger}, E., {Kasen}, D., {et~al.} 2017, \apjl, 848, L18,
  \dodoi{10.3847/2041-8213/aa9029}

\bibitem[{{Noeske} {et~al.}(2007){Noeske}, {Weiner}, {Faber}, {Papovich},
  {Koo}, {Somerville}, {Bundy}, {Conselice}, {Newman}, {Schiminovich}, {Le
  Floc'h}, {Coil}, {Rieke}, {Lotz}, {Primack}, {Barmby}, {Cooper}, {Davis},
  {Ellis}, {Fazio}, {Guhathakurta}, {Huang}, {Kassin}, {Martin}, {Phillips},
  {Rich}, {Small}, {Willmer}, \& {Wilson}}]{Noeske07}
{Noeske}, K.~G., {Weiner}, B.~J., {Faber}, S.~M., {et~al.} 2007, \apjl, 660,
  L43, \dodoi{10.1086/517926}

\bibitem[{{Oke} \& {Gunn}(1983)}]{Oke83}
{Oke}, J.~B., \& {Gunn}, J.~E. 1983, \apj, 266, 713, \dodoi{10.1086/160817}

\bibitem[{{Palmese} {et~al.}(2017){Palmese}, {Hartley}, {Tarsitano},
  {Conselice}, {Lahav}, {Allam}, {Annis}, {Lin}, {Soares-Santos}, {Tucker},
  {Brout}, {Banerji}, {Bechtol}, {Diehl}, {Fruchter}, {Garc{\'\i}a-Bellido},
  {Herner}, {Levan}, {Li}, {Lidman}, {Misra}, {Sako}, {Scolnic}, {Smith},
  {Abbott}, {Abdalla}, {Benoit-L{\'e}vy}, {Bertin}, {Brooks}, {Buckley-Geer},
  {Carnero Rosell}, {Carrasco Kind}, {Carretero}, {Castander}, {Cunha},
  {D'Andrea}, {da Costa}, {Davis}, {DePoy}, {Desai}, {Dietrich}, {Doel},
  {Drlica-Wagner}, {Eifler}, {Evrard}, {Flaugher}, {Fosalba}, {Frieman},
  {Gaztanaga}, {Gerdes}, {Giannantonio}, {Gruen}, {Gruendl}, {Gschwend},
  {Gutierrez}, {Honscheid}, {Jain}, {James}, {Jeltema}, {Johnson}, {Johnson},
  {Krause}, {Kron}, {Kuehn}, {Kuhlmann}, {Kuropatkin}, {Lima}, {Maia}, {March},
  {Marshall}, {McMahon}, {Menanteau}, {Miller}, {Miquel}, {Neilsen}, {Ogando},
  {Plazas}, {Reil}, {Romer}, {Sanchez}, {Schindler}, {Smith}, {Sobreira},
  {Suchyta}, {Swanson}, {Tarle}, {Thomas}, {Thomas}, {Walker}, {Weller},
  {Zhang}, \& {Zuntz}}]{Palmese17}
{Palmese}, A., {Hartley}, W., {Tarsitano}, F., {et~al.} 2017, \apjl, 849, L34,
  \dodoi{10.3847/2041-8213/aa9660}

\bibitem[{{Pan} {et~al.}(2017){Pan}, {Kilpatrick}, {Simon}, {Xhakaj},
  {Boutsia}, {Coulter}, {Drout}, {Foley}, {Kasen}, {Morrell},
  {Murguia-Berthier}, {Osip}, {Piro}, {Prochaska}, {Ramirez-Ruiz}, {Rest},
  {Rojas-Bravo}, {Shappee}, \& {Siebert}}]{Pan17}
{Pan}, Y.~C., {Kilpatrick}, C.~D., {Simon}, J.~D., {et~al.} 2017, \apjl, 848,
  L30, \dodoi{10.3847/2041-8213/aa9116}

\bibitem[{{Peng} {et~al.}(2010){Peng}, {Ho}, {Impey}, \& {Rix}}]{GALFIT}
{Peng}, C.~Y., {Ho}, L.~C., {Impey}, C.~D., \& {Rix}, H.-W. 2010, \aj, 139,
  2097, \dodoi{10.1088/0004-6256/139/6/2097}

\bibitem[{{Pian} {et~al.}(2017){Pian}, {D'Avanzo}, {Benetti}, {Branchesi},
  {Brocato}, {Campana}, {Cappellaro}, {Covino}, {D'Elia}, {Fynbo}, {Getman},
  {Ghirlanda}, {Ghisellini}, {Grado}, {Greco}, {Hjorth}, {Kouveliotou},
  {Levan}, {Limatola}, {Malesani}, {Mazzali}, {Melandri}, {M{\o}ller},
  {Nicastro}, {Palazzi}, {Piranomonte}, {Rossi}, {Salafia}, {Selsing},
  {Stratta}, {Tanaka}, {Tanvir}, {Tomasella}, {Watson}, {Yang}, {Amati},
  {Antonelli}, {Ascenzi}, {Bernardini}, {Bo{\"e}r}, {Bufano}, {Bulgarelli},
  {Capaccioli}, {Casella}, {Castro-Tirado}, {Chassande-Mottin}, {Ciolfi},
  {Copperwheat}, {Dadina}, {De Cesare}, {di Paola}, {Fan}, {Gendre},
  {Giuffrida}, {Giunta}, {Hunt}, {Israel}, {Jin}, {Kasliwal}, {Klose}, {Lisi},
  {Longo}, {Maiorano}, {Mapelli}, {Masetti}, {Nava}, {Patricelli}, {Perley},
  {Pescalli}, {Piran}, {Possenti}, {Pulone}, {Razzano}, {Salvaterra},
  {Schipani}, {Spera}, {Stamerra}, {Stella}, {Tagliaferri}, {Testa}, {Troja},
  {Turatto}, {Vergani}, \& {Vergani}}]{Pian17}
{Pian}, E., {D'Avanzo}, P., {Benetti}, S., {et~al.} 2017, \nat, 551, 67,
  \dodoi{10.1038/nature24298}

\bibitem[{{Pooley} {et~al.}(2003){Pooley}, {Lewin}, {Anderson}, {Baumgardt},
  {Filippenko}, {Gaensler}, {Homer}, {Hut}, {Kaspi}, {Makino}, {Margon},
  {McMillan}, {Portegies Zwart}, {van der Klis}, \& {Verbunt}}]{Pooley03}
{Pooley}, D., {Lewin}, W. H.~G., {Anderson}, S.~F., {et~al.} 2003, \apjl, 591,
  L131, \dodoi{10.1086/377074}

\bibitem[{{Pop} {et~al.}(2018){Pop}, {Pillepich}, {Amorisco}, \&
  {Hernquist}}]{Pop17}
{Pop}, A.-R., {Pillepich}, A., {Amorisco}, N.~C., \& {Hernquist}, L. 2018,
  \mnras, 480, 1715, \dodoi{10.1093/mnras/sty1932}

\bibitem[{{Price} {et~al.}(2014){Price}, {Kriek}, {Brammer}, {Conroy},
  {F{\"o}rster Schreiber}, {Franx}, {Fumagalli}, {Lundgren}, {Momcheva},
  {Nelson}, {Skelton}, {van Dokkum}, {Whitaker}, \& {Wuyts}}]{Price14}
{Price}, S.~H., {Kriek}, M., {Brammer}, G.~B., {et~al.} 2014, \apj, 788, 86,
  \dodoi{10.1088/0004-637X/788/1/86}

\bibitem[{{Quinn}(1984)}]{Quinn84}
{Quinn}, P.~J. 1984, \apj, 279, 596, \dodoi{10.1086/161924}

\bibitem[{{Ramirez-Ruiz} {et~al.}(2019){Ramirez-Ruiz}, {Andrews}, \&
  {Schr{\o}der}}]{RamirezRuiz19}
{Ramirez-Ruiz}, E., {Andrews}, J.~J., \& {Schr{\o}der}, S.~L. 2019, \apjl, 883,
  L6, \dodoi{10.3847/2041-8213/ab3f2c}

\bibitem[{{Rodrigo} {et~al.}(2012){Rodrigo}, {Solano}, \& {Bayo}}]{SVO12}
{Rodrigo}, C., {Solano}, E., \& {Bayo}, A. 2012, {SVO Filter Profile Service
  Version 1.0}, IVOA Working Draft 15 October 2012,
  \dodoi{10.5479/ADS/bib/2012ivoa.rept.1015R}

\bibitem[{{Rodriguez} {et~al.}(2016){Rodriguez}, {Chatterjee}, \&
  {Rasio}}]{Rodriguez16}
{Rodriguez}, C.~L., {Chatterjee}, S., \& {Rasio}, F.~A. 2016, \prd, 93, 084029,
  \dodoi{10.1103/PhysRevD.93.084029}

\bibitem[{{Ryan} {et~al.}(2020){Ryan}, {van Eerten}, {Piro}, \&
  {Troja}}]{Ryan20}
{Ryan}, G., {van Eerten}, H., {Piro}, L., \& {Troja}, E. 2020, \apj, 896, 166,
  \dodoi{10.3847/1538-4357/ab93cf}

\bibitem[{{Sanderson} \& {Helmi}(2013)}]{Sanderson13}
{Sanderson}, R.~E., \& {Helmi}, A. 2013, \mnras, 435, 378,
  \dodoi{10.1093/mnras/stt1307}

\bibitem[{{Schlafly} \& {Finkbeiner}(2011)}]{Schlafly11}
{Schlafly}, E.~F., \& {Finkbeiner}, D.~P. 2011, \apj, 737, 103,
  \dodoi{10.1088/0004-637X/737/2/103}

\bibitem[{{Smartt} {et~al.}(2017){Smartt}, {Chen}, {Jerkstrand}, {Coughlin},
  {Kankare}, {Sim}, {Fraser}, {Inserra}, {Maguire}, {Chambers}, {Huber},
  {Kr{\"u}hler}, {Leloudas}, {Magee}, {Shingles}, {Smith}, {Young}, {Tonry},
  {Kotak}, {Gal-Yam}, {Lyman}, {Homan}, {Agliozzo}, {Anderson}, {Angus},
  {Ashall}, {Barbarino}, {Bauer}, {Berton}, {Botticella}, {Bulla}, {Bulger},
  {Cannizzaro}, {Cano}, {Cartier}, {Cikota}, {Clark}, {De Cia}, {Della Valle},
  {Denneau}, {Dennefeld}, {Dessart}, {Dimitriadis}, {Elias-Rosa}, {Firth},
  {Flewelling}, {Fl{\"o}rs}, {Franckowiak}, {Frohmaier}, {Galbany},
  {Gonz{\'a}lez-Gait{\'a}n}, {Greiner}, {Gromadzki}, {Guelbenzu},
  {Guti{\'e}rrez}, {Hamanowicz}, {Hanlon}, {Harmanen}, {Heintz}, {Heinze},
  {Hernandez}, {Hodgkin}, {Hook}, {Izzo}, {James}, {Jonker}, {Kerzendorf},
  {Klose}, {Kostrzewa-Rutkowska}, {Kowalski}, {Kromer}, {Kuncarayakti},
  {Lawrence}, {Lowe}, {Magnier}, {Manulis}, {Martin-Carrillo}, {Mattila},
  {McBrien}, {M{\"u}ller}, {Nordin}, {O'Neill}, {Onori}, {Palmerio},
  {Pastorello}, {Patat}, {Pignata}, {Podsiadlowski}, {Pumo}, {Prentice}, {Rau},
  {Razza}, {Rest}, {Reynolds}, {Roy}, {Ruiter}, {Rybicki}, {Salmon}, {Schady},
  {Schultz}, {Schweyer}, {Seitenzahl}, {Smith}, {Sollerman}, {Stalder},
  {Stubbs}, {Sullivan}, {Szegedi}, {Taddia}, {Taubenberger}, {Terreran}, {van
  Soelen}, {Vos}, {Wainscoat}, {Walton}, {Waters}, {Weiland}, {Willman},
  {Wiseman}, {Wright}, {Wyrzykowski}, \& {Yaron}}]{Smartt17}
{Smartt}, S.~J., {Chen}, T.~W., {Jerkstrand}, A., {et~al.} 2017, \nat, 551, 75,
  \dodoi{10.1038/nature24303}

\bibitem[{{Soares-Santos} {et~al.}(2017){Soares-Santos}, {Holz}, {Annis},
  {Chornock}, {Herner}, {Berger}, {Brout}, {Chen}, {Kessler}, {Sako}, {Allam},
  {Tucker}, {Butler}, {Palmese}, {Doctor}, {Diehl}, {Frieman}, {Yanny}, {Lin},
  {Scolnic}, {Cowperthwaite}, {Neilsen}, {Marriner}, {Kuropatkin}, {Hartley},
  {Paz-Chinch{\'o}n}, {Alexander}, {Balbinot}, {Blanchard}, {Brown}, {Carlin},
  {Conselice}, {Cook}, {Drlica-Wagner}, {Drout}, {Durret}, {Eftekhari}, {Farr},
  {Finley}, {Foley}, {Fong}, {Fryer}, {Garc{\'\i}a-Bellido}, {Gill}, {Gruendl},
  {Hanna}, {Kasen}, {Li}, {Lopes}, {Louren{\c{c}}o}, {Margutti}, {Marshall},
  {Matheson}, {Medina}, {Metzger}, {Mu{\~n}oz}, {Muir}, {Nicholl}, {Quataert},
  {Rest}, {Sauseda}, {Schlegel}, {Secco}, {Sobreira}, {Stebbins}, {Villar},
  {Vivas}, {Walker}, {Wester}, {Williams}, {Zenteno}, {Zhang}, {Abbott},
  {Abdalla}, {Banerji}, {Bechtol}, {Benoit-L{\'e}vy}, {Bertin}, {Brooks},
  {Buckley-Geer}, {Burke}, {Carnero Rosell}, {Carrasco Kind}, {Carretero},
  {Castander}, {Crocce}, {Cunha}, {D'Andrea}, {da Costa}, {Davis}, {Desai},
  {Dietrich}, {Doel}, {Eifler}, {Fernandez}, {Flaugher}, {Fosalba},
  {Gaztanaga}, {Gerdes}, {Giannantonio}, {Goldstein}, {Gruen}, {Gschwend},
  {Gutierrez}, {Honscheid}, {Jain}, {James}, {Jeltema}, {Johnson}, {Johnson},
  {Kent}, {Krause}, {Kron}, {Kuehn}, {Kuhlmann}, {Lahav}, {Lima}, {Maia},
  {March}, {McMahon}, {Menanteau}, {Miquel}, {Mohr}, {Nichol}, {Nord},
  {Ogando}, {Petravick}, {Plazas}, {Romer}, {Roodman}, {Rykoff}, {Sanchez},
  {Scarpine}, {Schubnell}, {Sevilla-Noarbe}, {Smith}, {Smith}, {Suchyta},
  {Swanson}, {Tarle}, {Thomas}, {Thomas}, {Troxel}, {Vikram}, {Wechsler},
  {Weller}, {Dark Energy Survey}, \& {Dark Energy Camera GW-EM
  Collaboration}}]{Soares-Santos17}
{Soares-Santos}, M., {Holz}, D.~E., {Annis}, J., {et~al.} 2017, \apjl, 848,
  L16, \dodoi{10.3847/2041-8213/aa9059}

\bibitem[{{Speagle}(2020)}]{Speagle20}
{Speagle}, J.~S. 2020, \mnras, 493, 3132, \dodoi{10.1093/mnras/staa278}

\bibitem[{{Stetson}(1987)}]{Stetson87}
{Stetson}, P.~B. 1987, \pasp, 99, 191, \dodoi{10.1086/131977}

\bibitem[{{Tanvir} {et~al.}(2017){Tanvir}, {Levan},
  {Gonz{\'a}lez-Fern{\'a}ndez}, {Korobkin}, {Mandel}, {Rosswog}, {Hjorth},
  {D'Avanzo}, {Fruchter}, {Fryer}, {Kangas}, {Milvang-Jensen}, {Rosetti},
  {Steeghs}, {Wollaeger}, {Cano}, {Copperwheat}, {Covino}, {D'Elia}, {de Ugarte
  Postigo}, {Evans}, {Even}, {Fairhurst}, {Figuera Jaimes}, {Fontes}, {Fujii},
  {Fynbo}, {Gompertz}, {Greiner}, {Hodosan}, {Irwin}, {Jakobsson},
  {J{\o}rgensen}, {Kann}, {Lyman}, {Malesani}, {McMahon}, {Melandri},
  {O'Brien}, {Osborne}, {Palazzi}, {Perley}, {Pian}, {Piranomonte}, {Rabus},
  {Rol}, {Rowlinson}, {Schulze}, {Sutton}, {Th{\"o}ne}, {Ulaczyk}, {Watson},
  {Wiersema}, \& {Wijers}}]{Tanvir17}
{Tanvir}, N.~R., {Levan}, A.~J., {Gonz{\'a}lez-Fern{\'a}ndez}, C., {et~al.}
  2017, \apjl, 848, L27, \dodoi{10.3847/2041-8213/aa90b6}

\bibitem[{{Troja} {et~al.}(2017){Troja}, {Piro}, {van Eerten}, {Wollaeger},
  {Im}, {Fox}, {Butler}, {Cenko}, {Sakamoto}, {Fryer}, {Ricci}, {Lien}, {Ryan},
  {Korobkin}, {Lee}, {Burgess}, {Lee}, {Watson}, {Choi}, {Covino}, {D'Avanzo},
  {Fontes}, {Gonz{\'a}lez}, {Khandrika}, {Kim}, {Kim}, {Lee}, {Lee}, {Kutyrev},
  {Lim}, {S{\'a}nchez-Ram{\'\i}rez}, {Veilleux}, {Wieringa}, \&
  {Yoon}}]{Troja17}
{Troja}, E., {Piro}, L., {van Eerten}, H., {et~al.} 2017, \nat, 551, 71,
  \dodoi{10.1038/nature24290}

\bibitem[{{Troja} {et~al.}(2018){Troja}, {Piro}, {Ryan}, {van Eerten}, {Ricci},
  {Wieringa}, {Lotti}, {Sakamoto}, \& {Cenko}}]{Troja18}
{Troja}, E., {Piro}, L., {Ryan}, G., {et~al.} 2018, \mnras, 478, L18,
  \dodoi{10.1093/mnrasl/sly061}

\bibitem[{{Troja} {et~al.}(2021){Troja}, {O'Connor}, {Ryan}, {Piro}, {Ricci},
  {Zhang}, {Piran}, {Bruni}, {Cenko}, \& {van Eerten}}]{Troja21}
{Troja}, E., {O'Connor}, B., {Ryan}, G., {et~al.} 2021, arXiv e-prints,
  arXiv:2104.13378.
\newblock \doarXiv{2104.13378}

\bibitem[{{Utsumi} {et~al.}(2017){Utsumi}, {Tanaka}, {Tominaga}, {Yoshida},
  {Barway}, {Nagayama}, {Zenko}, {Aoki}, {Fujiyoshi}, {Furusawa}, {Kawabata},
  {Koshida}, {Lee}, {Morokuma}, {Motohara}, {Nakata}, {Ohsawa}, {Ohta},
  {Okita}, {Tajitsu}, {Tanaka}, {Terai}, {Yasuda}, {Abe}, {Asakura}, {Bond},
  {Miyazaki}, {Sumi}, {Tristram}, {Honda}, {Itoh}, {Itoh}, {Kawabata},
  {Morihana}, {Nagashima}, {Nakaoka}, {Ohshima}, {Takahashi}, {Takayama},
  {Aoki}, {Baar}, {Doi}, {Finet}, {Kanda}, {Kawai}, {Kim}, {Kuroda}, {Liu},
  {Matsubayashi}, {Murata}, {Nagai}, {Saito}, {Saito}, {Sako}, {Sekiguchi},
  {Tamura}, {Tanaka}, {Uemura}, \& {Yamaguchi}}]{Utsumi17}
{Utsumi}, Y., {Tanaka}, M., {Tominaga}, N., {et~al.} 2017, \pasj, 69, 101,
  \dodoi{10.1093/pasj/psx118}

\bibitem[{{Valenti} {et~al.}(2017){Valenti}, {Sand}, {Yang}, {Cappellaro},
  {Tartaglia}, {Corsi}, {Jha}, {Reichart}, {Haislip}, \&
  {Kouprianov}}]{Valenti17}
{Valenti}, S., {Sand}, D.~J., {Yang}, S., {et~al.} 2017, \apjl, 848, L24,
  \dodoi{10.3847/2041-8213/aa8edf}

\bibitem[{{Villar} {et~al.}(2017){Villar}, {Guillochon}, {Berger}, {Metzger},
  {Cowperthwaite}, {Nicholl}, {Alexander}, {Blanchard}, {Chornock},
  {Eftekhari}, {Fong}, {Margutti}, \& {Williams}}]{Villar17}
{Villar}, V.~A., {Guillochon}, J., {Berger}, E., {et~al.} 2017, \apjl, 851,
  L21, \dodoi{10.3847/2041-8213/aa9c84}

\bibitem[{{Williams} {et~al.}(2011){Williams}, {de Geus}, \&
  {Blitz}}]{clumpfind}
{Williams}, J.~P., {de Geus}, E.~J., \& {Blitz}, L. 2011, {Clumpfind:
  Determining Structure in Molecular Clouds}.
\newblock \doeprint{1107.014}

\bibitem[{{Wu} \& {MacFadyen}(2018)}]{Wu18}
{Wu}, Y., \& {MacFadyen}, A. 2018, \apj, 869, 55,
  \dodoi{10.3847/1538-4357/aae9de}

\bibitem[{{Wu} \& {MacFadyen}(2019)}]{Wu19}
---. 2019, \apjl, 880, L23, \dodoi{10.3847/2041-8213/ab2fd4}

\bibitem[{{Ye} {et~al.}(2020){Ye}, {Fong}, {Kremer}, {Rodriguez}, {Chatterjee},
  {Fragione}, \& {Rasio}}]{Ye20}
{Ye}, C.~S., {Fong}, W.-f., {Kremer}, K., {et~al.} 2020, \apjl, 888, L10,
  \dodoi{10.3847/2041-8213/ab5dc5}

\end{thebibliography}
